# Polarity Free Magnetic Repulsion and Magnetic Bound State

**Hamdi Ucar**

Independent Researcher, Istanbul Turkey, *jxucar@gmail.com*

This is a report on a dynamic autonomous magnetic interaction which does not depend on polarities resulting in short ranged repulsion involving one or more inertial bodies and a new class of bound state based on this interaction. Both effects are new to the literature, found so far. Experimental results are generalized and reported qualitatively. Working principles of these effects are provided within classical mechanics and found consistent with observations and simulations. The effects are based on the interaction of a rigid and finite inertial body (an object having mass and moment of inertia) endowed with a magnetic moment with a cyclic inhomogeneous magnetic field which does not require to have a local minimum. Such a body having some degrees of freedom involved in driven harmonic motion by this interaction can experience a net force in the direction of the weak field regardless of its position and orientation or can find stable equilibrium with the field itself autonomously. The former is called polarity free magnetic repulsion and the latter magnetic bound state. Experiments show that a bound state can be obtained between two free bodies having magnetic dipole moment. Various schemes of trapping bodies having magnetic moments by rotating fields are realized as well as rotating bodies trapped by a static dipole field in presence of gravity. Also, a special case of bound state called bipolar bound state between free dipole bodies is investigated.

## 1 Introduction

Things can be bound classically by force fields they possess and inertial forces balancing them. Despite some force fields have attractive and repulsive actions, it is not possible to obtain stable equilibrium through force fields alone as proven by Earnshaw's theorem. Therefore classically, everything to be bound should have some dynamics in order to incorporate inertial forces. The common mechanism is the orbital motion and can explain bound states of celestial bodies. Orbital bound state is also possible within electrically charged bodies and used in Bohr model of the atom based on inverse square law, but stability cannot be obtained within magnetic forces where their dependence to the distance is out of range of the power figure to obtain stable orbital motion within the central force problem [1]. This precludes orbital bound state of nucleons based on magnetic forces. While classical physics fails historically to propose a binding mechanism of nucleons, nuclear forces and additional mechanisms are introduced as part of quantum physics addressing this problem. The orbital motion based on dipole-dipole interaction was later investigated by several authors including Kozorez [2], Shironosov [3].

Here is introduced a new mechanism to obtain bound state of entities having magnetic moment, using again inertial behaviors but in a different manner. It is based on magnetic forces, using attraction force to keep entities having magnetic moments together and balance this attraction with dynamically created repulsive magnetic action which might be classified as a force too. This repulsive force or the interaction is called *polarity free magnetic repulsion* (PFR) and the resulting bound state as *magnetic bound state* (MBS). Although it is possible to evaluate MBS as a dynamic equilibrium based on magnetic interaction directly without introducing PFR, generalization might be difficult. MBS might be considered as versatile because it allows entities having mass, moment of inertia and magnetic dipole moment to be bound without precise requirements. These entities can be compact bodies; that is, one body is not surrounded by the other.

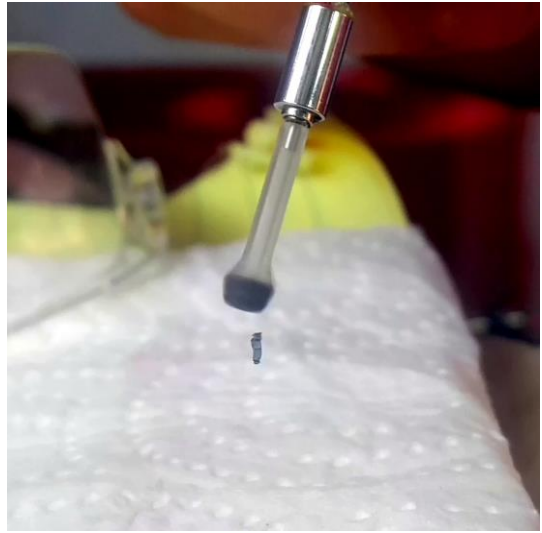

Fig. 1.1. A scaled-down MBS realization where a tiny fragment of a NdFeB magnet weighing about 5 mg is levitated at the tip of rotator assembly consisting of an irregular magnet fragment of ~3 mm in size coupled to a micro motor having dimensions ⌀4×10 mm running at 105000 RPM (11000 rad/s).

Dynamically created repulsive action based on magnetism this way has a short range; that is, it diminishes about twice as fast by distance compared to magnetic forces between dipoles. This allows to obtain stable equilibrium between PFR and magnetic or other attractive forces in various schemes and configurations. While the mathematical model of the PFR is applied to specific configurations in this work, experiments show it is present regardless of magnetic orientations of magnetic moments of entities. Stability and other characteristics of PFR and MBS are evaluated analytically and by simulations for basic cases. These cases are extended experimentally with numerous schemes and configurations, basically by bounding two permanent magnets in terrestrial environment without a physical contact, except the ambient air. It was not possible to test the bound state of two free magnets truly (in free fall) in terrestrial environment but performed in an approximated manner. On the other hand, gravity was helpful to test the strength of bound states and some configurations allowed to obtain bound state regardless of the direction of gravity. MBS of two free bodies is also evaluated through simulations and gives confidence that it can as well be experimentally realized in space and under friction free conditions. Scalability of MBS is tested experimentally by varying mass of the trapped magnets from 0.01 to 100 g in various configurations. It should be noted that for the bound state of two free bodies, selection of moment of inertia of one body have some limitations.

A known method for trapping magnetic bodies with cyclic fields [4] is based on parametric excitation and this way, the body can be kept at the local minimum of a quadrupole field. This solution is equivalent to well-known rolling marble on a rotating saddle and its dynamics can be modeled by the Mathieu equation [5]. This model is also used to trap charged particles [6].

It is also possible to trap magnetic bodies at local minimum of static or quasi-static quadrupole fields while trying to keep bodies in correct orientation in order they are pushed in the direction of the weak field. This is the main mechanism for trapping neutral atoms and particles [6, 7]. Since magnetic moments of these bodies are associated with quantum spin, they can keep their orientations parallel or anti-parallel to the field they are exposed. However, their aimed orientations can be lost or get flipped if the field becomes zero or too weak at the local minimum. This issue is addressed with different schemes. In a recently proposed scheme [8] to trap neutral particles using rotating magnetic field of an electromagnetic wave, the rotating field has a similar role of the rotating electric field in the Paul trap [6]. Solutions obtained there might be also related to [9].

Another scheme is known as Levitron® [10, 11, 12] and a variation called Horizontal Levitron [13]. In the original Levitron, a rotating dipole body around its dipole axis is levitated by a repulsive magnetic interaction having positive stiffness balanced by gravity and negative stiffness introduced in angular degrees of freedom is balanced by the positive dynamic stiffness caused by the rotation and precession of the body. In the horizontal axis Levitron, the body is trapped between multiple dipole magnets. In this solution, precession of the top has a similar role. A detailed analysis of this solution can be found here [14]. Superconductivity also allows to trap objects or obtain bound state directly by handling magnetic field belonging to the object known as flux pinning; however, classical physics does not cover this phenomenon.

Another mechanism to induce thrust using cyclic fields is ponderomotive force. With this scheme, it is possible to induce a net force in direction of the weak field on electrically charged particles regardless of their polarities by exposing them to an alternative electric field having a gradient. This is also a parametric excitation mechanism where a characteristics of the driven harmonic motion (DHM) called phase lag is present and the net force is produced while the particles oscillates through the gradient. Phase lag is the condition where the action and reaction are in opposite phases in a periodic excitation. An example of this in mechanics is a mass spring system driven through the spring. This mechanism also constitutes the principle of bass reflex speakers. While PFR and ponderomotive force have some similarities, the acted force in ponderomotive force has alternating cycle and a body needs to oscillate along the applied force direction but PFR can be obtained by a static interaction (including fields and forces) and within the local reference frame. PFR and MBS does not depend also on local minimum since they can be obtained between dipoles. In PFR, the repulsive force is basically produced by "antiparallel" orientation of a free body with an inhomogeneous magnetic field. Here the term antiparallel denotes an orientation where the angle between two vectors is greater than $\pi/2$. While such an orientation is unstable in magnetostatics, it can be stable within DHM under phase lag condition.

The way to obtain PFR is exposing an inertial body embedded with a magnetic moment to an inhomogeneous rotating field or by spinning such a body which creates a rotating field and exposing it to an inhomogeneous static field. PFR might be explained in the basic case by two steps in a simpler way. The first step is obtaining a suitable and stable angular motion of the body induced by the torque of the rotating field through DHM. As a characteristics of the harmonic motion, the phase of the motion could be in the opposite phase of the driving field, called *phase lag* condition. This allows to keep the magnetic orientation of the body in some antiparallel angle (>90°) with the field. Under proper conditions, body's motion can be fully synchronized with the field such that the system becomes static in the co-rotating reference frame. The angular motion is also accompanied by a synchronized small translational (lateral) motion which shifts the position on the co-rotating reference frame off the rotation axis in the direction favoring antiparallel orientation again governed by the phase lag.

In the second step, body experiences a magnetic force due to the gradient of the field. As the body orientation has an antiparallel angle, this force would have a repulsive component in direction of the axis (*z*) of the rotating field. In this two-step evaluation, the effect of translational motion on the torque is omitted. This approximation might be acceptable since the translational motion is absent on axis-*z* and the lateral motion has typically a small amplitude. Regarding experiments, the effect of the lateral translational motion is a shift of the center of the angular motion from the CM. In addition to this "antiparallel" scheme, body can also be in motion relative to the field, generating time dependent torques and forces. The phase lag mechanism can be also effective here and a net repulsive force can be induced similar to ponderomotive force. Experiments show that translation from one scheme to another by simply changing orientation or position of the body relative to the field is continuous; that is, the repulsion is sustained.

In MBS, there is an additional static field component, typically symmetric about the axis of the rotating field, providing the counteracting attractive or binding force. This can be obtained by rotating a dipole in an axis nearly orthogonal to the dipole axis. This deviation from orthogonal is called tilt angle and forms a time averaged dipole along the rotation axis which provides the attraction force with the logic of two parallel aligned dipoles. It is possible to obtain stable equilibrium between this attraction force and PFR, because the slope of the repulsion is steeper than the attraction at their cross point. In stiffness terms, PFR has positive stiffness and when combined with other force fields, it renders the total stiffness positive. Another effective scheme to obtain the binding force is to rotate a dipole on an axis orthogonal to the dipole axis while offsetting the dipole center from the rotation axis.

Another criterion about bound state is conservation of energy. Basically, a bound state should be based on non-dissipative interactions. This requires the bound state to be stable without damping. This stability is shown in simulations of motion of the free body exposed to a homogenous and then to inhomogeneous rotating fields and by experiments achieving bound state with magnets. Magnetic bound state is also found stable under external disturbances.

It is also observed that bodies under bound state or in a trapping solution can undergo complex angular motions with large amplitudes, whilst still conserving the stability. In some of these experiments, the magnetic moment vector of the free body spans almost a hemisphere and the motion exhibits some remarkable features. These features might extend the range of solutions that can be obtained using PFR and MBS.

In summary, PFR is a net force acting on an inertial body endowed with a magnetic moment and subjected to an inhomogeneous cyclic field in the direction of the weak field regardless of its position and its orientation. PFR has positive stiffness and stable equilibrium can be reached by combining it with static force fields provided externally or by components of the cyclic field itself. The latter is called magnetic bound state, abbreviated as MBS. Stable equilibriums based on magnetic interactions that do not fit to the bound state definition are also obtained experimentally in this work and presented as trapping solutions where PFR may be present.

The PFR is explained in Section 5 by the phase relation of the forced harmonic motion which favors antiparallel alignment of a free magnetic body with the driving field. This motion is formulated according to the experimental observations and verified by simulations. Stability and other characteristics of the PFR in various configurations observed in experiments are also obtained in simulations and detailed. This section also covers various trapping solutions realized in experiments. The MBS is covered in Section 6 where evaluations of experiments and simulations results can be found. Experiments also cover bound state of two free bodies in approximated realizations and simulations of bound state of two free dipole bodies. Various and complex motion characteristics of trapped bodies within experiments are also covered in this section.

## 2 Terms and definitions

Term *inertial body* or briefly *body* used in this work specify a physical rigid and compact object having finite dimensions, having mass and moment of inertia (MoI). It is usually endowed with a magnetic moment and having degrees of freedom (DoF).

Term *floating body* (*floator* for short) specifies an inertial body usually kept in the air as a result of magnetic interactions. It can be called *floating magnet* when the body consists of a permanent magnet. In detail, a floating body is required to have a mass, a MoI, a dipole or quadrupole moment and have full DoF. This can be simply a permanent magnet or a rigid assembly containing one or more such magnetic components. Body having full DoF is called also *free body.*

*Actuator magnet* is typically a rotating magnet or a compact magnetic assembly attached to a rotor. It can be called *actuator body* when its inertial properties play roles in case of presence of DoF. It can also be called *rotator* for short.

*Bound state* in its general meaning is the state of a body bound to another body/entity by a force field. When inertial properties of the entity have no relevance to the interaction, the interaction can be defined directly between the body and the force field associated with the entity. This allows to define a bound state between a body having full DoF and another body having limited DoF. In our context, these bodies should be compact, i.e. one body should not be surrounded by the other body, allowing to separate them by a straight plane. A bound state might comprise multiple free bodies resulting in a structure. A stable bound state should not be broken with infinitesimal work. External work required to break the bound state defines the binding energy as equal amount but having negative sign. In some circumstances, the work that can weaken or break the bound state can be supplied internally. In such a case, bound state can be categorized as quasi-stable.

Term *spin* describes a rotation of an object around itself.

# 3   Overview of realizations and experiments

Realizations consist mainly of interactions of rare earth (NdFeB) permanent dipole magnets of various sizes, shapes and strengths commercially available. These magnets usually have magnetic misalignments and inhomogeneities besides variations of strength and these factors are taken into account. The actual dimensions of most of the magnets (including the Ni coating) are found to be 0.12±0.05 mm less than their advertised dimensions and these were noted in experiments specifications. Arrangements of magnets are also used for obtaining desired field profiles and for obtaining quadrupoles and other field geometries. In realizations of interaction of two bodies having full DoF, the freedom of one body is approximated by mechanical methods in order to prevent its free fall under gravity.

Effects are obtained by following schemes:

- By interaction of a free body having dipole moment with a cyclic inhomogeneous magnetic field based on a rotating dipole moment.
- By interaction of a free body having quadrupole moment with a cyclic inhomogeneous magnetic field based on a rotating dipole moment.
- By interaction of a free body having a cyclic field with a static inhomogeneous magnetic field based on rotating a dipole or a quadrupole moment.
- By interaction of two free bodies, at least one of them has to generate a rotating inhomogeneous field based on a dipole moment.
- By interaction of a free body having dipole moment with two or more rotating dipole fields.
- By interaction of a free body having dipole moment with a cyclic field having quadrupolar component.

These interactions cause cyclic forces and torques and can accelerate bodies in cyclic manner. In all schemes, cyclic fields are obtained by rotation of a magnetic moment. As a result of these interactions, dipoles or quadrupoles experience cyclic forces and torques and at least a body involves in a cyclic motion. The cyclic field can also have a static component which is required for the MBS.

- In realizations, magnetic fields are obtained by using dipole permanent rare earth magnets of NdFeB types.
- The condition called bound state in realizations in this work allows two parties to be separated by a gap where a plane can be passed between.
- Realized bound states have various strengths and stiffness and can anticipate static and dynamic external forces.
- Bodies or force fields may be involved in multiple concurrent interactions.
- Presence of gravitational force is considered as an external disturbing force acting on bodies under bound state except in some cases it contributes to the stability. The magnetic field of the Earth is not isolated although it is found to have no effect on obtaining results. Some experiments involve obtaining stable equilibrium of a free body through multiple environments like liquids of various density and viscosities and while the environment is varied dynamically.

In details, experiments can be categorized as:
- Experiments showing basic properties of polarity free magnetic repulsion obtained between a cyclic field and a body having a magnetic moment. The body may have full or partial DoF and may be attached to a support in various ways.
- Experiments on polarity free magnetic repulsion where a body with full DoF is trapped by the field of one or multiple rotating dipoles and optionally by inclusion of a static dipole field. Gravity assistance is also required in some configurations.
- Experiments on magnetic bound state where the body with full DoF is trapped by the field of rotating dipole. In some realizations, the strength of the bound state is enough to anticipate force of the gravity regardless of its direction and in some other realizations, gravity direction should be in a given range. Bodies possess a dipole moment and additionally may have a quadrupole moment.
- Experiments approximating bound state of two free bodies having dipole moments in presence of gravity.

- Experiments trapping a body having a dipole moment and full DoF by a static magnetic field and with assistance of gravity.
- Experiments trapping a free body endowed with combined dipole and quadrupole moments by a static magnetic field and with assistance of gravity.
- Experiments realizing bound state condition where bound state consists of one body with full DoF and multiple rotating dipoles.
- Experiments realizing bound state condition and exposing some aspects of the interaction and optionally by an inclusion of external static fields, which can improve the stability, can extend the range of the bound state or compensate the gravitational force.

Observations are made directly without using instruments or with additional basic tools like ruler, weighing scale, tachometer, and measurement tools over photographic images. Stroboscopic recordings and laser beam tracing methods are also used for observing oscillation patterns and phase figures.

The working principles of these effects are given within classical mechanics and magnetism and found consistent with observations. Other mechanisms which can cause repulsion in presence of magnetic field like diamagnetism, induction based repulsion are ruled out. Even if they are present, their strengths would be one or more orders of magnitude weaker for obtaining the observed effects.

Experimental results are evaluated under classical magnetism and mechanics. Analyses are mostly qualitative, pointing to factors playing roles and specifying required conditions to allow generalizations or particularization. Configurations of experiments are mostly specified in related figure captions.

## 4 Force and torque between magnetic dipoles

Basic equations about magnetism used in this work according [15] are given below. Vectors are denoted by arrow notation or by bold characters, unit vectors by circumflexes, scalar values by plain characters. Magnetic field of a dipole moment $\vec{m}$ at a relative point defined by a vector $\vec{r}$ reads

$$\vec{B}(r) = \frac{\mu_0}{4\pi r^3}[3(\vec{m}\cdot\hat{r})\hat{r} - \vec{m}] \tag{4.1}$$

Force $F$ and torque $\tau$ acting on dipole moment $\vec{m}_b$ by another dipole moment $\vec{m}_a$ can be expressed as

$$\vec{F}_{ab} = \frac{3\mu_0 m_a m_b}{4\pi r^4}[\hat{r}(\hat{m}_a\cdot\hat{m}_b) + \hat{m}_a(\hat{r}\cdot\hat{m}_b) + \hat{m}_b(\hat{r}\cdot\hat{m}_a) - 5\hat{r}(\hat{r}\cdot\hat{m}_a)(\hat{r}\cdot\hat{m}_b)] \tag{4.2}$$

$$\vec{\tau}_{ab} = \frac{\mu_0 m_a m_b}{4\pi r^3}[3(\hat{m}_a\cdot\hat{r})(\hat{m}_b\times\hat{r}) + \hat{m}_a\times\hat{m}_b] \tag{4.3}$$

where $\mu_0$ is the permeability of vacuum equal to $4\pi\times 10^{-7}$ T·m/A, $\vec{m}_a, \vec{m}_b$ being dipole moments which can be expressed as A·m$^2$ or as N·m/T and $\vec{r}$ the vector going $a$ to $b$, in meter units. From these equations, it can be seen the force between two point dipoles is inversely proportional to fourth power to their distance and the torque is inversely proportional to third power. Another referenced equation is

$$\vec{\tau} = \vec{m}\times\vec{B} \tag{4.4}$$

which gives the torque $\vec{\tau}$ that a magnetic moment $\vec{m}$ receives from a magnetic field $\vec{B}$ This relation can be written in scalar form as

$$\tau = m B \sin\phi \tag{4.5}$$

where $\phi$ is the angle between vectors $\vec{m}$ and $\vec{B}$ and plain symbols represent magnitudes of entities. In this work, distances between dipoles are close to dipole size and force/torque calculations based on point dipole approximation can have significant deviations from real figures. It might be possible to reduce these deviations by varying power factors of the distance on related equations within ranges. This method is used for obtaining better analytical figures and matching them to FEM based simulation results. On the other hand, point dipole approximation is found sufficient to obtain models of evaluated interactions and above power factors are not critical since most experimental results are evaluated quantitatively. The consistency between analytical, simulation and experimental results are checked also using spherical magnets whenever possible, since interactions of a uniformly magnetized spherical magnet can be modeled as interaction of a point dipole according to literature [16, 17].

# 5  Polarity free magnetic repulsion

Called here as *polarity-free magnetic repulsion* (PFR) can be considered as a basic novel effect generating a net force on a body having a magnetic moment and having some DoF exposed to a cyclic (alternating) magnetic field having a gradient. The generated force is in the direction of the weak field, as the name implies. If the cyclic field belongs to a dipole, the body is repelled from this dipole. It is found that the effect is present regardless of the body's magnetic orientation. It is also found that the effect is present regardless of the angle between the field vector and its gradient vector. If the field belongs to an alternating or a rotating dipole, the body is repelled from the dipole regardless of its position.

PFR has some similarities with ponderomotive force where a body having electrical charge is repelled from an alternating electric field having gradient. While the ponderomotive force depends on translational motion in direction of the generated force, PFR depends mainly on angular motion. Both effects are based on harmonic motion and depend on a condition called *phase lag* where acceleration and displacement are in opposite phases. This is essentially a trigonometric property where the second derivatives of sine or cosine functions are themselves with opposite signs. While ponderomotive force requires monopoles therefore only possible through Coulomb interaction, PFR can be obtained effectively with dipole-dipole or dipole-quadrupole interactions, available magnetically and possibly electrically (Coulomb based interactions are not evaluated in this work). The repulsion force arises through the angular motion because this motion keeps the dipole instantaneous orientation of body in an antiparallel angle with the rotating field due to phase lag (Fig. 5.1(R), Fig. 5.13(a)).

Such an antiparallel orientation is unstable under magnetostatic interactions but can be stable under DHM. Translational oscillation is also present, typically lateral and in small amplitudes, this also contributes to PFR. This characteristics allows to keep a body in a precise location in a stable equilibrium by balancing PFR by a counteracting force. In DHM, the above mentioned phase lag condition requires the driving frequency $\omega$ be above the frequency that system may oscillate in absence of the driving force when a recoiling factor is present. This frequency is called natural frequency $\omega_0$ and for a linear system, it can be calculated as

$$\omega_0 = \sqrt{C/Z} \quad (5.1)$$

where $C$ is the spring constant and $Z$ is the placeholder for inertial property which denotes mass for translational motion and MoI for angular motion. The equation governing a simple DHM can be written as

$$Z\ddot{x} + \beta\dot{x} + Cx + A\cos\omega t = 0 \quad (5.2)$$

where $x$ is the motion variable, $\beta$ the damping factor and term $A$ the amplitude of the driving force or the torque, depending whether the motion is translational or angular.

Another driving mechanism is called parametric excitation and in its simplest linear form it can be written as

$$Z\ddot{x} + \beta\dot{x} + (C_0 + C_1 \cos\omega t)x = 0 \quad (5.3)$$

It is called parametric excitation because the spring constant parameter $C$ in Eq. 5.2 is time varying in the above equation where its value oscillates around $C_0$ with amplitude $C_1$. Parametric excitation in a nonlinear form is present angular driving mechanism of PFR and has a role on stability of the motion but not on generation of the repulsion force except in presence of longitudinal motions. That is, when the motion has a component in direction of the generated force.

The basic configuration of the PFR is shown in Fig. 5.1. In this configuration, everything is static in a reference frame co-rotating with the driving field. Rotation center (RC) of the motion coincides with the center of mass (CM) when rotating field is homogenous and CM coincides with the dipole center (Fig. L). When magnetic forces are present (Fig. R), CM draws a small circle around the rotation axis in synchronized with the field and this shifts the RC on axis-$z$ in direction of the weak field but still everything is static in the co-rotating reference frame.

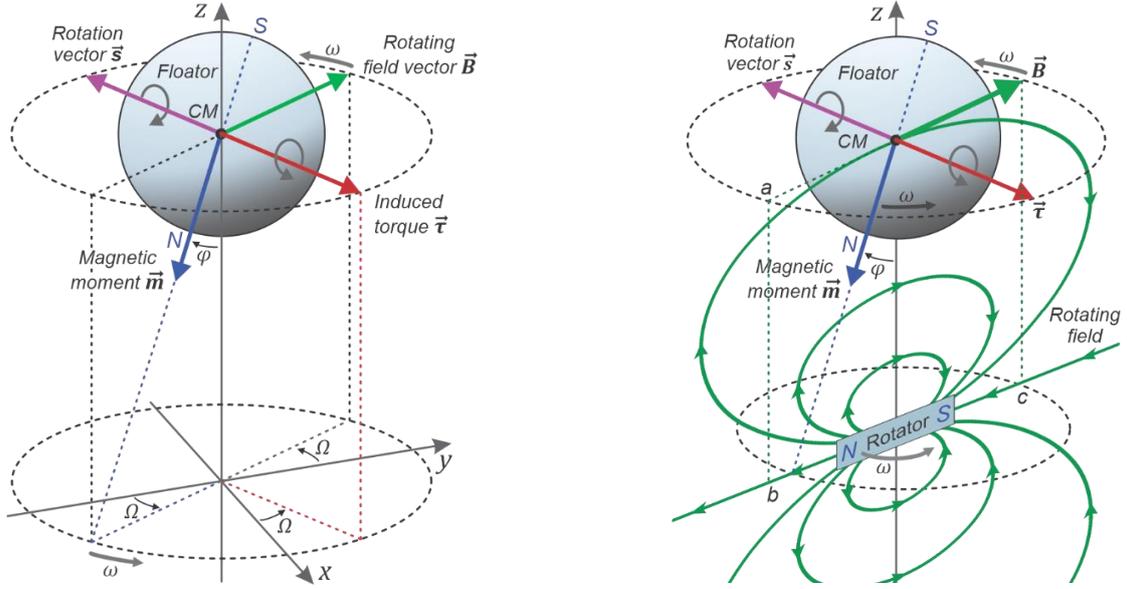

Fig. 5.1. (**L**) Motion of a free body (floator) endowed with a magnetic dipole moment subject to uniform rotating magnetic field around axis-$z$ with angular velocity $\omega$. Within suitable configuration parameters and initial conditions, it is expected that the floator draws a conical angular motion around the axis-$z$, synchronized with the rotating field. The torque experiences the floator can be calculated as $\vec{\tau} = \vec{m} \times \vec{B}$. According the model of the motion, the rotation vector $\vec{s}$ of the body becomes antiparallel to torque vector $\vec{\tau}$ which share the same unit vector with angular acceleration vector $\vec{\alpha}$. The angle $\varphi$ is the result of the dynamics and can have a constant value with suitable initial conditions. This value can be calculated according to the model (Eq. 5.24) and initial angular velocities using Eq. 5.12, ($t = 0$). Under this conical motion, azimuthal angles of vectors $\vec{m}, \vec{s}, \vec{B}$ and $\vec{\tau}$ can be expressed as $\omega t + A$ where $A_\tau = 0, A_B = \pi/2$, $A_s = \pi$, $A_m = -\pi/2$. Symbol $\Omega$ denotes $\omega t$. It is remarkable that this relation is forced by the DHM even these angles at initial condition are different. In this diagram, the torque vector is shown on the $xy$ plane; that is, it has no $z$ component. This is valid when azimuthal angle between **B** and **m** is zero or $\pi$. Otherwise, this component get a value proportional to sine of this angle. (**R**) Similar configuration but floator is exposed to inhomogeneous field of a rotating dipole (rotator) centered on axis-$z$. This ensures the vector B at the CM point rotates on $xy$ plane. Here it can be seen the floator pole *N* deviates from the axis-$z$ toward rotator pole *N* and can cause a repulsion between dipoles. It should be noted that under the inhomogeneous field of the rotator, the rotation center is not at CM since the body also involves in a translational motion on the $xy$ plane where the CM follows a small circle around axis-$z$. This shifts the RC away from rotator on axis-$z$. Rotator's force lines are shown on the plane defined by its dipole orientation (cb) and the axis-$z$. The characteristics of the motion and obtained torque do not depend on rotation direction of the rotating field.

In the following, the model of PFR and its analysis are proceeded in two steps. First step only takes account of the magnetic torque of the interaction and the model is based on the motion of a body having a magnetic dipole moment subject to a homogeneous rotating magnetic field. The second step analyzes magnetic forces the body receives by its interaction to the rotating magnetic dipole which also causes the torque in the first step. While the torque induced by homogeneous fields and by dipole fields have differences, the special alignment of the body with the rotating field in basic configurations makes this difference do not invalidate the first analysis. However, the two step analyses may not be applied generally for covering all experimental results where PFR and MBS are present. On the other hand, as all experimental observations have some common characteristics covered in these analyses, these characteristics are linked to the given model.

## 5.1 Evaluation of motion of a body subject to homogeneous rotating field

Here, motion of an inertial body embedding a magnetic dipole moment and having angular DoF subject to a rotating homogeneous magnetic field is evaluated. This covers some basic configurations common to experiments. Later, this homogeneous field is replaced by an inhomogeneous one in order to incorporate forces.

Due to complexities to express and obtain the solution of a differential equation system driven by this interaction, a reverse path is followed in the following such as choosing a model for the motion (the solution) by the guidance of experimental observations and validate this model by matching the expected motion to the defined magnetic interaction trough Euler's equations of motion.

The angular motion of a body subject to a rotating torque vector with constant amplitude and velocity around axis-$z$ and orthogonal to it can be described as a conical motion which having similarity with the motion of conical pendulum drawing a circular path. Angular motion of the pendulum arm should be considered for this similarity. In both of motions, under equilibrium, the vector defining the orientation of the body and the pendulum arm can rotate around axis-$z$ with fixed zenith angle and with fixed azimuthal velocity. The motion also includes the rotation of the pendulum blob on the axis of its arm and similarly, the rotation $\rho$ of the body around its orientation vector. This angular motion can be expressed by three Euler rotations ($\theta\ \varphi\ \rho$) using ZXZ convention. Visualizing a body as a soda can (a cylinder with distinguishable sides) where its orientation vector is defined as the axis of the cylinder, if one wants to keep same side of the can looking to a specific direction in accordance with experiments, it is needed to provide a third rotation with $\rho$ equal to $-\theta$, otherwise without $\rho$, a side will always face to axis-$z$. These three rotations and the Euler rotation matrix $R_E$ with abbreviations ($s$, $c$) for sine and cosine terms reads

$$R_\theta = \begin{bmatrix} c_\theta & -s_\theta & 0 \\ s_\theta & c_\theta & 0 \\ 0 & 0 & 1 \end{bmatrix}, \qquad R_\varphi = \begin{bmatrix} 1 & 0 & 0 \\ 0 & c_\varphi & -s_\varphi \\ 0 & s_\varphi & c_\varphi \end{bmatrix}, \qquad R_\rho = \begin{bmatrix} c_\rho & s_\rho & 0 \\ -s_\rho & c_\rho & 0 \\ 0 & 0 & 1 \end{bmatrix} \tag{5.5}$$

$$R_E = R_\theta R_\varphi R_\rho = \begin{bmatrix} c_\theta c_\rho - s_\theta s_\rho c_\varphi & -c_\theta s_\rho - s_\theta c_\rho c_\varphi & s_\theta s_\varphi \\ s_\theta c_\rho + c_\theta s_\rho c_\varphi & c_\theta c_\rho c_\varphi - s_\theta s_\rho & -c_\theta s_\varphi \\ s_\rho s_\varphi & c_\rho s_\varphi & c_\varphi \end{bmatrix} \tag{5.6}$$

By replacing $\theta$ by $\omega t$ and $\rho$ by $-\omega t$ in $R_E$ where $\omega$ is a real number and $t$ denotes the time, we obtain the rotation matrix of the actual motion as

$$R = \begin{bmatrix} c_{\omega t}^2 + s_{\omega t}^2 c_\varphi & c_{\omega t} s_{\omega t}(1 - c_\varphi) & s_{\omega t} s_\varphi \\ c_{\omega t} s_{\omega t}(1 - c_\varphi) & s_{\omega t}^2 + c_{\omega t}^2 c_\varphi & -c_{\omega t} s_\varphi \\ -s_{\omega t} s_\varphi & c_{\omega t} s_\varphi & c_\varphi \end{bmatrix} \tag{5.7}$$

Rotations on $x$, $y$ and $z$ axes can be obtained by the operation

$$[\vec{X}]_\times = \frac{R - R^T}{2} = \begin{bmatrix} 0 & 0 & s_{\omega t} s_\varphi \\ 0 & 0 & -c_{\omega t} s_\varphi \\ -s_{\omega t} s_\varphi & c_{\omega t} s_\varphi & 0 \end{bmatrix} \tag{5.8}$$

where $[\,.\,]_\times$ denotes skew symmetric operator. Resulting vector $X$ describes the rotation of the body, where the unit vector is the axis of the rotation and its magnitude is sine of rotation amount.

$$\vec{X} = s_\varphi \begin{bmatrix} c_{\omega t} \\ s_{\omega t} \\ 0 \end{bmatrix} \tag{5.9}$$

This shows that at any time, the body rotates around an axis rotating around axis-$z$ and orthogonal to it with an angle equal to $\varphi$. The vector $X$ is sketched as $\vec{s}$ at Fig. 5.1 in that configuration. Time dependent angular velocity vector $\vec{v}(t)$ of the motion can be obtained by multiplying time derivative of $R$ with its transpose as

$$R'R^T = S = \omega \begin{bmatrix} 0 & c_\varphi - 1 & c_{\omega t} s_\varphi \\ 1 - c_\varphi & 0 & s_{\omega t} s_\varphi \\ -c_{\omega t} s_\varphi & -s_{\omega t} s_\varphi & 0 \end{bmatrix} \tag{5.10}$$

It should be noted that in the time derivation of $R$, the angle $\varphi$ is assumed constant. Resulting skew-symmetric matrix $S$ contains the angular velocity components according skew symmetric operator $[\,.\,]_\times$

$$[\vec{v}]_\times = S = \begin{bmatrix} 0 & -\omega_3 & \omega_2 \\ \omega_3 & 0 & -\omega_1 \\ -\omega_2 & \omega_1 & 0 \end{bmatrix} \tag{5.11}$$

This way, angular velocity vector $\vec{v}$ and the angular acceleration vector $\vec{a}$ derived from it, read

$$\vec{v} = \omega \begin{bmatrix} -s_{\omega t} s_\varphi \\ c_{\omega t} s_\varphi \\ 1 - c_\varphi \end{bmatrix} \tag{5.12}$$

$$\vec{a} = \vec{v}' = \omega^2 s_\varphi \begin{bmatrix} -c_{\omega t} \\ -s_{\omega t} \\ 0 \end{bmatrix} \tag{5.13}$$

Here, it can be seen the vector $\vec{a}$ is antiparallel to the rotation vector defined in Eq. 5.9 and have constant magnitude equal to $\omega^2 s_\varphi$. This allows to obtain the angle $\varphi$ if $\vec{a}$ is known. This angular acceleration should be induced by the torque the body experiences. If the body has uniform MoI $I$, the relation can be defined by Newton's second law

$$\vec{\tau} = I \vec{a} \tag{5.14}$$

Otherwise Euler's equations need be used

$$\vec{\tau} = \mathbf{I_T}\,\vec{a} + \vec{v} \times \mathbf{I_T}\,\vec{v}, \qquad \mathbf{I_T} = R\,\mathbf{I}\,R^T \tag{5.15}$$

where $\mathbf{I}$ denotes the tensor of inertia of the body and $\mathbf{I_T}$ is its derivation in the global coordinate system. Using the angular velocity vector $\vec{u}$ valid in body fixed coordinate system, Euler's equations can be written in component form as

$$\begin{aligned}
\tau_1 &= I_1 u_1' + (I_3 - I_2) u_3 u_2 \\
\tau_2 &= I_2 u_2' + (I_1 - I_3) u_1 u_3 \\
\tau_3 &= I_3 u_3' + (I_2 - I_1) u_2 u_1
\end{aligned} \tag{5.16}$$

where terms $I_{1,2,3}$ denote principal moments of inertia on numbered axes. We consider first a case where the torque is independent of the orientation of the body. This cannot be applied to magnetic interactions; however, could be useful for evaluating basic characteristics of the motion. According Eq. 5.14, the torque and the angular acceleration should have a common unit vector. Eq. 5.13 satisfy the postulated requirements of constant magnitude, constant angular velocity and orthogonality to axis-$z$. This way, a torque vector can be defined as

$$\vec{\tau}(t) = \tau \begin{bmatrix} -c_{\omega t} \\ -s_{\omega t} \\ 0 \end{bmatrix} \tag{5.17}$$

By substituting $\vec{a}$ from Eq. 5.13 in Eq. 5.14, we obtain the torque coefficient $\tau$ defined above as

$$\tau = I \omega^2 s_\varphi \tag{5.18}$$

where the angle $\varphi$ can be calculated and gives the torque upper limit in terms of $I$ and $\omega$. Series of simulations corresponding to a range of $\varphi$ between 0 to 1.57 rad ($\pi/2$ is the limit) shows stability of this model.

When the required torque has to be magnetic, it should also depend on the orientation of the body. We define this orientation by the unit vector $\vec{\psi}$ having same orientation of its magnetic dipole moment $\vec{m}$, such as

$$\vec{m} = m \vec{\psi} \tag{5.19}$$

If the vector $\vec{\psi}$ prior to a rotation is defined along axis-$z$, time dependent orientation can be written as

$$\vec{\psi}(t) = R \begin{bmatrix} 0 \\ 0 \\ 1 \end{bmatrix} = \begin{bmatrix} s_{\omega t} s_\varphi \\ -c_{\omega t} s_\varphi \\ c_\varphi \end{bmatrix} \tag{5.20}$$

In order to obtain a torque having the same unit vector of angular acceleration at Eq. 5.13, we can define a rotating uniform magnetic field with opposite phase of $\vec{\psi}$ as

$$\vec{B} = \vec{B}_C + \vec{B}_S = B_C \begin{bmatrix} -s_{\omega t} \\ c_{\omega t} \\ 0 \end{bmatrix} + B_S \begin{bmatrix} 0 \\ 0 \\ 1 \end{bmatrix} \tag{5.21}$$

where $B_C$ is the rotating field on the $xy$ plane, around axis-$z$ and a static field $B_S$ in the direction of axis-$z$.

Experiments and simulations show that such a static field is needed to obtain stable motion of the body conforming Eq. 5.20. Resulting magnetic torque can be obtained as

$$\vec{\tau} = \vec{m} \times \vec{B} = m\,\vec{\psi} \times \vec{B} = \vec{\tau}_C + \vec{\tau}_S = (\tau_C c_\varphi + \tau_S s_\varphi) \begin{bmatrix} -c_{\omega t} \\ -s_{\omega t} \\ 0 \end{bmatrix} \quad (5.22)$$

where $\tau_C$ and $\tau_S$ correspond to coefficients of torques caused by cyclic and static components of the magnetic field and can be defined as product of norm of vectors $\vec{m}$ and $\vec{B}_C$ or $\vec{B}_S$. Here it can be seen the magnetic torque vector have same unit vector of $\vec{\alpha}$ defined in Eq. 5.13. By applying Newton's second law (Eq. 5.14), we obtain

$$\tau_C c_\varphi + \tau_S s_\varphi = I\omega^2 s_\varphi \quad (5.23)$$

where unit vectors are eliminated. By arranging terms, the angle $\varphi$ satisfying the equilibrium can be derived as

$$\varphi = \tan^{-1} \frac{\tau_C}{I\omega^2 - \tau_S} \quad (5.24)$$

Therefore, it can be concluded that the conical motion defined by the rotation matrix $R$ can be obtained by the described magnetic interaction. This also shows the equilibrium is independent of the sign of $\omega$.

Eq. 5.24 also defines an upper limit for the static field. That is, $\tau_S$ should be less than $I\omega^2$ in order to obtain repulsion. Despite this system is nonlinear, the term $\tau_S/I$ corresponds to square of natural frequency $\omega_0$ according to Eq. 5.1 and fits this definition since phase lag condition requires $\omega_0 < \omega$. Increasing $\tau_S$ above $I\omega^2$ corresponds to condition $\omega_0 > \omega$ where angle $\varphi$ changes sign according Eq. 5.24. While the motion still conforms to the rotation matrix $R$ for negative values of $\varphi$, this sign change is reflected on the z-component of the force the magnetic moment experiences when rotating field is associated with a dipole as shown in Fig. 5.1(R) and the repulsive interaction becomes attractive. Two values of $\tau_S$ giving same $\varphi$ (0.536 rad) but with opposite signs are calculated as 4.633e-3 and 8e-3 Nm for $I$ = 1e-7 kgm$^2$, $\omega$ = 80$\pi$ rad/s, $\tau_C$ = 1e-3 Nm and motion simulations based on these values result in identical motions except a phase difference equal to $\pi$.

The torque defined at Eq. 5.22 can be obtained by interaction of a rotating moment $\vec{m}_R$ with body moment $\vec{m}_F$. These vectors and the spatial vector $\vec{r}$ going $\vec{m}_R$ to $\vec{m}_F$ can be defined according above equations as

$$\vec{m}_R = m_R \begin{bmatrix} s_{\omega t} c_\gamma \\ -c_{\omega t} c_\gamma \\ s_\gamma \end{bmatrix}, \qquad \vec{m}_F = m_F \begin{bmatrix} s_{\omega t} s_\varphi \\ -c_{\omega t} s_\varphi \\ c_\varphi \end{bmatrix}, \qquad \vec{r} = r \begin{bmatrix} 0 \\ 0 \\ 1 \end{bmatrix} \quad (5.25)$$

Here, the term $\gamma$ is called tilt angle, corresponding the angle between $\vec{m}_R$ and the xy plane. By applying Eq. 4.3, we obtain

$$\vec{\tau} = \frac{\mu_0\, m_R\, m_F}{4\pi r^3} (c_\gamma c_\varphi + 2 s_\gamma s_\varphi)\vec{\Omega}, \qquad \vec{\Omega} = \begin{bmatrix} -c_{\omega t} \\ -s_{\omega t} \\ 0 \end{bmatrix} \quad (5.26)$$

We can separate $\vec{m}_R$ into cyclic and to static components as

$$\vec{m}_R = \vec{m}_C + \vec{m}_S, \qquad \vec{m}_C = m_R c_\gamma \begin{bmatrix} s_{\omega t} \\ -c_{\omega t} \\ 0 \end{bmatrix}, \qquad \vec{m}_S = m_R s_\gamma \begin{bmatrix} 0 \\ 0 \\ 1 \end{bmatrix} \quad (5.27)$$

By applying the torque equation for each components, the cyclic $\vec{\tau}_C$ and the static $\vec{\tau}_S$ torques can be associated with $\vec{m}_C$ and $\vec{m}_S$, respectively.

$$\vec{\tau}_C = \frac{\mu_0\, m_R\, m_F}{4\pi r^3} c_\gamma c_\varphi \vec{\Omega}, \qquad \vec{\tau}_S = \frac{\mu_0\, m_R\, m_F}{2\pi r^3} s_\gamma s_\varphi \vec{\Omega} \quad (5.28\ \text{a, b})$$

Since $\vec{\tau}_C$ is $\tau_C c_\varphi \vec{\Omega}$ and $\vec{\tau}_S$ is $\tau_S s_\varphi \vec{\Omega}$ according Eq. 5.22, by their substitutions in the above equation, we obtain the ratio of torque coefficients as

$$\tau_S/\tau_C = 2 \tan \gamma \quad (5.29)$$

Using above relations, the angle $\varphi$ can be also expressed as

$$\varphi = \tan^{-1}\left(\frac{\tau \cos\gamma}{I\omega^2 - 2\tau \sin\gamma}\right), \qquad \tau = \frac{\mu_0\, m_R\, m_F}{4\pi r^3} \tag{5.30}$$

Interaction of two magnetic moments also induces force; therefore, body is also subject to translational motion beside angular. This is the scope of Section 5.2. Here, translational motions are not taken into account.

By using rotation matrix $R_E$ for general cases, the orientation vector of the body reads

$$\vec{\psi}(t) = R_E \begin{bmatrix} 0 \\ 0 \\ 1 \end{bmatrix} = \begin{bmatrix} s_\theta s_\varphi \\ -c_\theta s_\varphi \\ c_\varphi \end{bmatrix} \tag{5.31}$$

Following this, by using magnetic field defined in Eq. 5.21 and the definition of magnetic moment given in Eq. 5.19, the magnetic torque can be expressed as

$$\vec{\tau} = \vec{m} \times \vec{B} = \begin{bmatrix} -\tau_c c_{\omega t} c_\varphi - \tau_s c_\theta s_\varphi \\ -\tau_c s_{\omega t} c_\varphi - \tau_s s_\theta s_\varphi \\ \tau_c s_{(\theta-\omega t)} s_\varphi \end{bmatrix} \tag{5.32}$$

which reduces to Eq. 5.22 when the azimuthal angle $\theta$ equal to $\omega t$. Rotation $\rho$ is absent in this equation since rotation of a magnetic moment around its axis has no effect. For an axisymmetric body, the inertia tensor is defined on body frame as

$$I = \begin{bmatrix} I_R & 0 & 0 \\ 0 & I_R & 0 \\ 0 & 0 & I_A \end{bmatrix} \tag{5.33}$$

The torque required for the motion could be obtained using Eq. 5.1 using vectors $\vec{v}$ and $\vec{\alpha}$ are defined in Eq. 5.12 and 5.13, and inertia tensor defined above.

$$\vec{\tau} = I_T\,\vec{\alpha} + \vec{v} \times I_T\,\vec{v} = \big(I_A + (I_R - I_A)c_\varphi\big)\vec{\alpha} \tag{5.34}$$

where vectors $\vec{v}$ and $\vec{\alpha}$ are defined in Eq. 5.12 and Eq. 5.13. Here again, the required torque vector to obtain the conical motion have the same unit vector of magnetic torque defined in Eq. 5.22. By equating these vectors, an equation where the angle $\varphi$ can be calculated is obtained as

$$\tau_c c_\varphi + \tau_s s_\varphi = \omega^2 s_\varphi \big(I_A + (I_R - I_A)c_\varphi\big) \tag{5.35}$$

Since the logic above which links the magnetic interaction to the motion is not based on a differential equation solution, stability criterion of the motion might not be derived from it. Therefore, we rely basically on simulation results to obtain stability figures.

It should be noted that circular conical motion can be obtained only with bodies having equal principal MoI or axisymmetric with respect to $\vec{m}$. Simulations based on bodies having unequal principal MoI ($I_1 \neq I_2 \neq I_3$) where the conical motion is elliptic, therefore the model based on circular motion is not applicable, stable motion still can be obtained by providing a static field. A series of simulations is made for evaluating these cases and found that motions are stable in presence of static field of various strengths. This work is reported in Section 5.1.4.

### 5.1.1 Heuristic Stability Criterion

In simulations and experiments, when the static field is absent, the axis of motion of the body deviates from axis-$z$ and the conical motion gradually becomes elliptical and becomes linear when angle $\varphi$ reaches $\pi/2$. This behavior cannot be fixed by ratios of MoI; that is, it happens regardless of the body's geometry. However, simulations show that the axis of the motion can remain aligned with axis-$z$ without requirement of a static field when the body has a spin as evaluated in Section 5.1.6. According to simulations, deviation from axis-$z$ is caused by the dependence of the magnetic torque on the angle $\varphi$ with the term $\cos\varphi$ since the motion becomes stable in absence of this term; however, this term is unavoidable in magnetic interactions. On the other hand, the static field $B_S$ in the $z$ direction (Eq. 5.21) by counteracting to this deviation factor, can keep the motion symmetric around axis-$z$ as it is observed experimentally and in simulations. By evaluation of simulation results, a related stability criterion is found for bodies having uniform MoI $I$. A function $F(\varphi)$ based on Eq. 5.23 is defined as

$$F(\varphi) = s_\varphi(\omega^2 - Js) - c_\varphi J_C \tag{5.36}$$

where terms $J_C$ and $J_S$ denote $\tau_C/I$ and $\tau_S/I$, respectively. This equation corresponds to the equilibrium in presence of an external torque $\tau_E$ equal to $F(\varphi)I$. This function crosses zero with maximum slope at $\varphi$ equal to the equilibrium angle $\varphi_E$. This characteristics can be shown explicitly by defining variables $u$ and $v$ such as

$$u = \left(J_C^2 + (\omega^2 - Js)^2\right)^{1/2}, \quad v = tan^{-1}\left(\frac{J_C}{\omega^2 - Js}\right) \tag{5.37}$$

and expressing Eq. 5.36 using these terms as

$$F(\varphi) = u(s_\varphi c_v - c_\varphi s_v) = u s_{(\varphi-v)} \tag{5.38}$$

since sine function has maximum slope when crossing zero. The first derivative of $F(\varphi)$ can be written as

$$F'(\varphi) = c_\varphi(\omega^2 - Js) + s_\varphi J_C = u c_{(\varphi-v)} \tag{5.39}$$

Since $\varphi_E$ is equal to $v$, the right cosine term above becomes one when $\varphi = \varphi_E$, therefore we can write

$$F'(\varphi_E) = u \tag{5.40}$$

By analyzing simulation results, it is found that the axis of conical motion gets fixed to axis-$z$ when the condition below is satisfied.

$$F'(\varphi_E) < \omega^2 \tag{5.41}$$

That is, maximum slope of the $F(\varphi)$ should be less than $\omega^2$. This equation can be expressed as

$$\omega^2(1 - c_{\varphi_E}) + c_{\varphi_E} Js - s_{\varphi_E} J_C > 0 \tag{5.42}$$

By using Eq. 5.40, this condition can be written free of $\varphi_E$ as

$$\omega^4 > J_C^2 + (\omega^2 - Js)^2 \tag{5.43}$$

By expanding it, we obtain

$$J_C^2 + Js^2 - 2\omega^2 Js < 0 \tag{5.44}$$

By solving this equation for $J_S$, we obtain

$$Js > \omega^2 \pm (\omega^4 - J_C^2)^{1/2} \tag{5.45}$$

The solution with minus root fits other calculations and simulation results with bodies having uniform MoI.

$$\begin{aligned} Js &> \omega^2 - (\omega^4 - J_C^2)^{1/2} \\ \tau_S &> I\omega^2 - ((I\omega^2)^2 - \tau_C^2)^{1/2} \end{aligned} \tag{5.46}$$

This criterion is further tested through numerous simulations with ranges of parameters and found that it gives correct results (Table 5.3). These results also include axisymmetric bodies by referring their MoI in the radial direction ($I_R$).

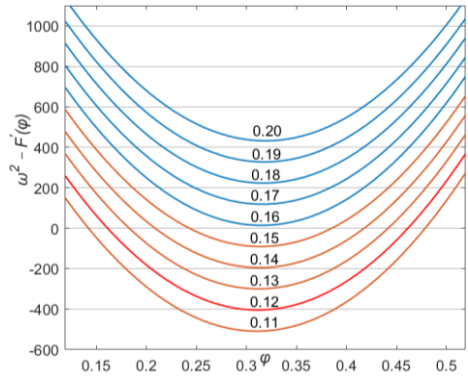

Fig. 5.2. (**a**) Plot of stability criterion term $C = \omega^2 - F'(\varphi)$ covering configurations based cases 13 to 19 in Table 5.3 where ω is $60\pi$ rad/s, $J_C$ = 11000. Identifiers 0.20 to 0.11 correspond to $\tau_S/\tau_C$ ratios. Ratios 0.16 to 0.20 correspond to stable configurations as the criterion term is always above zero. These configurations can perform stable conical motion

around axis-$z$ with fixed angle $\varphi = \varphi_E$. Below this ratio, the axis of the conical motion drifts and the angle $\varphi$ cannot settle to $\varphi_E$, but the body motion may still have stability depending on initial conditions.

Fig. 5.2 shows plots of function $C(\varphi) = \omega^2 - F'(\varphi)$ for a configuration where curves are obtained for a range of $\tau_S/\tau_C$ ratios. When this condition is not met, the motion could be still stable while the axis of the conical motion having a fixed angle $\lambda$ with the axis-$z$, depending on initial conditions. A simulation result for this case is shown at Fig. 5.4. Stable equilibriums where the axis of the motion is off the axis-$z$ are common to experiments. These experimental solutions are extended in presence of off $z$-axis static fields where cycling and static fields are generated by rotating dipole magnets.

The requirement of the static field (therefore the factor which deviates the axis of the motion) might be explained by non-zero time integral of the $x$ component of the torque $\tau_{cx} = \tau_c \cos \omega t \cos \varphi(t)$ defined in Eq. 5.32, due the angle $\varphi$ is not constant when conical motion axis does not meet the axis-$z$. Although the motion can have a constant $\varphi$ for a while, this equilibrium is not stable. The time integral of $\tau_{cx}$ is not zero when $\varphi$ varies, except the phase of $\varphi(t)$ with rotating torque $\vec{\tau}_c$ is $\pm\pi/2$. According simulation results, $\varphi(t)$ can be expressed as

$$\varphi(t) = a_0 + \sum_{i=1..n} a_i \cos(i\omega t) \tag{5.47}$$

where coefficients $a_i$ decreases by about one order of magnitude on each step. The net torque caused by the cyclic torque $\tau_{Cx}$ can be expressed as

$$\langle \tau_{Cx} \rangle = \frac{1}{T} \int_0^T \tau_{cx}(t)\, dt = \frac{\omega}{2\pi} \tau_c \int_0^{\omega/2\pi} \cos \omega t \cos\left(a_0 + \sum_i a_i \cos(i\omega t)\right) dt \tag{5.48}$$

This torque will be called *torque-phi* then. Since this torque is caused by variation of angle $\varphi$ and the amplitude of $\varphi$ (size of the ellipse) is approximately proportional to the inverse square of rotation field velocity (Eq. 5.24), it is expected that it also depends on $\omega$ by the same relation. Calculation of this torque based on a simulation data is given in Table 5.1. Although, the torque ($\tau_{ESx}$) of the static field $B_S$ is very close to $\langle \tau_{Cx} \rangle$ and acts as a counter torque, the equilibrium also involves the inertial torque as it expressed at Eq. 5.23 for a motion with constant angle $\varphi$.

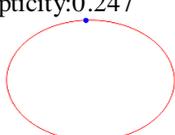

| Simulation model: | Motion characteristics: | General model Fourier-3: |
|---|---|---|
| Body: cylindrical, $I_1=I_2=I_3=0.997\text{e-}7$ Nm$^2$ | $\lambda = \pi/4$ rad | $f(t) = a_0 + \sum_{i=1,2,3} a_i \cos(i\omega t) + b_i \sin(i\omega t)$ |
| $\omega = 60\pi$ rad/s | $\varphi_x = 0.2006$ rad | |
| $\tau_C = |\mathbf{B_C}|\cdot|\mathbf{m}| = 1.0000\text{e-}3$ Nm | $\varphi_y = 0.2664$ rad | Coefficients: |
| $\tau_S = |\mathbf{B_S}|\cdot|\mathbf{m}| = 1.0162\text{e-}4$ Nm | Ellipticity: 0.247 | $\omega = 188.5$ |
| Initial conditions: | | $a_0 = -0.7883$ |
| $\varphi_0 = \pi/4 + 0.2011$ rad | | $a_1 = -0.2005 \quad b_1 = 8.2\text{e-}6$ |
| $\upsilon_0 = [0, 50, 5.28]$ rad/s | | $a_2 = 0.00237 \quad b_2 = -3.5\text{e-}6$ |
| | | $a_3 = -5.326\text{e-}5 \quad b_3 = -1.9\text{e-}6$ |
| Estimated time averaged torque $x$ component due to the $\tau_C$ | | Goodness of fit: |
| $\langle \tau_{Cx} \rangle = \frac{\omega}{2\pi}\tau_C \int_0^{\frac{2\pi}{\omega}} \cos \omega t \cos \varphi\, dt = 0.0707\ \tau_C = 7.07\text{e-}5$ Nm | | SSE: 8.832e-8 |
| Estimated time averaged torque $x$ component due to the $\tau_S$ | | RMSE: 1.501e-5 |
| $\langle \tau_{Sx} \rangle = \frac{\omega}{2\pi}\tau_S \int_0^{\frac{2\pi}{\omega}} \sin \varphi\, dt = 0.702\ \tau_S = 0.0713\ \tau_C = 7.13\text{e-}5$ Nm | | |

Table 5.1. Estimation of $x$ component of the net torque $\langle \tau_{Cx} \rangle$ due to the cyclic torque $\tau_C$. In this simulation, the strength of the static torque $\tau_S$ is not sufficient to keep the body's conical motion axis on $z$ ($\lambda = 0$), but allows to find a stable equilibrium at $\lambda = \pi/4$. At this angle, it is expected that $\langle \tau_{Cx} \rangle$ is compensated by the average static torque $\langle \tau_{Sx} \rangle$. Calculation results support this model by providing these torque figures with a close match with an error less than 1%. The elliptical figure here shows the trajectory of the top center point of the body is seen from the direction of the axis of the conical motion. Figures on the right are Fourier analysis of the zenith angle $\varphi$ obtained from the simulation (top left) of the motion which consist of harmonics with cosine terms where coefficient $a_0$ is the angle $\lambda$ and with negligible sine terms. It should be noted that only axisymmetric bodies are found to have stable motion when angle $\lambda$ is not zero. In this simulation, principal moments of inertia are equal.

### 5.1.2 Reduction of the conical motion to one dimension

The conical motion obtained by rotating field defined in Eq. 5.21 can be reduced to a single DoF where driving torques and the angular motion of the body are confined to a single axis and the zenith angle $\varphi$ becomes the motion variable and the azimuthal variable $\theta$ of the original motion vanishes. This reduction can be obtained by zeroing $x$ or $y$ component of the rotating field. In this case, by zeroing the $x$, the magnetic field vector reads

$$\vec{B} = B_C \begin{bmatrix} 0 \\ c_{\omega t} \\ 0 \end{bmatrix} + B_S \begin{bmatrix} 0 \\ 0 \\ 1 \end{bmatrix} \tag{5.481}$$

This is also realizable by generating this field by an AC coil. A magnetic moment $\vec{m}$ lying on plane-$yz$ can only receive a torque on the axis-$x$; that is, it forces to rotate $\vec{m}$ in the same plane. Such a moment associated with a body can be defined on this plane using the rotation matrix $R_\varphi$ defined in Eq. 5.5 which corresponds to a rotation on the axis-$x$. The orientation vector sharing the same unit vector of $\vec{m}$ reads

$$\vec{\psi}(t) = R_\varphi \begin{bmatrix} 0 \\ 0 \\ 1 \end{bmatrix} = \begin{bmatrix} 0 \\ -s_\varphi \\ c_\varphi \end{bmatrix} \tag{5.482}$$

Resulting magnetic torque can be calculated as

$$\vec{\tau} = \vec{m} \times \vec{B} = \vec{\tau}_C + \vec{\tau}_S = -(\tau_C c_\varphi c_{\omega t} + \tau_S s_\varphi)i \tag{5.483}$$

Here, the rotation and the torque are on the same axis-$x$ and we can drop the vector form. If the body have equal MoI, Euler's equations reduces to Newton's second law resulting in an equation spanning one DoF as

$$\varphi'' = -\frac{\tau_S}{I}\sin\varphi - \frac{\tau_C}{I}\cos\omega t \cos\varphi \tag{5.484}$$

This defines an equation of motion based nonlinear parametric excitation. This equation allows to obtain stable harmonic motion with phase lag condition. Within a numerical solution where the motion amplitude is about 0.25 rad, the motion follows

$$\varphi(t) = a\cos\omega t + b\cos 3\omega t, \quad a = 0.248, \quad b = -2.06e\text{-}4 \tag{5.485}$$

which is clearly characterized by the first harmonic. Here we can also obtain a net force in $z$ direction if an oscillating dipole moment $\vec{m}_c$ is associated with $\vec{B}_C$. This can be realized similar to Fig. 5.1(R) by replacing the rotating dipole by a fixed AC coil. In this configuration, $\vec{m}_c$ would have opposite orientation with $\vec{B}_C$ that floator's magnetic moment $\vec{m}_F$ interacts. By omitting higher harmonics of $\varphi$, $\vec{m}_F$ can be expressed as

$$\vec{m}_F = m_F \begin{bmatrix} 0 \\ -\sin(a\cos\omega t) \\ \cos(a\cos\omega t) \end{bmatrix}, \quad 0 < a < \pi/2 \tag{5.486}$$

By defining the cyclic moment $\vec{m}_c$ and the spatial vector $\vec{r}$ between moments ($\vec{m}_c$ to $\vec{m}_F$) as

$$\vec{m}_c = m_c \begin{bmatrix} 0 \\ -\cos\omega t \\ 0 \end{bmatrix}, \quad \vec{r} = \begin{bmatrix} 0 \\ 0 \\ r \end{bmatrix} \tag{5.487}$$

the force and torque exerted on the body calculated through Eq. 4.2, 4.3 reads

$$\vec{F}_F = \frac{3\mu_0 m_c m_F}{4\pi r^4} \begin{bmatrix} 0 \\ -\cos\omega t \cos(a\cos\omega t) \\ \cos\omega t \sin(a\cos\omega t) \end{bmatrix} \tag{5.488}$$

$$\vec{\tau}_F = \frac{\mu_0 m_c m_F}{4\pi r^3} \begin{bmatrix} -\cos\omega t \cos(a\cos\omega t) \\ 0 \\ 0 \end{bmatrix} \tag{5.489}$$

In these equations, integrals of 'all cosine' terms over one cycle are zero, corresponding to zero average torque and force components when the $\vec{r}$ is constant; that is, when the body has no translational motion. On the other hand, the component-*z* of the force containing a sine term is always positive and it cycles at twice velocity of the rotating field. Therefore, a net thrust is generated in direction away from the coil corresponding to PFR. This thrust can be balanced by a static dipole oriented in *z* direction, generating the $\vec{B}_S$ field defined in Eq. 5.481.

It is also found that numerical solutions conform to the stability criterion Eq. 5.46. Stable numerical solutions also show that the amplitude $\Phi$ of the motion $\varphi$ satisfies the Eq. 5.23. Simulation of this model is tested with a body having full DoF and observed that the motion is stable, fitting to its numerical solution and also the angular motion is confined to the axis-*x*. On the other hand, this configuration allows to simultaneously observe the driven motion on axis-*x* and non-driven pendulum motion on the axis-*y* by providing a small angular velocity as 0.5 rad/s on this axis although they are dependent. As a result, the body obtains an angular motion with amplitude $\Phi_x$ = 0.196 rad and with the driving frequency $\omega_x$ = $80\pi$ rad/s, $\Phi_y$ = 0.0141 rad and $\omega_y$=35.32 rad/s on axis-*y*. This $\omega_y$ velocity differs from the natural frequency ($\omega_0$ = 35.14 rad/s) of the system based on the equation $\omega_0^2 = \tau_S/I$, valid for linear systems. Small angle approximation holds here since the amplitude $\Phi_y$ is 0.0141 rad. This small angle also allows to implement Eq. 5.483 since $\cos\varphi_y$ is very close to one, anytime. Indeed, the system oscillates at this exact frequency $\omega_0$ in absence of motion on axis-*x* (in absence of driving torque). This deviation does not follow the reduction of $\omega_0$ by the increase of amplitude of the motion in this system nor in real pendulum. On the axis-*x*, one can expect a similar effect since the field $B_S$ (associated with $\tau_S$) is aligned with axis-*z*. By evaluating this deviation by a series of simulations by varying $\tau_C$ between 1e-3 and 2e-3 Nm and by providing suitable $\tau_S$ values for stability, it is observed that the relation between $\omega_y$ and $\omega_0$ fits to

$$\omega_y \sim \frac{\omega_0}{\cos(\Phi_x)}, \qquad 0 \leq \Phi_x < 0.4 \tag{5.490}$$

where $\Phi_x$ denotes the amplitude of driven motion on axis-*x*. In these tests, $\omega_y$ varied between 35 and 79 rad/s, far below driving frequency 251.3 rad/s, body is a cylinder having equal MoI of 8.097e-8 kgm$^2$.

Within small angle approximation, Eq. 5.484 reduces to simple DHM where term *sin φ* reduces to $\varphi$ and *cos φ* to one. The solution of this equation can be expressed as

$$\varphi(t) = -\frac{\tau_C}{I(\omega^2 - \omega_0^2)} \cos\omega t + c_1 \sin(\omega_0 t) + c_2 \cos\omega_0 t, \qquad \omega_0 = \sqrt{\tau_S/I} \tag{5.491}$$

where $c_1$ and $c_2$ are some constants, the term $\omega_0$ denotes the natural frequency of the system. This approximated solution, however, is not useful for evaluating stability of the actual motion since the driving torque becomes independent of the motion variable (similar to the case of Eq. 5.17). On the other hand, this system (Eq. 5.484) in absence of the driving field, is equivalent to common pendulum. The frequency of the pendulum can be calculated with small angle approximation as $\omega_0 = \sqrt{g/l}$ where *g* denotes gravitational acceleration and *l* the pendulum length. The exact calculation of $\omega_0$ is also available [18, 19, 20] but rather complex.

### 5.1.3 Evaluation of the angular motion through simulations

In the following, the dynamics of an inertial body having full DoF and subject to the torque defined in Eq. 5.32 are evaluated based on simulations results obtained using FEM based commercial application Comsol Multiphysics. In these simulations, magnetic interactions are based on dipole moments. Simulations work by defining the geometry of the body and its density and then the torque vector M (moment) in global coordinate system corresponding to the vector defined at Eq. 5.32 are implemented as shown in the table 5.2.

| | |
|---|---|
| M$_X$ : | Tc*cos(omega*t)*mbd.rd1.rotzz + Ts*mbd.rd1.rotyz |
| M$_Y$ : | Tc*sin(omega*t)*mbd.rd1.rotzz − Ts*mbd.rd1.rotxz |
| M$_Z$ : | −Tc*(sin(omega*t)*mbd.rd1.rotyz + cos(omega*t)*mbd.rd1.rotxz) |
| mbd.rd1.rotxz = $\sin\theta \sin\varphi$, | mbd.rd1.rotyz = $-\cos\theta \sin\varphi$,    mbd.rd1.rotzz = $\cos\varphi$ |

Table 5.2. Definition of applied moment to a free body in the simulation where the vector M corresponds to the torque received by a magnetic moment $\vec{m}$ from a homogenous rotating field *B* in accordance with Eq. 5.22 and its related definitions. Term mbd.rd1 is the identifier of the body, Tc and Ts are coefficients of cyclic ($\vec{\tau}_C$) and static ($\vec{\tau}_S$) torques defined in Eq. 5.22, omega denotes the angular velocity $\omega$ of $\vec{\tau}_C$ and t, the time. Terms rotxz, rotyz and rotzz are named elements of internally calculated time dependent rotation matrix of the body in the ZXZ convention.

The definition of moment M given here reflects Eq. 5.32 by reversing the direction of the cyclic torque. Variables affected by this change are the sign of angle $\varphi$ and sign of initial angular velocity components which all of them become negative in this implementation. Relevant initial conditions for these simulations are tangential angular velocity $\omega_T$ of the body which is calculated according to Eq. 5.12 at $t = 0$, the orientation of the body ($\theta_0$, $\varphi_0$) by setting $\theta_0 = 0$ and calculating angle $\varphi_0$ which satisfy Eq. 5.24 or Eq. 5.35.

Results of some simulations regarding stability of the axis of the conical motion of axisymmetric (cylindrical) bodies are summarized in Table 5.3.

| No | $I_A$ (axial) (kg m$^2$)+ | $I_R$ (radial) (kg m$^2$) | $\Omega$ (rad/s) | $\tau_C$ (Nm) | $\tau_S/\tau_C$ | Min $\tau_S/\tau_C$ | Bias $\upsilon_Y$ (rad/s) | $\varphi_0$ (rad) | $\varphi$-calc (rad) | $\lambda$ (rad) | Criterion |
|---|---|---|---|---|---|---|---|---|---|---|---|
| 1 | 8.099E-8 | 1.058E-7 | 60π | 7.5e-4 | 0.30 | 0.101 | 0.5 | 0.2100 | 0.2102 | 0 | 1 |
| 2 | 8.099E-8 | 1.058E-7 | 60π | 7.5e-4 | 0.25 | 0.101 | 0.5 | 0.2080 | 0.2080 | 0 | 1 |
| 3 | 8.099E-8 | 1.058E-7 | 60π | 7.5e-4 | 0.25 | 0.101 | 0.5 | 0.2069 | 0.2069 | 0 | 1 |
| 4 | 8.099E-8 | 1.058E-7 | 60π | 7.5e-4 | 0.22 | 0.101 | 0.5 | 0.2066 | 0.2067 | 0 | 1 |
| 5 | 8.100E-8 | 1.058E-7 | 60π | 7.5e-4 | 0.20 | 0.101 | 0.5 | 0.2055 | 0.2059 | 0 | 1 |
| 6 | 8.099E-8 | 1.058E-7 | 60π | 7.5e-4 | 0.18 | 0.101 | 0.5 | 0.2049 | 0.2050 | 0 | 1 |
| 7 | 8.099E-8 | 1.058E-7 | 60π | 7.5e-4 | 0.10 | 0.101 | 0.5 | 0.2013 | 0.2018 | 0 | 1 |
| 8 | 8.099E-8 | 1.058E-7 | 60π | 7.5e-4 | 0.08 | 0.101 | 0.5 | 0.2008 | 0.2010 | >0 | 5 |
| 9 | 8.099E-8 | 1.058E-7 | 60π | 7.5e-4 | 0.05 | 0.101 | 0.5 | 0.1996 | 0.1998 | >0 | 3 |
| 10 | 8.099E-8 | 1.058E-7 | 80π | 7.5e-4 | 0.06 | 0.056 | 0.2 | 0.1123 | 0.1127 | 0 | 1 |
| 11 | 8.099E-8 | 1.058E-7 | 80π | 7.5e-4 | 0.05 | 0.056 | 0.0 | 0.1121 | 0.1125 | >0 | 5 |
| 12 | 1.000e-7 | 1.000e-7 | 60π | 1.5e-3 | 0.55 | 0.221 | 0.0 | 0.5027 | 0.5027 | >0 | 4 |
| 13 | 1.000e-7 | 1.000e-7 | 60π | 1.1e-3 | 0.20 | 0.159 | 0.0 | 0.3188 | 0.3188 | 0 | 1 |
| 14 | 1.000e-7 | 1.000e-7 | 60π | 1.1e-3 | 0.17 | 0.159 | 0.0 | 0.3158 | 0.3158 | 0 | 1 |
| 15 | 1.000e-7 | 1.000e-7 | 60π | 1.1e-3 | 0.155 | 0.159 | 0.0 | 0.3144 | 0.3144 | 0 | 1 |
| 16 | 1.000e-7 | 1.000e-7 | 60π | 1.1e-3 | 0.150 | 0.159 | 0.0 | 0.3139 | 0.3139 | >0 | 2 |
| 17 | 1.000e-7 | 1.000e-7 | 60π | 1.1e-3 | 0.145 | 0.159 | 0.0 | 0.3133 | 0.3134 | >0 | 2 |
| 18 | 1.000e-7 | 1.000e-7 | 60π | 1.1e-3 | 0.140 | 0.159 | 0.5 | 0.313 | 0.3133 | >0 | 2 |
| 19 | 1.000e-7 | 1.000e-7 | 60π | 1.1e-3 | 0.15 | 0.159 | 0.0 | 0.3137 | 0.3139 | >0 | 2 |
| 20 | 1.000e-7 | 1.000e-7 | 60π | 1.2e-3 | 0.40 | 0.174 | 0 | 0.3721 | 0.3723 | 0 | 1 |
| 21 | 1.000e-7 | 1.000e-7 | 60π | 1.2e-3 | 0.18 | 0.174 | 1.0 | 0.3455 | 0.3452 | 0 | 1 |
| 22 | 1.000e-7 | 1.000e-7 | 60π | 1.2e-3 | 0.17 | 0.174 | 0 | 0.3431 | 0.3440 | 0 | 1 |
| 23 | 1.000e-7 | 1.000e-7 | 60π | 1.2e-3 | 0.16 | 0.174 | 0 | 0.3429 | 0.3429 | >0 | 2 |
| 24 | 1.000e-7 | 1.000e-7 | 60π | 1.2e-3 | 0.115 | 0.174 | 1.0 | 0.4095 | n/a | >0 | 2 |
| 25 | 1.000e-7 | 1.000e-7 | 60π | 2.5e-4 | 0.04 | 0.035 | 0.2 | 0.070 | 0.070 | 0 | 1 |
| 26 | 1.000e-7 | 1.000e-7 | 60π | 2.5e-4 | 0.02 | 0.035 | 0.2 | 0.070 | 0.070 | >0 | 5 |
| 27 | 2.500E-8 | 3.750E-8 | 60π | 2.5e-4 | 0.20 | 0.095 | 0.0 | 0.1938 | 0.1938 | 0 | 1 |
| 28 | 2.500E-8 | 3.750E-8 | 60π | 3.75e-4 | 0.20 | 0.144 | 0.0 | 0.2940 | 0.2940 | 0 | 1 |
| 29 | 2.500E-8 | 3.750E-8 | 60π | 3.75e-4 | 0.15 | 0.144 | 0.5 | 0.2896 | 0.2898 | 0 | 1 |
| 30 | 2.500E-8 | 3.750E-8 | 60π | 3.75e-4 | 0.13 | 0.144 | 0.5 | 0.2874 | 0.2881 | >0 | 2 |
| 31 | 2.500E-8 | 3.750E-8 | 60π | 3.75e-4 | 0.10 | 0.144 | 0.0 | 0.2854 | 0.2856 | >0 | 3 |
| 32 | 2.000E-7 | 1.125E-7 | 60π | 1.0e-3 | 0.13 | 0.127 | 0.0 | 0.2472 | 0.2472 | 0 | 1 |
| 33 | 2.000E-7 | 1.125E-7 | 60π | 1.0e-3 | 0.12 | 0.127 | 0.0 | 0.2466 | 0.2466 | >0 | 5 |
| 34 | 2.000E-7 | 1.125E-7 | 60π | 1.0e-3 | 0.11 | 0.127 | 0.0 | 0.2461 | 0.2461 | >0 | 5 |
| 35 | 1.000e-7 | 1.000e-7 | 80π | 3.0e-3 | 0.25 | 0.253 | 0.0 | 0.4943 | 0.4943 | >0 | 4 |
| 36 | 1.000e-7 | 1.000e-7 | 80π | 3.0e-3 | 0.267 | 0.253 | 0.0 | 0.498 | 0.498 | 0 | 1 |

Table 5.3. Summary of stability tests of conical motion of the body through simulations. Most of the configurations are tested using multiple static field strengths in order to evaluate their stability using this factor. Angular motion deviates from the equilibrium angle $\varphi$ when the static torque coefficient $\tau_S$ is below a certain value. In this case, the motion axis still finds an equilibrium at a zenith angle $\lambda$ with proper initial conditions. However, in these tests, initial conditions are given for the motion for $\lambda = 0$. For this reason, the motion cannot settle to the equilibrium angle $\lambda$, but oscillates around it since there is no friction. These oscillations are considered as an instability factor here. The data fits very well to the criterion for obtaining symmetric conical motion around axis-$z$ given by Eq. 5.46 where the values at column "Min. $\tau_S/\tau_C$" correspond to this equation. Results are characterized by the following criteria:

(1) $\varphi$ is constant ($\lambda$=0) or has a sinusoidal curve.
(2) $\varphi$ varies with a complex cyclic curve.
(3) $\varphi$ ever increases.
(4) $\varphi$ oscillates with ever increasing amplitude.
(5) $\varphi$ varies but the simulation ran short for obtaining exact criterion.

Fig. 5.3 shows motion characteristic of cylindrical body having uniform MoI and subject to homogenous rotating field obtained from a simulation. Here, parameters (Table 5.3, No. 21) are chosen for obtaining a large angular motion in order to test the compliance of results with Eq. 5.7 and with the stability criterion Eq. 5.46. With these arguments, $\tau_S/\tau_C$ ratio is chosen as 0.18, just above the minimum ratio of 0.174 for obtaining symmetric conical motion around axis-$z$. The compliance of the simulation to the model (Eq. 5.7) is shown visually in Fig. 5.3 by trajectories of reference points on the cylinder with local coordinates $p_1$, $p_2 = (0, 0, \pm h/2)$ and $p_3 = (r, 0, 0)$ where $h$ and $r$ are height and the radius of the cylinder. The trace (b) corresponds a zoomed section of circular trace belong to $p_1$ and (c) to the zoomed trajectory of $p_3$. These traces show that trajectories are kept converged as a sign of stability. The transformed coordinates $p'_3$ which following the motion are calculated by applying rotation matrix $R$ with angle $\varphi = -0.3455$ rad as

$$p'_3(t) = R\, p_3 = r \begin{bmatrix} 1 - (1 - \cos 2\omega t)(1 - \cos\varphi)/2 \\ \sin 2\omega t\, (1 - \cos\varphi)/2 \\ -\sin\omega t \sin\varphi \end{bmatrix} = r \begin{bmatrix} 0.02955 \cos 2\omega t + 0.9705 \\ 0.02955 \sin 2\omega t \\ 0.33867 \sin\omega t \end{bmatrix} \quad (5.49)$$

and found to fit simulation data with a precision of four significant digits. This equation also shows the reason of eight shaped trace belong to $p_3$ by the presence of terms $2\omega t$ in $x$ and $y$ components.

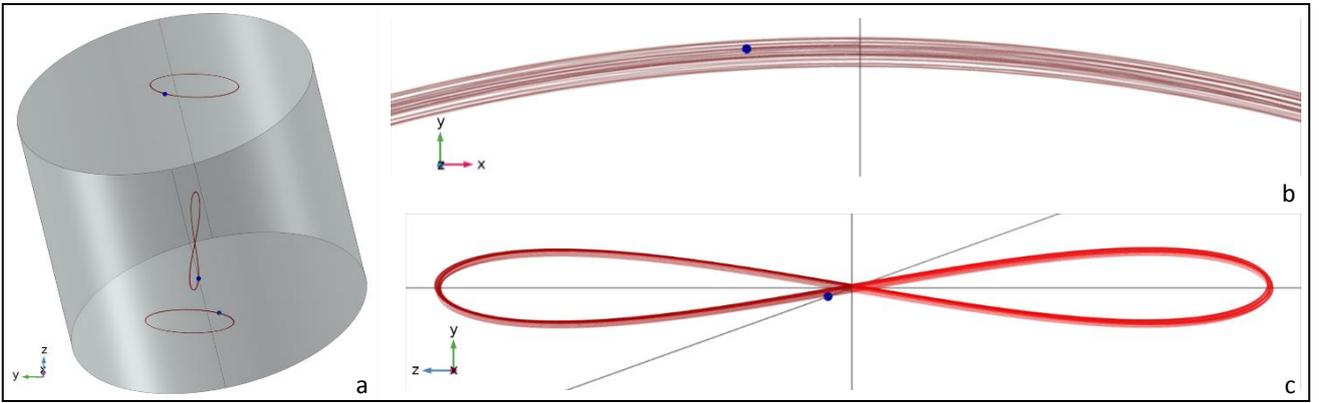

Fig. 5.3. Simulations of motion a cylindrical body having a dipole moment $\vec{m}$ in its axial direction subject a rotating field $B_C$ around axis-$z$ and orthogonal to it with constant velocity equal to $60\pi$ rad/s and with constant magnitude. A static field $B_S$ in $z$ direction having strength of $0.18\,B_C$ is also present. Torque figures are $\tau_C = \|B_C\|\cdot\|m\| = $ 1.2e-3 Nm, $\tau_S = 0.18\,\tau_C$. This ratio is slightly above the required ratio 0.174 to obtain conical motion with fixed zenith angle according Eq. 5.46. Initial condition are given in order to obtain a smooth conical motion, but not finely tuned. Angle $\varphi$ set initially to 0.3455 rad and kept almost constant. As a result, body's main axis draws conical circular motions around axis-$z$ with zenith angle $\varphi = 0.345$. Body dimensions are chosen in order to obtain principal moments are inertia are equal to 1e-7 Nm² (0.999e-7 Nm² used by simulation) has MoI equal to 1e-7 kgm² on all its principal axes. Its mass is 6.68 g, radius is 5.4711 mm, height is 9.4763 mm and has uniform density of 7.5 g/cm³. Angular velocity of the applied torque is $60\pi$ = 188.5 rad/s. Left figure shows an instance of the body with traces of top and bottom center points (1, 2) and middle side point (3) of the body. Trace 3 have 'eight' shape because presence of angular motion second harmonic in $z$ direction in accordance of Eq. 5.49. Right figures show zoomed view of traces where the top figure corresponds to the arc from the circular trace of top point projected on $xy$ plane at instance $t = 0.767$ ms where the motion completes its 23'th cycle. Bottom figure similarly shows the rotated 'eight' figure on $zy$ plane. Vertical and horizontal hairlines in this figure are $z$ and $y$ axes and the diagonal is the main axis of the cylinder.

Fig. 5.4 shows a similar simulation result where static field strength is insufficient for obtaining a conical motion with fixed zenith angle $\varphi$. By providing proper initial conditions, it is observed that the motion obtains stability while the axis of the motion obtains an angle ($\lambda$) with the axis-$z$. This deviation also introduces ellipticity to the motion. In the second run of this simulation, an extra angular velocity about 1 rad/s is provided on axis-$y$. This causes the body orientation to slowly rotate around axis-$z$. Simulation shows a small speed variation in this course, later found as a simulation artefact and eliminated by lowering the time-step parameter.

Fig. 5.5 shows $xyz$ components of the torque vector belong to the simulation 8 in Fig. 5.6 where angle $\lambda$ is 0.114 rad and $\tau_S/\tau_C$ ratio is 0.06. It is remarkable that the stability is maintained while the magnetic torque varies in large extents within a cycle and deviates significantly from sinusoidal shape.

Based on simulations with bodies having equal MoI, motions are found stable without lower limit of static field strength and no stability issues are encountered with axisymmetric bodies. Relations of the angle $\lambda$ and ellipticity of the motion to the ratio $B_S/B_C$ is shown in Fig. 5.6 for a given configuration. On the other hand, it is found that the stability might not be obtained with non-axisymmetric bodies with when angle λ is not zero.

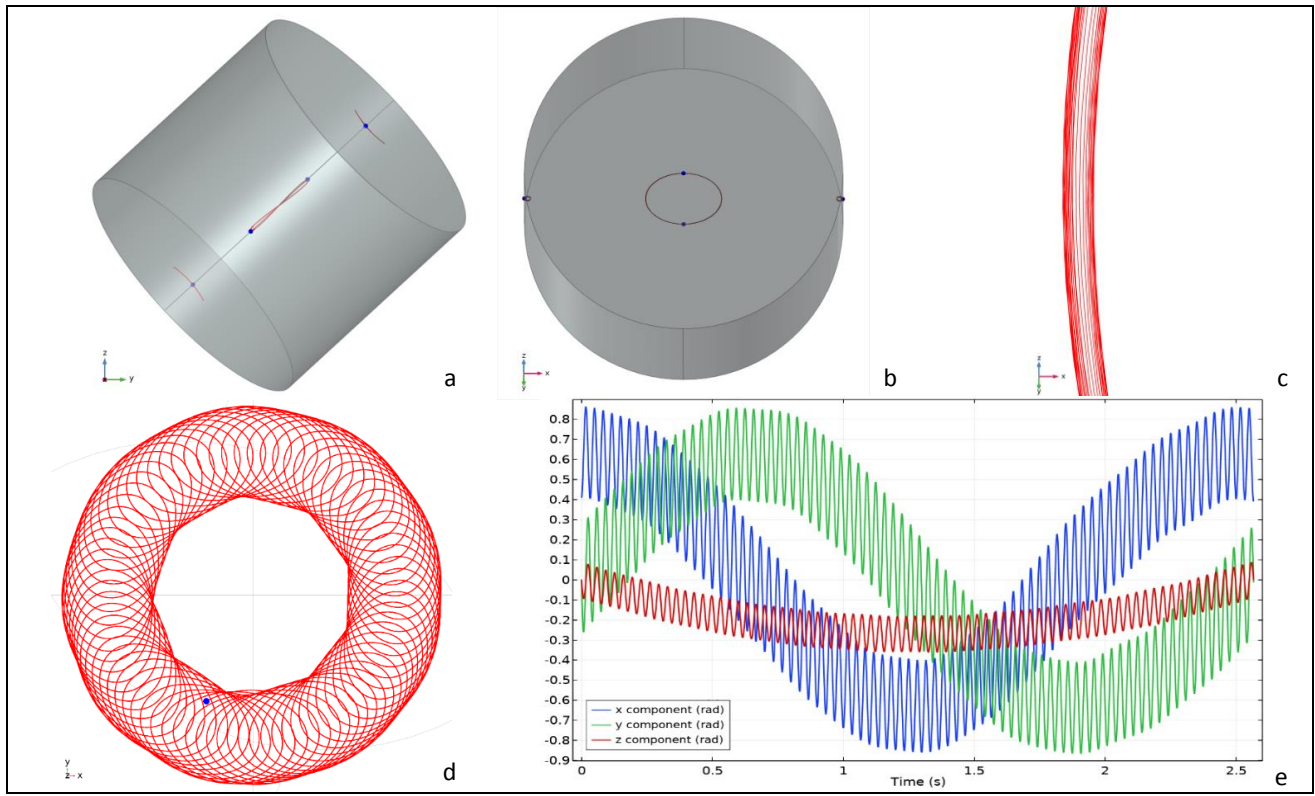

Fig. 5.4. Simulations of motion of a cylindrical body with equal principal MoI where static field strength $B_S$ is below the minimum $B_S/B_C$ ratio (0.158) based by Eq. 5.46 for the body to have stable motion around the axis-$z$. This condition shifts the axis of the conical motion from axis-$z$ by angle called λ and the circular angular motion becomes elliptic. In the first simulation (top figures), $B_S/B_C$ is 0.0944 and the motion finds an equilibrium at λ = 0.852 rad (48.8°). Simulation run about 0.5 s and motion is found stable without any drift thanks to precise initial conditions. (**a**) The body shown in an instance where rotation λ is aligned with axis-$x$. (**b**) Same as (a) but with a camera angle close to λ where traces of top and bottom center points overlap. (**c**) Zoomed part of the elliptical trace at (b). This shows the characteristics of the stability where traces are bounded in a narrow limit. Amplitude of the elliptical motion on axes $x$ and $y$ are 0.186 and 0.263 rad, giving an ellipticity about 0.292. In the second simulation (bottom figures), $B_S/B_C$ is 0.115 and the motion finds an equilibrium at λ = 0.64 rad. Here, an excess angular velocity of 1 rad/s on axis-$y$ is given on initial condition to obtain the drift of the motion around axis-$z$ in CCW direction. (**d**) The trace of the top center point of the body on projection to the $xy$ plane. Here the crossed hairlines mark $xy$ axes. This snapshot is taken at $t = 2570$ ms. The ring has a decagon shape (10 sides) due to nutation of the motion due to initial conditions. It is observed that the angular drift of the axis of conical motion has a small acceleration which progressively decreases and switches to deceleration at about $t = 1760$ ms. This non-zero acceleration is found to be a simulation artefact and vanishes when time-step is reduced from 0.4 ms to 0.2 ms. (**e**) Plot of orientation of the body in terms of rotations around axes $x$, $y$ and $z$. Parameters and initial conditions of simulation 1 are:
$I_1 = I_2 = I_3 = 0.999$e-7 Nm$^2$, $\tau_C$ = 1e-3 Nm, $\tau_S$ = 9.94e-5 Nm, $\omega = 60\pi$ rad/s, $\varphi_0 = -1.038$ rad, $\vec{v}_0 = [0,-47.9,-4.891]$ rad/s.

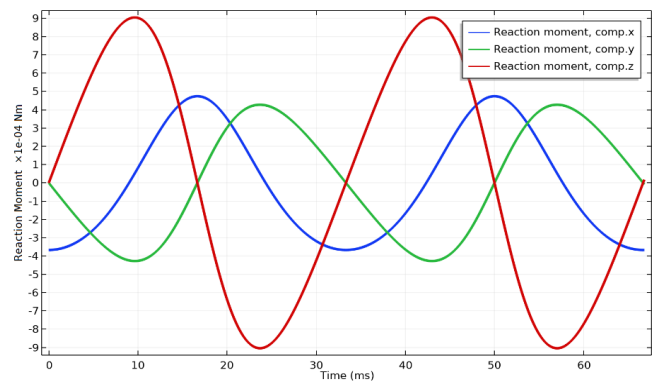

Fig. 5.5. Simulation result showing reaction moment receives a body for two cycles of the rotation field. In this specific case, body's conical motion is elliptical (ellipticity = 0.52) and motion axis is off the axis-z ($\lambda$ = 0.114 rad) due to static field strength not being sufficient for keeping motion axis on z ($B_S/B_C$ = 0.06) while the minimum ratio calculated from Eq. 5.46 is 0.144 and this value should be between 0.14 and 0.15 according to simulations.

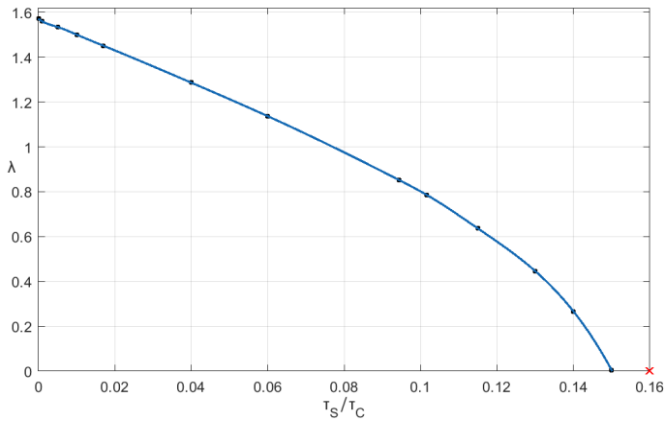

| No | $\tau_S/\tau_C$ | $\lambda$ (rad) | $R_X$ | $R_Y$ | Ellipticity |
|----|-----|--------|---------|-------|-------------|
| 1  | 0.160 | 0.0001 | 0.287 | 0.287 | 0 |
| 2  | 0.150 | 0.0045 | 0.286 | 0.286 | 3e-4 |
| 3  | 0.140 | 0.2666 | 0.275 | 0.283 | 0.0279 |
| 4  | 0.130 | 0.4461 | 0.256 | 0.279 | 0.082 |
| 5  | 0.115 | 0.6362 | 0.227 | 0.268 | 0.154 |
| 6  | 0.1016 | 0.7859 | 0.2006 | 0.266 | 0.247 |
| 7  | 0.0944 | 0.8518 | 0.186 | 0.263 | 0.291 |
| 8  | 0.060 | 1.1369 | 0.119 | 0.247 | 0.519 |
| 9  | 0.040 | 1.2862 | 0.0792 | 0.237 | 0.666 |
| 10 | 0.017 | 1.4510 | 0.0337 | 0.226 | 0.851 |
| 11 | 0.010 | 1.5004 | 0.0198 | 0.225 | 0.912 |
| 12 | 0.005 | 1.5346 | 0.00985 | 0.219 | 0.955 |
| 13 | 0.001 | 1.5594 | 0.00645 | 0.217 | 0.970 |
| 14 | 0.000 | 1.5708 | 0 | 0.216 | 1 |

Fig. 5.6. Variation of angle $\lambda$ and the ellipticity of angular motion of the body by the strength of the static field according simulations results where the body has uniform MoI $I$ = 0.998E-7 Nm², $\tau_C$ = 1e-7 Nm and rotating field velocity $\omega$ = 60 $\pi$ rad/s. Semi-axis lengths are denoted by $R_X$ and $R_Y$ in radians. Ellipticity is calculated by formula 1–$R_X/R_Y$. It should be noted the elliptical shape gains some asymmetry with respect to the major axis by increasing the angle $\lambda$. Performed simulations are similar to given in Fig. 5.4.

### 5.1.4 Characteristics of the motion of a body having unequal moments of inertia

Stability of motion of bodies having unequal MoI or not axisymmetric with respect to dipole axis is evaluated by simulations. Providing proper initial conditions for obtaining a motion suitable for analysis is challenging here because present formulas for these figures cover only axisymmetric bodies. Simulations show that these non-axisymmetric bodies can obtain a stable angular elliptical motion around axis-z when static field strength is above a limit similar to axisymmetric bodies. The Eq. 5.46 still be useful for estimation of minimum $\tau_S$ for obtaining a motion around axis-z by using an interpolated radial MoI from two radial directions.

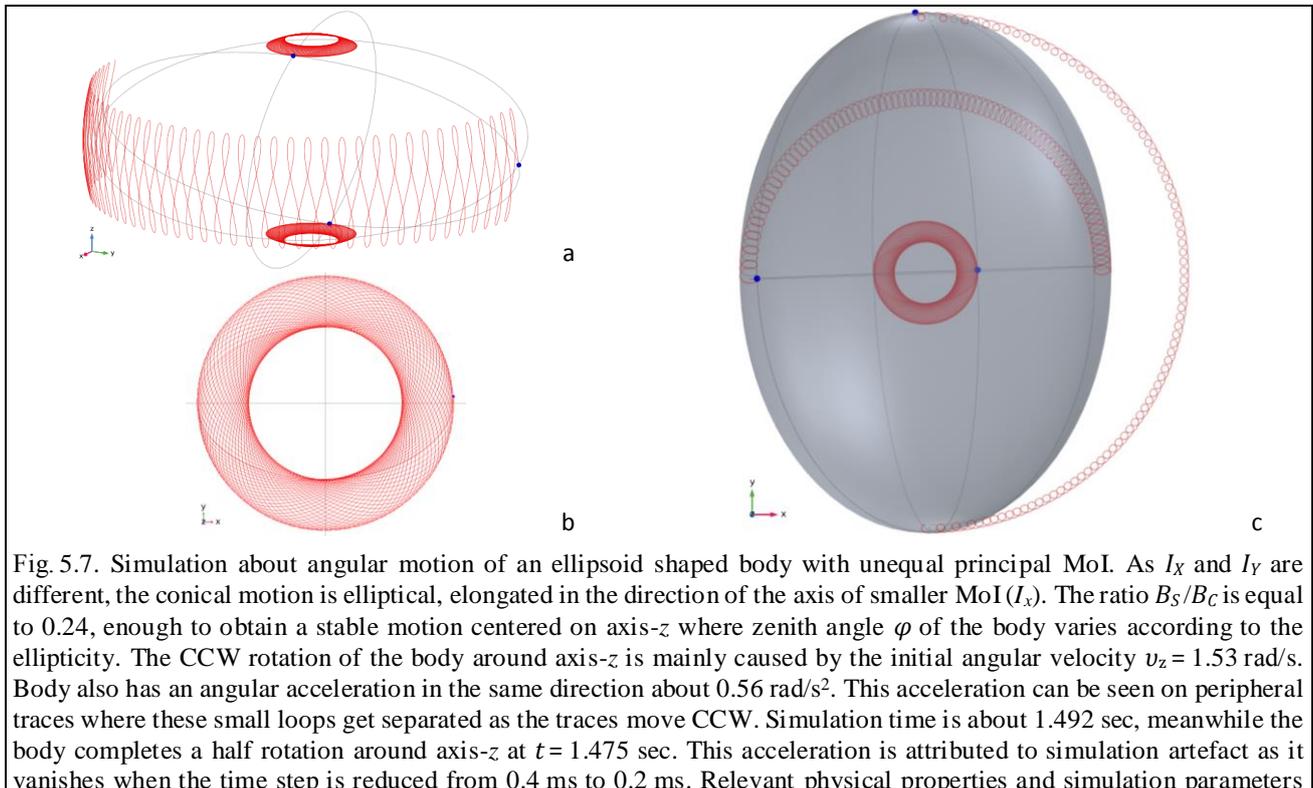

Fig. 5.7. Simulation about angular motion of an ellipsoid shaped body with unequal principal MoI. As $I_X$ and $I_Y$ are different, the conical motion is elliptical, elongated in the direction of the axis of smaller MoI ($I_x$). The ratio $B_S/B_C$ is equal to 0.24, enough to obtain a stable motion centered on axis-z where zenith angle $\varphi$ of the body varies according to the ellipticity. The CCW rotation of the body around axis-z is mainly caused by the initial angular velocity $v_z$ = 1.53 rad/s. Body also has an angular acceleration in the same direction about 0.56 rad/s². This acceleration can be seen on peripheral traces where these small loops get separated as the traces move CCW. Simulation time is about 1.492 sec, meanwhile the body completes a half rotation around axis-z at $t$ = 1.475 sec. This acceleration is attributed to simulation artefact as it vanishes when the time step is reduced from 0.4 ms to 0.2 ms. Relevant physical properties and simulation parameters

are: Semi axes lengths (*x*, *y*, *z*) : 6, 8.5, 4 mm; Mass : 6.385 g; Principal moments of inertia: 1.123736e-7, 6.623701e-8, 1.3780954e-7 Nm$^2$; Torque coefficients ($\tau_C$, $\tau_S$): 1e-3 Nm, 2.4e-4 Nm; Rotating field velocity ($\omega$): $60\pi$ rad/s; Initial zenith angle ($\varphi_0$): –0.254 rad, Initial angular velocity vector ($\vec{v}$): –[0.002 80.95 6.858] rad/s.

In the first simulation (Fig. 5.7), the static field strength is sufficient ($\tau_S = 0.24\,\tau_C$) to keep the axis of the conical motion on axis-*z* and it is found that the motion is stable. In the second, $\tau_S$ is lowered to $0.215\,\tau_C$ and a shift is observed on the motion axis toward an equilibrium angle. When this equilibrium angle is reached, it does not settle there, instead oscillates around because there is no damping to absorb the energy gained on this shift. By the help of other simulations, it found that the minimum $\tau_S$ for obtaining stable motion centered on axis-*z* is close to $0.22\,\tau_C$ where $\tau_C$ is kept as 1e-3 Nm for these series. Here, the body is an ellipsoid with axes lengths as 6×8.5×4 mm where its dipole is aligned with the shortest axis. This ellipsoid body is also tested while the dipole is aligned with the longest axis. In this case, it is found that the motion is centered on axis-*z* for $\tau_S \geq 0.14\,\tau_C$ and follows a tilted axis about 0.54 rad when $\tau_S = 0.09\,\tau_C$. This latter simulation duration is not enough to obtain a net stability figure.

### 5.1.5 Characteristics of the motion with zenith angle π/2

As mentioned above, in absence of static field $B_S$ and the spin, the angle $\varphi$ goes to $\pi/2$. That is, the orientation vector $\psi$ and the motion get confined to the *xy* plane. Following this scheme, by defining the initial vector $\psi_0$ as

$$\vec{\psi}_0 = \begin{bmatrix} 0 \\ -1 \\ 0 \end{bmatrix} \quad (5.50)$$

and the rotation matrix as

$$R = \begin{bmatrix} c_\theta & -s_\theta & 0 \\ s_\theta & c_\theta & 0 \\ 0 & 0 & 1 \end{bmatrix} \quad (5.51)$$

the orientation, angular velocity and angular acceleration vectors reads

$$\vec{\psi} = R\vec{\psi}_0 = \begin{bmatrix} s_\theta \\ -c_\theta \\ 0 \end{bmatrix} \quad (5.52)$$

$$\vec{v} = \begin{bmatrix} 0 \\ 0 \\ \theta' \end{bmatrix}, \quad \vec{\alpha} = \begin{bmatrix} 0 \\ 0 \\ \theta'' \end{bmatrix} \quad (5.53)$$

By using cyclic component of the magnetic field defined in Eq. 5.21, the magnetic torque reads

$$\vec{\tau} = \vec{m} \times \vec{B} = m\,\vec{\psi} \times \vec{B} = m \begin{bmatrix} s_\theta \\ -c_\theta \\ 0 \end{bmatrix} \times B_C \begin{bmatrix} -s_{\omega t} \\ c_{\omega t} \\ 0 \end{bmatrix} = \tau_C \begin{bmatrix} 0 \\ 0 \\ s_{(\theta - \omega t)} \end{bmatrix} \quad (5.54)$$

By applying Newton's second law for angular motion to a body having uniform MoI as

$$\vec{\tau} = I\,\vec{\alpha} \rightarrow \tau \begin{bmatrix} 0 \\ 0 \\ s_{(\theta - \omega t)} \end{bmatrix} = I \begin{bmatrix} 0 \\ 0 \\ \theta'' \end{bmatrix} \quad (5.55)$$

the equation of the motion reads

$$\theta'' = \frac{\tau}{I} \sin(\theta - \omega t) \quad (5.56)$$

Numerical solutions of Eq. 5.56 for parameters $J_C$ = 1e4 m/s$^2$, and $\omega$ = 188.5 rad/s are obtained for various initial conditions. Curves obtained by this method are sinusoidal in presence of small harmonics. Functions of $\theta(t)$ obtained using curve fitting method are summarized in Table 5.4. By neglecting small cosine terms, this function can be expressed as

$$\theta(t) = \sum_{i=1..n} a_i \sin(i\omega t) \quad (5.57)$$

where residuals of curve fitting are not periodic beyond n = 4. Generally, oscillations are stable, but the angle $\theta$ drifts slowly and its direction and its profile is sensitive to initial condition. Similar results are obtained from simulations and Fig. 5.8 shows a curve fitting result based on simulation data using Fourier series with three harmonics terms. The amplitude of the motion fits well to Eq. 5.24 by replacement of the angle $\varphi$ by $\theta$ as

$$|\theta| \cong tan^{-1}\frac{J_C}{\omega^2} \tag{5.58}$$

Using the numerical data for Table 5.4, case 1, the plot of the azimuthal acceleration $\theta''(t)$ is shown in Fig. 5.9. The change in angular velocity within one cycle of rotating field can be used to obtain the net angular acceleration. Using the angular displacement data from simulation or from numerical solution and putting this in Eq. 5.56, the instantaneous angular acceleration $\theta''$ can be obtained.

$$\Delta\theta' = \theta'_{T1} - \theta'_{T0} = J_C \int_{T0}^{T1} \theta''(t)\,dt \tag{5.59}$$

$$\theta''_{aver} = \frac{\Delta\theta'}{\Delta T} = \frac{J_C}{2\pi/\omega}\int_0^{2\pi/\omega} sin(\theta(t) - \omega t)\,dt \tag{5.60}$$

By implementing data from a numerical solution for ($J_C$ = 1e4, $\omega$ = 60$\pi$), the average angular acceleration is found as $-1.29\text{e}{-}6 J_C$. Such a small figure can be expected since the integral of a function $f(x)$ yields always zero when it is defined as

$$f(x) = sin\left(\sum_i a_i sin(k_i x) - x\right) \tag{5.61}$$

where $k_i$ is a positive integer.

| No | $J_C$ | $\theta_0$ | $\theta'_0$ | $\theta(t)$ |
|---|---|---|---|---|
| 1 | -1e4<br>1e4 | 0<br>$\pi$ | -55.664 | $-0.2748\,sin\,\omega t - 9.55\text{e-}3\,sin\,2\omega t - 4.3\text{e-}4\,sin\,3\omega t$ |
| 2 | 1e4<br>-1e4 | 0<br>$\pi$ | 48.45 | $0.2748\,sin\,\omega t - 9.55\text{e-}3\,sin\,2\omega t + 4.3\text{e-}4\,sin\,3\omega t$ |
| 3 | 1e4 | $\pi/2$-0.206 | 0 | $1.64 - 0.2748\,cos(\omega t - 0.069) - 9.58\text{e-}3\,sin(2\omega t - 0.14)$<br>$+ 4.5\text{e-}4\,sin(3\omega t - 0.28)$ |

Table 5.4. Some numerical solutions of Eq. 5.56.

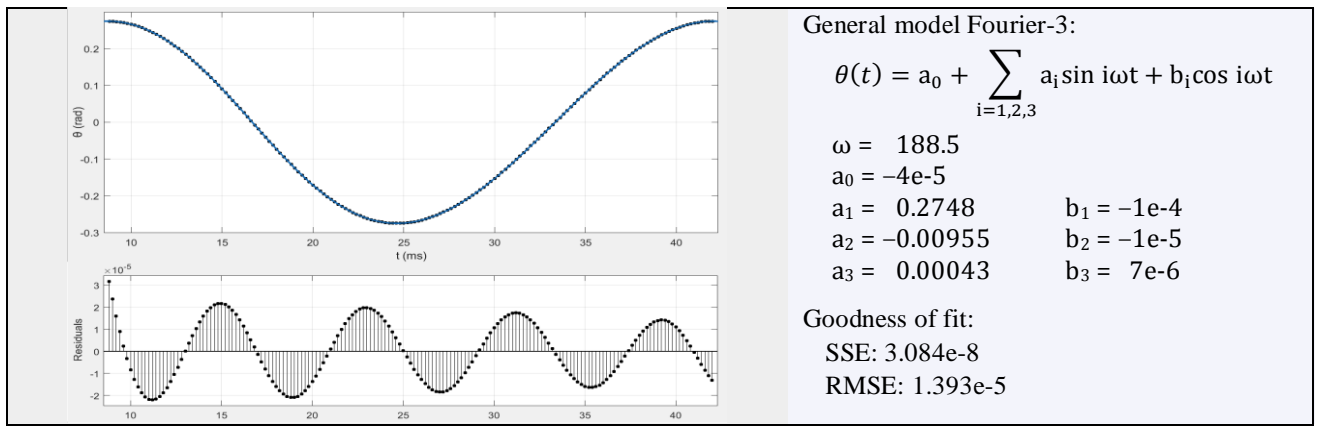

General model Fourier-3:

$$\theta(t) = a_0 + \sum_{i=1,2,3} a_i sin\,i\omega t + b_i cos\,i\omega t$$

$\omega$ = 188.5
$a_0$ = −4e-5
$a_1$ = 0.2748    $b_1$ = −1e-4
$a_2$ = −0.00955  $b_2$ = −1e-5
$a_3$ = 0.00043   $b_3$ = 7e-6

Goodness of fit:
SSE: 3.084e-8
RMSE: 1.393e-5

Fig. 5.8. Curve fitting of angular motion of a body obtained from a simulation data over one cycle using Fourier analysis. In this simulation, a cylindrical body having magnetic moment in axial direction and having MoI as $I_R = I_A = 0.999\text{e-}7$ kg m$^2$ with main axis on $xy$ plane is subject a homogeneous rotating magnetic with constant amplitude and velocity $\omega$ = 60$\pi$ rad/s on axis-$z$. The torque with maximum amplitude of 1e-3 Nm induced by the magnetic interaction forces the body to a periodic motion. Resulting motion is sinusoidal with additional small harmonics. The top trace has cosine appearance (instead of sine) because sample data are evaluated about 8 ms after the simulation starts.

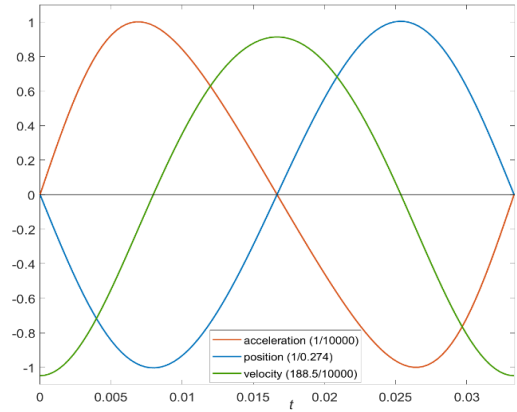

Fig. 5.9. Normalized plot of azimuthal position, angular velocity and acceleration of the body over one cycle of rotating field derived from data of numerical solution 1 from Table 5.4.

A simulation of this scheme presented in Fig. 5.10 shows compliance with the numerical solutions of Eq. 5.56 and also generates stable motion under full DoF. In this simulation, the angular motion is confined around the axis-$z$. In similar simulations where small initial angular velocities are given on axis-$x$, these only add oscillations along this axis but do not destabilize the system. Frequency of these oscillations are obtained around 42 rad/s. Since the mechanism which enforces the orientation of the body orthogonal to rotating field axis is identified as torque-phi effect, this oscillation could be associated with a harmonic motion. This way, torque-phi corresponds to a spring constant $C$ having value between 1.72 and 1.76e-4 Nm/rad using natural frequency formula Eq. 5.1 under small angle approximation where MoI of the body is 0.998e-7 kgm$^2$. Here, lower $C$ values are obtained with tests having larger oscillation amplitudes. These figures are not far from minimum static torque coefficient $\tau_S$ = 1.44e-4 to obtain stable conical motion around axis-$z$ based on heuristic Eq. 5.46.

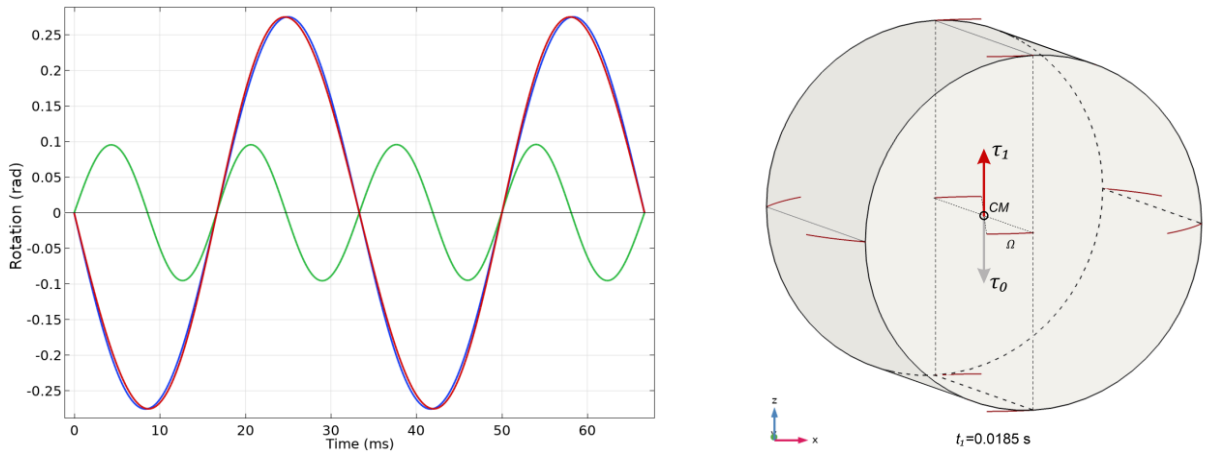

Fig. 5.10. (**L**) Angular motion ($\theta$) of a body around the axis-$z$ subject to a rotating field defined in Eq. 5.54 with velocity $\omega$ = 88.5 rad/s covering two cycles of the rotating field obtained through a simulation. The amplitude $\tau_C$ of the torque is 1e-3 Nm. Body is a cylinder with equal MoI of 0.998e-7 kgm$^2$ endowed with a magnetic moment at CM along the cylinder axis. Body rotation is shown by the red trace having amplitude 0.2756 rad. Blue trace corresponds to a pure sine curve with amplitude 0.27565 rad and the green curve is their differences amplified by 10 times. This shows the presence of a second harmonic with an amplitude 0.03453 of the first. This fits well to the ratio (0.03475) obtained through numerical solution where body's MoI is 1e-3 kgm$^2$ instead of 0.998e-3. (**R**) An overview of a similar simulation where the motion of the cylindrical body is shown with red colored traces. Here, the rotating field axis, body rotation axis and torque vector are aligned with the axis-$z$. The torque vector is shown at instances $t_0$ and $t_1$ near to the half cycle of the rotating field.

### 5.1.6 Motion characteristics of a spinning body under PFR

A body conforming the angular motion defined in Fig. 5.1 can also have a spin around its magnetic axis. This spin can be defined in rotation matrix Eq. 5.6 through the rotation variable $\rho$. Setting $\rho$ as opposite and equal to $\theta$ produces zero spin and any other value introduces a spin. Since $\theta$ is defined as $\omega t$, $\rho$ can be defined as $k\omega t$ where $k$ is a real constant. Therefore $k = -1$ corresponds to zero spin as is set in Eq. 5.7. The spin can be introduced by initial conditions or by a driving mechanism. This frequently happens in experiments naturally (see Section 6.5) and may lead to complex motions.

By substituting terms $\theta$ and $\rho$ with $\omega t$ and $k\omega t$ in rotation matrix $R_E$ defined in Eq. 5.6 and denoting it as $R_k$, angular velocity $\vec{v}$ of the body and its angular acceleration $\vec{\alpha}$ are obtained by following the logic of equations through 5.10 to 5.13.

$$\vec{v} = \omega \begin{bmatrix} k s_{\omega t} s_\varphi \\ -k c_{\omega t} s_\varphi \\ 1 + k c_\varphi \end{bmatrix} \qquad (5.62)$$

$$\vec{\alpha} = \vec{v}' = k\omega^2 s_\varphi \begin{bmatrix} c_{\omega t} \\ s_{\omega t} \\ 0 \end{bmatrix} \qquad (5.63)$$

Since this spin is on the magnetic axis, it does not affect magnetic interactions and Eq. 5.22 is also valid here. Also, the magnetic torque vector derived from there and the angular acceleration vector above can be expressed using the same unit vector. This gives the possibility to obtain the motion defined by $R_k$ through the present magnetic interaction. By applying Euler's equations of motion Eq. 5.15 using transformed tensor of inertia as $I_T = R_k I R_k^T$ and by equaling the result to the magnetic torque from Eq. 5.22, we obtain

$$\tau_C c_\varphi + \tau_S s_\varphi = I_T \vec{\alpha} + \vec{v} \times I_T \vec{v} = \omega^2 s_\varphi \left( -k I_A + (I_R - I_A) c_\varphi \right) \qquad (5.64)$$

which is different from Eq. 5.35 by the addition of multiplier $-k$ for $I_A$. This equation is time independent because magnetic and the inertial torques share the same unit vector which rotating on $xy$ plane around $z$ with velocity $\omega$. Maybe the interesting part of this equation is the component $-k\omega^2 s_\varphi I_A$ since it carries the polarity of $k$. Simulations based on specific parameters show that most of stable solutions lie on negative values of $k$. As stated above, $k = -1$ corresponds to zero spin and angular velocity vector for this case is given in Eq. 5.12. In this condition, a laser tachometer registers zero when directed to a patch on this body. On the other hand, the body performs a simple rotation around axis-$z$ as shown in simulation Fig. 5.11(b) for $k = 0$ which is reflected by the vector $\vec{v} = \omega \mathbf{k}$ by Eq. 5.62. In this motion, applied torque and the rotation vector of the body are in the same phase (the phase lag is absent) which is reflected by negative sign of the angle $\varphi$ which corresponds to tilt of the cylinder. Here, we define the term $v_S$ as *spin velocity* which correspond to a tachometer reading (rad/s) as

$$v_S = (k + 1)\omega \qquad (5.65)$$

and the *spin ratio q* as

$$q = v_S/\omega = k + 1 \qquad (5.66)$$

For bodies having uniform MoI, Eq. 5.64 can be expressed as

$$\tau_C c_\varphi + (\tau_S + k\omega^2 I) s_\varphi = 0, \qquad I = I_R = I_A \qquad (5.67)$$

This equation indicates that the motion is no longer a harmonic motion when $k$ is zero, since neither MoI nor $\omega$ do not enter the equation and this merely gives a static equilibrium of $\tau_C$ and $\tau_S$ on a co-rotating reference frame.

Angle $\varphi$ can be derived from equation above as

$$\varphi = -\tan^{-1} \frac{\tau_C}{kI\omega^2 + \tau_S} \qquad (5.68)$$

This shows that the parameter $k$ related to the spin has a primary effect on the angle $\varphi$ since the term $\tau_S$ is typically small compared to the term with $k$. In turn, this varies the strength of the PFR proportional to $\sin \varphi$ as it is shown in Eq. 5.71 in Section 5.2.

It is previously shown that the symmetrical motion ($\lambda = 0$) of the body around axis-$z$ cannot be sustained in absence of static field $B_S$ due to phi-torque effect; however, this changes in presence of spin. According to simulations, it is found that the body might keep its symmetrical motion around axis-$z$ when it spins. Fig. 5.11 shows some results from these simulations. No precession observed on motions of first two simulations (motion of the third is too complex for to determine the precession) and this can be explained by torque-phi is zero when angle $\lambda$ is zero.

While Eq. 5.64 and Eq. 5.67 give the equilibrium condition for obtaining circular angular motion of the body, they don't tell about stability of the motion. Stability characteristics are explored by a series of simulations where stability results are mapped into ($k$, $\tau_S$) space while keeping other parameters constant. A cylindrical body having equal principal MoI is used in these simulations and initial conditions are calculated based on Eq. 5.67. Simulations results are summed in Fig. 5.12. In this map, the stable zone narrows while $k$ goes to positive direction and terminates at about $k$ = +0.1. The map also gives the upper limit of $\tau_S$ as function of $k$ which is a straight line crossing the top right corner of the map. This limit might be associated with term ($\tau_S + k\omega^2 I$) of Eq. 5.67 which should less than −1.6e-3 for $\omega$ = 80$\pi$, $\tau_C$ = 1e-3, $I$ = 1e-7. The Eq. 5.67 can be also used to obtain interdependency between $k$ and $\omega$ when the right side of the below equation is kept constant.

$$-k\omega^2 = \frac{\tau_C c_\varphi + \tau_S s_\varphi}{I s_\varphi} \quad (5.69)$$

This equation is used to evaluate experimental test results in Section 6.5. Simulations show that a stable motion around axis-$z$ can be obtained with both signs of $v_S$ and requires a minimum spin velocity when the static field is absent. For the case of the simulation presented in Fig. 5.11(a), it is found that the motion is stable for $q \geq 0.30$ in the same direction and $q \geq 0.27$ in the opposite direction where $\omega$ is $80\pi$ rad/s.

The spin also allows to obtain stable equilibrium when the static field is reversed; that is, the torque induced by this field tries to deviate the body dipole from the rotation axis, corresponding to a negative stiffness. This stability offered by the spin, which is available even the body has uniform MoI needs further evaluation. According simulation results shown Fig. 5.12, the range of the static field strength and its polarity for obtaining stable angular motion varies by spin velocity and this range is extended to cover negative values of the static torque, that is, presenting a negative stiffness. Similarly, under DHM with parametric excitation, it is possible to obtain stability through negative static stiffness like the well-known inverted pendulum solution [21].

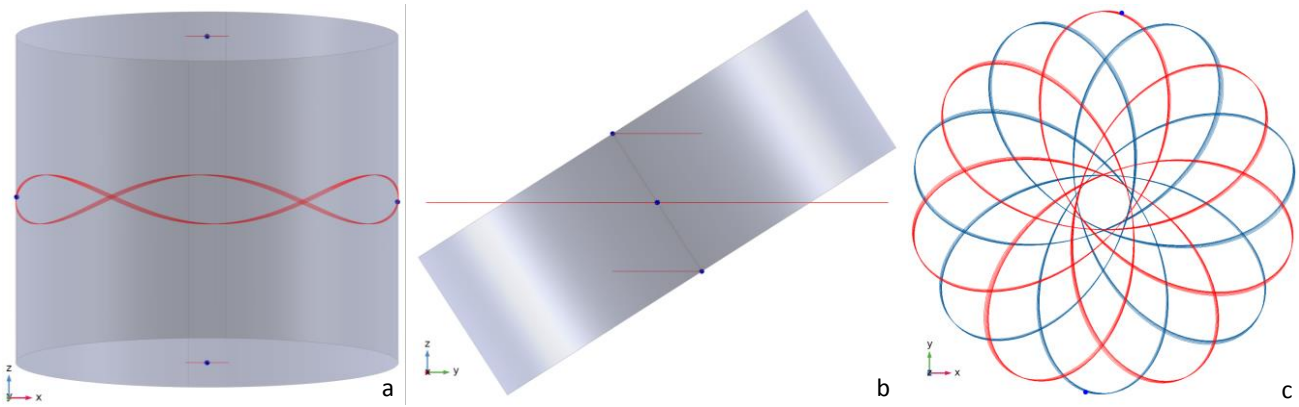

Fig. 5.11. (**a**) A simulation showing motion of a spinning cylindrical body (⌀5.245×9.087 mm) having magnetic moment aligned on axis-$z$ and having uniform MoI (8.1e-8 kg m$^2$) is subject a rotating field ($\omega$ = 80 $\pi$ rad/s CCW) having no $z$ component ($\tau_C$ = 1e-3 Nm, $\tau_S$ = 0). Body's initial conditions are: $\varphi$ = 0.12962 rad, $\vec{v}$ = *48.71 **j** +122.28 **k***. The vector $\vec{v}$ is chosen in order the body obtains a symmetric angular motion around axis-$z$ while its spin is half of the rotating field but in the opposite direction ($q$ = −0.5). The ribbon like trace corresponds to overlapped trajectories of middle side points (blue dots) of the cylinder. Traces of top and bottom centers are also visible as flat lines which are precise circles on the *xy* projection. This simulation runs about 450 ms, corresponding to 22.5 cycles of the rotating field. Simulation shows that the body obtains a symmetric angular motion around axis-$z$ despite absence of static component of the rotating field. (**b**) A similar simulation where body size is ⌀7×5 mm, MoI's $I_A$ = 1.4127e-7, $I_R$ = 8.2654e-8, $v$ = $\omega$ = 80 $\pi$ rad/s CCW. Motion is found stable and symmetric around axis-$z$ where angle $\varphi$ is −0.5706 rad. Simulation ran for 275 ms. Trajectories of selected points (actually any point belong to the body) draw circles parallel to the *xy* plane since the body has a simple motion around axis-$z$ ($q$ = 1). (**c**) Motion of the same body but with different initial conditions where $\vec{v}$ = *−30.28 **j** −373.29 **k*** (rad/s), $\varphi_0$ = 0.24337 rad and $\omega$ = 253.15 rad/s. While the motion appears stable (run time = 348 ms) , the body speeds up about 0.5 rad/s during this period. Red and blue traces correspond to top and bottom centers points, respectively. In both simulations, the stability of the motion around axis-$z$ is obtained by the spin of the body since the rotating field has no static component.

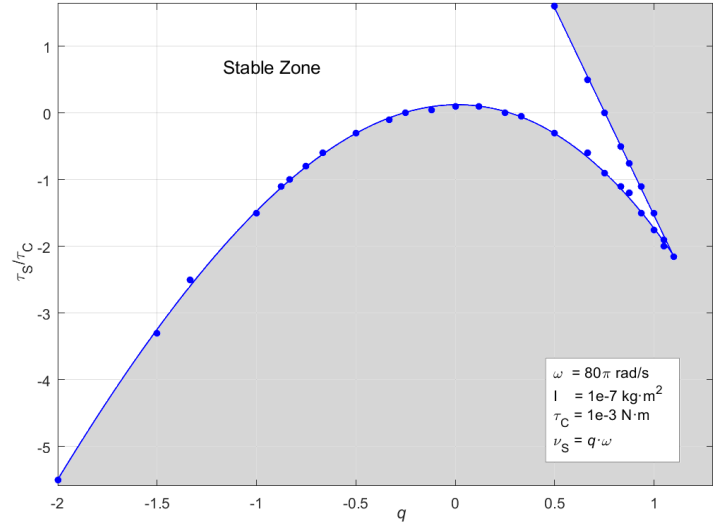

Fig. 5.12. Dependence of the motion stability of a rotating body subject a rotating torque to the body spin and to the static component of the torque for a configuration where the body is a homogeneous cylindrical object having equal principal MoI equal to 1e-7 Nm$^2$. The rotating torque assumed be obtained by a homogeneous rotating magnetic field around axis-$z$ with an angular velocity 80 $\pi$ rad/s interacting with the magnetic moment of the body centered at origin. This torque is defined by cyclic and static torque coefficients $\tau_C$ and $\tau_S$. This chart is obtained by running the simulation for various values of parameter $\tau_S$ and body spin velocity $v_S$ which are presented by parameter $q$ corresponding to ratio $v_S/\omega$. Each data point on the curve which separates stable and unstable zones correspond to parameter pairs providing stable motion. These stable data points are paired with unstable data points at proximity but residing in the unstable region (not shown). This chart only covers cases where the angular motion of the body is in the opposite phase of the driving field, conforming the PFR scheme. There are also cases where body motion is stable but in the same phase of the rotating field where parameter $q$ is greater than one. These cases also cover negative $\tau_S$ values. The parabolic part of the curve fits to a polynomial function of third degree $f(x) = \sum a_i x^i$ with R-square better than 0.999 where coefficients $a_0$ to $a_3$ are (0.12, 0.04, -1.72, -0.17).

Solutions covering negative static torques may have further importance when the corresponding field is inhomogeneous and can provide a repulsive interaction through static field alone while the stability is provided by the rotating field and the body spin. A special case where the parameter $q$ = 1 which corresponds to a simple rotation of the body around axis-$z$ is found stable within a narrow range of $\tau_S/\tau_C$ ratio −1.5 to −1.8 within this configuration. While simulations show stable solutions with zero and negative static torque values in presence of spin, they are not experimentally tested. However, if achievable, such solutions may extend possibilities of applications like particle trapping and magnetic bearing. It might also be worth to evaluate the motion stability of a body having magnetic moment with both cyclic and static components and exposed to a static field. Such a trapping solution is presented in Section 5.7 with positive static stiffness but no experiment was designed for exploring the case where a static stiffness is negative.

### 5.1.7 Summary

In this section, it is shown that a body endowed with a dipole moment and subjected to a homogeneous rotating magnetic field can obtain a stable angular motion synchronized with the field and its magnetic alignment with the rotating field can be antiparallel (>90°) which corresponds of a phase difference equal to $\pi$ between these motions. This phase is called *phase lag* and is related to DHM where the driving frequency is above the natural frequency $\omega_0$ of the system. This natural frequency is determined by the strength of the static component of the field and also varies with amplitude of the motion in nonlinear systems. The minimum of this field for obtaining a stable conical motion around axis-$z$ in absence of body spin is given by Eq. 5.46 and Eq. 5.48 with different approaches and fits to simulation results. Presence of spin removes the requirement of the static field for the above condition. In simulations, the strength of the rotation field is chosen large in general for evaluating the stability of the motion in the limits where the amplitude of the angular motion $\varphi$ can be large as 0.3 rad. In these conditions, the strength of the static field needs to be large up to 1/4 of the rotating field. On the other hand, Eq. 5.24 gives the upper limit of $\tau_S$ by the condition $\tau_S < I\omega^2$. Natural frequency of the system goes above the driving frequency when this condition is not met and the phase lag condition vanishes. In experiments, the angle $\varphi$ is less than 0.2 rad in general and small angles like 0.05 rad are sufficient to obtain a stable levitation. Simulations and slow-motion visuals from experiments validate the presence of phase lag condition. This condition is the key requirement for obtaining PFR as explained in the following section. So far, simulations results are found in full accordance with the models defined in Eq. 5.7, Eq. 5.56 and Eq. 5.484.

## 5.2 Evaluation of magnetic forces on a body subject to inhomogeneous rotating field

This is the second step of analysis of the PFR model. Here, it is assumed that the conical motion of a body when subjected to a homogenous rotating field is also valid when the rotating field is inhomogeneous and belongs to a rotating dipole. Briefly, the previous section considers magnetic torques and this one adds magnetic forces. Here, we evaluate the simple case where the body draws a conical motion with a fixed angle $\varphi$; that is, the axis of the motion is aligned with axis-$z$. Since the motion is synchronized with the rotating field, this allows a static evaluation of the magnetic interaction on a co-rotating reference frame around axis-$z$ where the rotating field and the body stand still. Fig. 5.13(a) shows this picture. Item called *rotator* is a dipole magnet producing the rotating field and the *floator* is another dipole magnet interacting with this field. This is the same configuration given at Fig. 5.1(R) at top of Section 5. According to the equation of the motion, the floator is tilted CW on axis-$y$ by angle $\varphi$ while torque $\tau_C$ generated by the magnetic interaction forces the floator to rotate CCW on this axis. This alignment is fixed in the co-rotating reference frame on axis-$z$; that is, floator's S pole N never turn toward rotator's S pole despite the torque $\tau_C$. This unusual picture obviously cannot happen in magnetostatics but happens here because the angular acceleration of the body and its angular displacement are in opposite phases as the result of the phase lag condition.

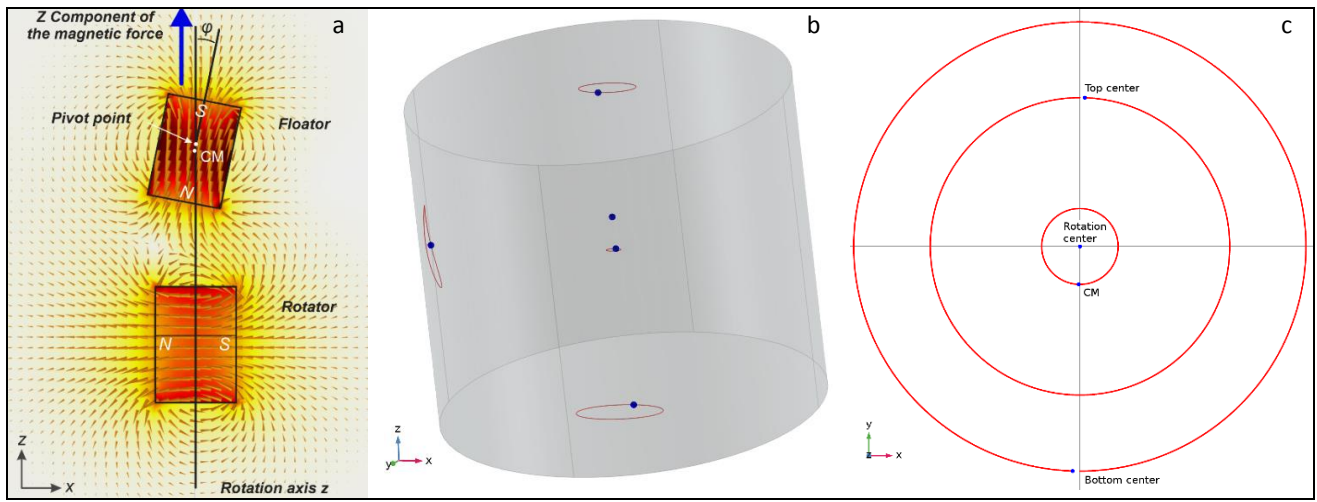

Fig. 5.13. (**a**) A visualization of magnetic field and alignment of the floator in the interaction with the rotator in a basic configuration shown on the plane where magnetic moments lies. Due to the synchronized angular motion of the floator with the rotator rotation, this alignment is constant while this plane rotates around axis-$z$. The floator angular motion is accompanied with a translational motion since the rotator field also induce a force. These combined motions cause the floator's rotation center (denoted as pivot point) to shift in $z$ direction. In this alignment, floator's N pole is always held toward rotator's N pole and cause a steady repulsion in $+z$ direction. Here, it can be also seen that effects of the angle $\varphi$ and the shift of the floator's CM (CM coincides by its magnetic moment) add up on this repulsion force. The motion of floator is called conical motion and the angle $\varphi$ is retained all the time when the motion is have axis-$z$ symmetry. This instance corresponds to $\omega t = \pi/2$ of Eq. 5.25, with reversed magnetic poles and to angle $\gamma = 0$. (**b, c**) Simulation result of a configuration similar to (a) where floator and rotator magnetic models are approximated as point dipoles. On this configuration, rotator is tilted around axis-$y$ with a relatively large angle $\gamma$ about 0.159 rad (9°), which induce a static field component ($B_S$ in direction of axis-$z$), corresponding to ratio $B_S/B_C$ equal to 0.16 according Eq. 5.27. This ratio is above the minimum ratio 0.107 (Eq. 5.46) to keep the axis of the conical motion on axis-$z$ Floator's conical motion is characterized by angle $\varphi$ equal to 0.2154 rad and by upward shift of the RC by 0.964 mm from CM. This shift can be seen also in these figures where CM circles around axis-$z$ and by circle of bottom center point being larger than of the top center point. Rotating field velocity is $60\pi\ rad$/s and torque coefficients $\tau_C$ and $\tau_S$ are 7.5e-4 and 1.2e-4 Nm, respectively. Body has equal principal moments of inertia as 0.999e-8 kg m². Initial angular velocity vector of the body for obtaining this motion is $\omega_F = 0.0045\mathbf{i} + 40.295\mathbf{j} + 4.355\mathbf{k}$ rad/s. The simulation ran about 683 ms. Since the position

and orientation of the floator's magnetic moment is constant in a co-rotating reference frame with the field, torque and force received from the field are also constant.

Since the magnetic field of the rotator has a gradient, floator also experiences a force additional to the torque. This force can be evaluated by interaction of two magnetic moments, rotator's $\vec{m}_R$ and floator's $\vec{m}_F$. By denoting the spatial vector connecting these moments as $\vec{r}_{RF}$, these vectors can be expressed according to Eq. 5.25 on a co-rotating reference frame by setting $\omega t = \pi/2$ and under condition $\gamma = 0$ as

$$\vec{m}_R = m_R \boldsymbol{i}, \qquad \vec{m}_F = m_F(\sin\varphi\, \boldsymbol{i} + \cos\varphi\, \boldsymbol{k}), \qquad \vec{r}_{RF} = r\boldsymbol{k} \tag{5.70}$$

In the actual motion, the vector $\vec{r}_{RF}$ have a small $x$ component as a result of the circular lateral motion caused by these forces. Here, this displacement is not taken into account in order to simplify the analysis aimed in obtaining a figure about the dependency of forces to the component-$z$ of the distance. Similarly, this factor is also not taken into account in the calculation of the angle $\varphi$. By entering above vectors into Eq. 4.2, a force figure is obtained as

$$\vec{F}_{RF} = \frac{3M}{r^4}(\cos\varphi\, \boldsymbol{i} + \sin\varphi\, \boldsymbol{k}), \qquad M = \frac{\mu_0\, m_R\, m_F}{4\pi} \tag{5.71}$$

Here, the component-$z$ of the force is in direction of $\vec{r}_{RF}$, means repulsive and is constant. Therefore, it can be balanced by a counteracting static force. While the angle $\varphi$ can be obtained from Eq. 5.24, we need to calculate torques from dipole-dipole interaction using Eq. 4.3 based on point dipole approximation. By using the configuration at Eq. 5.70, the torque received by the floator reads

$$\vec{\tau}_C = -\frac{M}{r^3}\cos\varphi\, \boldsymbol{j} \tag{5.72}$$

This figure is in accordance with torque obtained in Eq. 5.22 by following the setting $\omega t = \pi/2$ above. Here, the fractional term corresponds to cyclic torque coefficient $\tau_C$ as

$$\tau_C = \frac{M}{r^3} \tag{5.73}$$

For instance, we may keep the gradient free static field as is, without associating with a dipole. This way, we avoid the force from this static component since this calculation aims to find the force caused by the rotating dipole. By substituting $\tau_C$ from above to the Eq. 5.24, we obtain

$$\varphi = \tan^{-1}\frac{M}{r^3(I\omega^2 - \tau_S)} \tag{5.74}$$

The term $\sin\varphi$ which required for calculating $z$ component of the force can be written as

$$\sin\varphi = \left(\left(\frac{r^3(I\omega^2 - \tau_S)}{M}\right)^2 + 1\right)^{-1/2} \tag{5.75}$$

Similarly the term $\cos\varphi$ can be written as

$$\cos\varphi = \left(\left(\frac{M}{r^3(I\omega^2 - \tau_S)}\right)^2 + 1\right)^{-1/2} \tag{5.76}$$

This way, the Eq. 5.71 reads

$$\vec{F}_{RF} = \frac{3}{r\sqrt{r^6/M^2 + 1/(I\omega^2 - \tau_S)^2}}\boldsymbol{i} + \frac{3M^2}{r^7\sqrt{M^2/r^6 + (I\omega^2 - \tau_S)^2}}\boldsymbol{k} \tag{5.77}$$

For obtaining a clearer figure about the dependency of the PFR (component-$z$ of $\vec{F}_{RF}$) on the distance between rotator and floator, some distance ranges are defined and effective power factors for these ranges are obtained using curve fitting. The first range starts with covering largest angles $\varphi$ obtained in experiments. The second range covers large angles, the third, common angles and the fourth, about small angles which correspond to significantly weak repulsion figures. The procedure consists of obtaining data points of $sin\varphi$ and distance $r$ for selected ranges of $\varphi$ where $r$ values are calculated by the inverse relation obtained from Eq. 5.74 as

$$r(\varphi) = (C \tan \varphi)^{-1/3}, \qquad C = \frac{(I\omega^2 - \tau_S)}{M} \tag{5.78}$$

Then curve fitting is applied by setting $r$ data for $x$ and $sin\varphi$ data for $f(x)$ and coefficients $a$ and $b$ are obtained for each range using equation

$$f(x) = \frac{a}{C} x^{-b} \tag{5.79}$$

Here, the value $C$ can be chosen arbitrarily, it is set as $C =$ cotangent $(0.1)$ in order to obtain $r = 1$ for $\varphi = 0.1$ rad in Eq. 5.78, as a reference distance. Results are given in Table 5.5. As we can write

$$\sin \varphi \cong \frac{a}{Cr^b} \tag{5.80}$$

according to the curve fitting model, by substituting this term in Eq. 5.71, this allows to express the force-$z$ as

$$F_Z \cong \frac{M}{r^4} \frac{a}{Cr^b} = \frac{M^2 a}{(I\omega^2 - \tau_S)r^{4+b}} \tag{5.81}$$

In the equation above, it is noteworthy that the repulsive force component $F_Z$ becomes proportional to $M^2$ except for very short distances, in turn, to proportional to square of factored magnetic moments. When these magnetic moments belong to permanent magnets, this gives a relation between PFR strength and magnetic field density of the material by the fourth power. In the Table 5.5, the force factor is calculated by assigning the term $I\omega^2 - \tau_S$ equal to $C^2$. Regarding experiments on PFR, the $\varphi$ range between 0.15 and 0.05 (No. 5) covers most of the tests. It should be noted that on the dependency of $F_Z$ on the distance between magnetic moments, the static torque is assumed constant. However it will be not constant if the static field $B_S$ is generated by the tilt angle $\gamma$ of the $\vec{m}_R$ as given in Eq. 5.28. As $\tau_S$ is proportional to $r^{-3}$, this has a small effect in increasing the power factor of the $F_Z$. This formula gives close figures with simulation parameters in Table 5.6 when the simulations exclude the lateral motion which causes a negative offset on $x$ position of the floator. This is expected because on the first hand this offset is not present in the definition of the vector $\vec{r}_{RF}$ in Eq. 5.70.

| No | $\varphi$ Range (rad) | $\varphi$ Range (deg) | Multiplier $a$ | Exponent $b$ | Relative distance | Force Factor ($z$) (Eq. 5.77) | Force Factor ($z$) (Eq. 5.81) |
|---|---|---|---|---|---|---|---|
| 1 | 0.60 – 0.40 | 34.4 – 22.9 | 1.3 | 2.30 | 0.53 – 0.62 | 71 – 26 | 71 – 26 |
| 2 | 0.40 – 0.25 | 22.9 – 14.3 | 1.072 | 2.69 | 0.62 – 0.73 | 26.3 – 8.8 | 26.2 – 8.8 |
| 3 | 0.25 – 0.15 | 14.3 – 8.6 | 1.007 | 2.88 | 0.73 – 0.87 | 8.8 – 2.6 | 8.8 – 2.6 |
| 4 | 0.25 – 0.10 | 14.3 – 5.73 | 1.001 | 2.91 | 0.73 – 1.00 | 8.8 – 1.0 | 8.8 – 1.0 |
| 5 | 0.15 – 0.05 | 8.6 – 2.86 | 1.000 | 2.97 | 0.87 – 1.26 | 2.6 – 0.2 | 2.6 – 0.2 |
| 6 | 0.10 – 0.02 | 5.73 -1.15 | 1.000 | 2.99 | 1.00 – 1.71 | 1.0 – 0.0235 | 1.0 – 0.0235 |

Table 5.5. Results of curve fitting based on Eq. 5.79 where $x$ and $f(x)$ data are respectively the distance $r$ and $sin \varphi$ for data points corresponding to $\varphi$ range. Force factor columns give results for the exact formula and the approximate formula for comparison. Overall, the deviations are less than 0.5%.

Despite the displacements of the body in lateral directions are ignored in above equations, the lateral component $F_X$ of the force is substantial (Table 5.6) and also causes another DHM having phase lag. That is, the displacement and the acceleration are in opposite phases. As the lateral force pushes the floator in $+x$ direction referring Fig. 5.13, the result is displacement $\beta$ on $-x$. This brings the N poles further closer to each other, causing an increase of the force on axis-$z$ about 20%. This mechanism is evaluated in Section 5.4.6. These lateral and angular motions combined have an effect to shift the RC of the angular motion away from CM in $+z$ direction as seen in the Fig. 5.13 and in the simulation (Fig. 5.14) where the inhomogeneous field is taken into account . As mentioned in the previous section, the static field $B_S$ can be identified as the component-$z$ of the rotating field. Here, by associating the rotating field with a rotating dipole or a magnet, $B_S$ can be obtained by tilting the dipole axis from the rotation plane. In Fig. 5.14(a), the bottom entity is a rotating magnet on the vertical axis ($z$) while its poles are aligned with the rotation plane ($xy$). In (b), the magnet is tilted and its dipole axis has angle $\gamma$ with the $xy$ plane. Due to this tilt, the magnetic field at any point on axis-$z$ has non-zero $z$- component while it is zero in the configuration at (a). It can be shown that the average of the field over one cycle of rotation around axis-$z$ corresponds to a field of a dipole aligned with axis-$z$ having strength equal to sin $\gamma$ of the rotating dipole. This equivalent dipole obtained by time integral is called *virtual dipole* here. This way, field of the rotating dipole can be separated in two components as

$$\vec{B} = \vec{B}_C + \vec{B}_S = B \begin{bmatrix} -c_{\omega t} c_\gamma \\ -s_{\omega t} c_\gamma \\ 0 \end{bmatrix} + B \begin{bmatrix} 0 \\ 0 \\ s_\gamma \end{bmatrix} \quad (5.82)$$

which is valid at any point in space. According Eq. 4.1, the relation between magnetic field and magnetic moment is linear and allows to write

$$\vec{B}(\vec{m}_1 + \vec{m}_2, \vec{r}) = \vec{B}(\vec{m}_1, \vec{r}) + \vec{B}(\vec{m}_2, \vec{r}) \quad (5.83)$$

as long as $\vec{m}_1$ and $\vec{m}_2$ originate from the same coordinates. Therefore we can separate a magnetic moment $\vec{m}$ to two components as

$$\vec{m} = \begin{bmatrix} c_{\omega t} c_\gamma \\ s_{\omega t} c_\gamma \\ s_\gamma \end{bmatrix} m = \vec{m}_c + \vec{m}_s, \qquad \vec{m}_c = m c_\gamma \begin{bmatrix} c_{\omega t} \\ s_{\omega t} \\ 0 \end{bmatrix}, \qquad \vec{m}_s = m s_\gamma \begin{bmatrix} 0 \\ 0 \\ 1 \end{bmatrix} \quad (5.84)$$

This allows to separately calculate magnetic interactions for $\vec{m}_c$ and $\vec{m}_s$, and add them to obtain same result of the $\vec{m}$. This way, the cyclic torque $\vec{\tau}_C$ and the static torque $\vec{\tau}_S$ where equations of motion are based on previous section can be associated with $\vec{m}_c$ and $\vec{m}_s$, respectively (Eq. 5.27, 5.28). Here, the term $\gamma$ denotes the constant angle between the rotating field moment $\vec{m}$ and the *xy* plane and called *tilt angle*.

### 5.2.1 Integrated magnetic and rigid body dynamics simulations of PFR

In experimental configurations, the distance between two interacting magnets is in the same order of magnets dimensions, so some errors can be expected in calculations based on point dipole model. For this reason, FEM based magnetic simulations are used to obtain better force and torque figures. This way, physical properties of magnets and figures from magnetic simulations used as input in motion simulations and spatial results from motion simulation are used for configuring the magnetic simulation in a mutual relation. Using a series of integrated magnetic and rigid body dynamics simulations, it was possible to obtain more precise dependency figures allowing some generalizations. Fig. 5.14 shows an overview of an integrated simulations pair. Data of this simulation is given in Table 5.6, no 5. In figures (a) and (b), the bottom magnet is the rotator and the top one is the floator. In the first figure, the angle $\gamma$ is zero therefore the rotator's static moment component $\vec{m}_s$ is zero too. Poles of magnets are not marked there, but it can be seen that the floator's bottom pole and rotator's right pole have the same polarity through the field density and from arrows directions of the field. The force experienced by the floator is shown by the blue arrow. This force pushes the floator to the left (3.15 N) and up (0.852 N). Regarding the horizontal force, this would cause a large acceleration about 472 m/s². Despite this, the floator involves a circular motion around axis-$z$ with a radius $\beta$ equal to 0.362 mm. In the second simulation, rotator's tilt angle is chosen as $\gamma = -0.151$ rad in order the component-$z$ of the force becomes zero. In this condition, floator is kept stable while performing conical motion around axis-$z$ as shown in Fig. 5.14(c, f), obtained from motion simulation. In figure (c), floator is shown from a side with trajectories of top and bottom centers, RC, CM and a middle point on the side. These trajectories are also shown (except the side point) as projected on the *xy* plane in figure (f). Floator - rotator distance is kept constant due the stability is ensured by the negative slope of the force versus distance as seen in figure (d). This curve is the characteristic of the magnetic bound state and fits to evaluated model in form of

$$F_z(r) = a r^{-b} - c r^{-d} \quad (5.85)$$

where the term $r$ denotes the distance between magnet centers in $z$ direction, the first term corresponds to PFR and the second to the attraction force between $z$ components of magnetic moments. Here, all parameters have positive values and power factors have values $b = 7.5 \pm 0.4$, $d = 3.8 \pm 0.4$, covering simulation results and experimental observations in this work. The power factor $b = 7.37$ obtained in this simulation is slightly larger than the power factors given in Table 5.5 based on point dipole model in Eq. 5.77 and Eq. 5.81. Similar to the force component-$z$, the component-$x$ for a specific configuration fits the model of Eq. 5.85 as

$$F_x(r) = 6.7e\text{-}7 \, r^{-4.16} - 6e\text{-}13 \, r^{-7.34} \tag{5.86}$$

where $r$ is the $z$ distance in meters between magnet centers. Here, the second power term is related to dependence of the angle $\varphi$ and the offset $\beta$ on distance according to data from motion simulation but only effective at very short distance where magnets are about in touch or stability might not be possible.

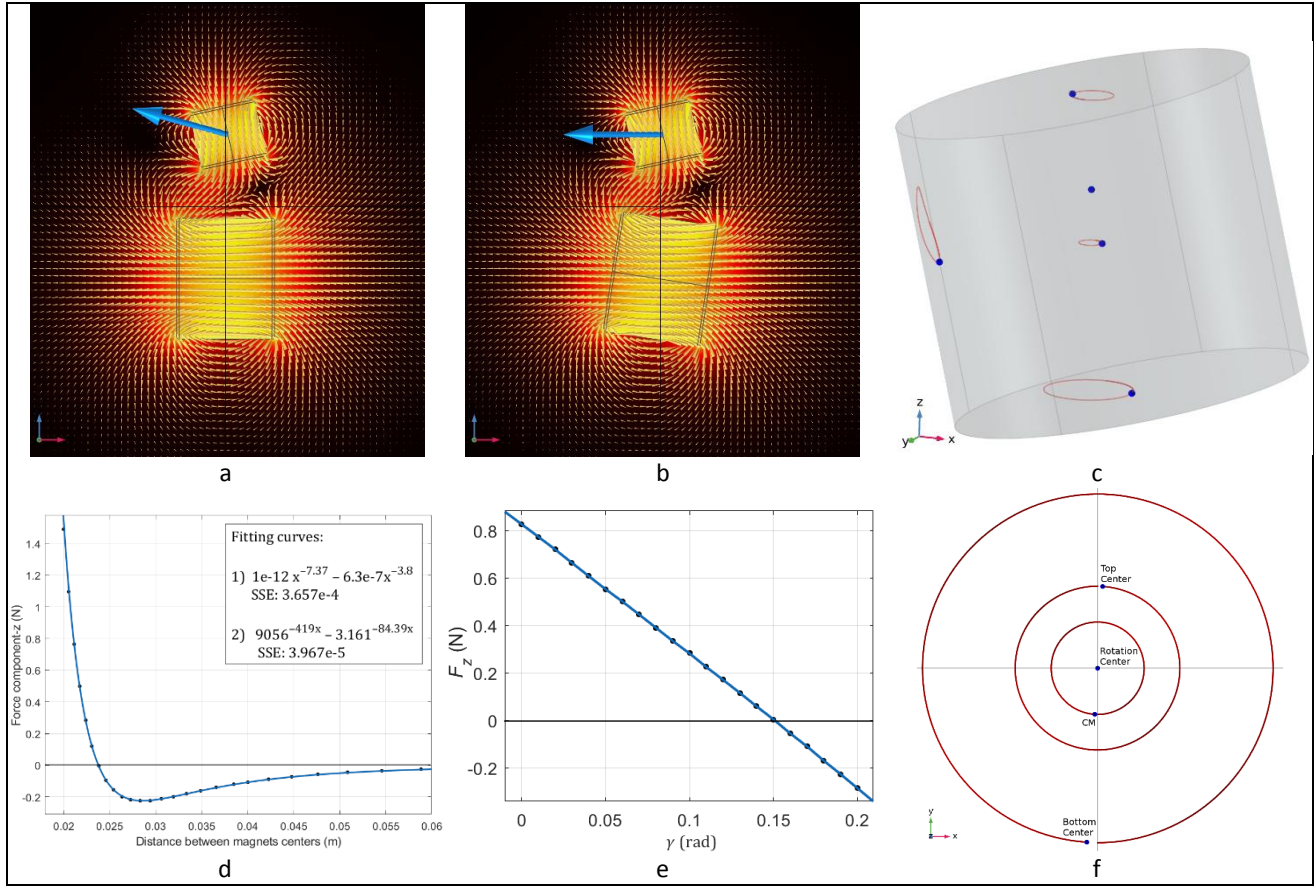

Fig. 5.14. An integrated magnetic (a, b, d, e) and rigid body dynamics simulation (c, f) of PFR and MBS. Alignment of bodies, torque and component-$x$ of the forces are shared between simulations. (**a**) Floator (top magnet) experiences torque and force (blue arrow) from the rotator (bottom magnet) which is a dipole on the $xy$ plane supposed to rotate around axis-$z$. The induced rotating torque generates the conical motion with angle $\varphi$=0.213 rad. This motion also has a lateral component $\beta$ (0.37 mm in amplitude) due the force component rotating on the $xy$ plane. This has an effect to shift the center of the angular motion (RC) away from rotator about 0.18 times of the body height from CM as 1.7 mm. The phase lag condition can be seen here within lateral and angular positions and corresponding acceleration vectors where the bottom pole of the floator approaches the pole of the rotator having the same polarity despite forces and the torque trying the opposite. As a result, the component-$z$ of the force get positive sign; that is, away from rotator. (**b**) The requirement of static field for alignment of the axis of the conical motion with axis-$z$ is obtained by the tilt (angle $\gamma$) of the rotator, causing one pole to look slightly up and the other down. The effect of this tilt is the generation of a magnetic moment in $z$ direction, constant in time. This moment generates the static field $B_S$ responsible for static torque $\tau_S$ and also generates an attractive force in $z$ direction when the floator is aligned to it in parallel. In the simulation (b), the angle $\gamma$ is set to 0.151 rad (6.3°) in order to obtain component-$z$ of the force between magnets as zero on this specific distance ($r$ = 2.38 mm) between magnets centers. Complete parameters of this simulation can be found in Table 5.6, No. 5. (**c, f**) Motion figures of the floator. The torque and the force responsible for this motion is obtained by its interaction with the rotator through magnetic simulation where orientation and position data is carried from simulation of rigid body dynamics. Red traces are trajectories of selected points (blue dots) on the floator projected on $xz$ and $xy$ planes. (**d**) Variation of component-$z$ of the magnetic force versus distance between body centers in configuration b where the tilt angle and $x$ offset of the floator satisfying both simulations. (**e**) Variation of component-$z$ of the force with the tilt angle $\gamma$ of the rotator where configurations corresponding to $\gamma$ = 0 and 0.151 rad are shown in figures (a) and (b), respectively.

Regarding simulation results given in Table 5.6, $\tau_S/\tau_C$ ratios in all simulations satisfy their minimum to obtain aligned axis of the conical motion with axis-$z$ (angle $\lambda$ = 0) by a margin of two or three times of this minimum ($R_{MIN}$). This margin allows smaller tilt angles of rotator without stability issues, in order to produce larger net repulsion force at the same distance which can balance the weight of the floator when it hovers over the rotator.

| | | 1 | 2 | 3 | 4 | 5 | 6 | 7 | 8 | 9 | 10 | 11 | 12 | 13 | 14 | 15 |
|---|---|---|---|---|---|---|---|---|---|---|---|---|---|---|---|---|
| | | | R×H (mm) | Mass (g) | I (Nm²) | Moment (Am²) | $\omega$ (rad/s) | Dist. (mm) | $\varphi, \gamma_e$ (rad) | $\beta$ (mm) | $\tau_C, \tau_S$ (N·m) | $\tau_S/\tau_C$, $R_{MIN}$ | $F_X$ $\gamma=0, \gamma_e$ (N) | $F_Z$ $\gamma=0$ (N) | Ref. F (N) | $F_Z$ (DP) (N) |
| 1 | Flo | | 5.47×9.48 | 6.67 | 1e-7 | 0.8 | 589.2 | 33 | 0.2 | 0.25 | 6.74e-3 | 0.2 | -0.622 | 0.145 | 0.972 | 0.131 |
| | Rot | | 10×10 | 23.5 | - | 2.83 | | | 0.104 | | 1.35e-3 | 0.1 | -0.632 | | | |
| 2 | Flo | | 5.47×9.48 | 6.67 | 1e-7 | 0.8 | 1100 | 28 | 0.0946 | 0.156 | 1.135e-2 | 0.1027 | -1.264 | 0.148 | 1.76 | 0.129 |
| | Rot | | 10×10 | 23.5 | - | 2.83 | | | 0.0513 | | 1.166e-3 | 0.0471 | -1.272 | | | |
| 3 | Flo | | 5.47×9.48 | 6.67 | 1e-7 | 0.8 | 837.33 | 38 | 0.1 | 0.113 | 6.94e-3 | 0.114 | -0.5213 | 0.059 | 1.00 | 0.052 |
| | Rot | | 10×16 | 37.7 | - | 4.52 | | | 0.0569 | | 7.91e-4 | 0.050 | -0.5243 | | | |
| 4 | Flo | | 5.47×9.48 | 6.67 | 1e-7 | 0.8 | 995.6 | 28 | 0.18 | 0.27 | 1.728e-2 | 0.229 | -1.75 | 0.3715 | 3.17 | 0.407 |
| | Rot | | 10×16 | 37.7 | - | 4.52 | | | 0.114 | | 3.957e-3 | 0.088 | -1.71 | | | |
| 5 | Flo | | 5.47×9.48 | 6.67 | 1e-7 | 0.8 | 1129 | 23.8 | 0.213 | 0.362 | 2.58e-2 | 0.304 | -3.15 | 0.852 | 5.69 | 0.919 |
| | Rot | | 10×16 | 37.7 | - | 4.52 | | | 0.151 | | 7.85-3 | 0.102 | -3.28 | | | |

Table 5.6. Integration data of simulations of interaction of rotator and floator within conical motions. Col. 14 gives the force between magnets in coaxial configuration as a reference figure for a comparison with the PFR force (Col. 13). Col.15 gives $F_Z$ values based on point dipole formula. Col. 12 gives values for the *x* component of the force for two rotator tilt configurations such as shown in Fig. 5.14(a, b).

### 5.2.2 PFR within the motion with zenith angle π/2

In absence of the static field $B_S$ and the spin, the motion can obtain stability at angle $\varphi = \pi/2$ as evaluated in Section 5.1.5. In experiments where the floator is held in air in this scheme, the floator cannot be kept centered on axis-*z* in absence of $B_S$. For this reason, a dipole magnet having horizontal orientation is placed above in such setups as shown in Fig. 5.17(a). This stability mechanism is explained later in this section. In this configuration, by omitting the presence of this magnet, force acted on the floator can be calculated as follows: In compliance with definitions Eq. 5.25 and Eq. 5.52, rotator's moment $\vec{m}_R$, floator's moment $\vec{m}_F$, and the spatial vector $\vec{r}_{RF}$ connecting these moments can be chosen as

$$\vec{m}_R = m_R \begin{bmatrix} s_{\omega t} \\ -c_{\omega t} \\ 0 \end{bmatrix}, \qquad \vec{m}_F = -m_F \begin{bmatrix} s_\theta \\ -c_\theta \\ 0 \end{bmatrix}, \qquad \vec{r}_{RF} = \begin{bmatrix} 0 \\ 0 \\ 1 \end{bmatrix} \qquad (5.87)$$

where the time dependence of the angle $\theta$ can be obtained from numerical solution of the equation of motion defined by Eq. 5.57. As the floator is centered on axis-*z*, the vector $\vec{r}_{RF}$ is orthogonal to magnetic moments, dot product of terms $\vec{r}$ and $\vec{m}$ becomes zero in force equation 4.2 and it simplify as

$$\vec{F}_{RF} = \frac{3\mu_0 m_R m_F}{4\pi r^4} \hat{r}(\hat{m}_R . \hat{m}_F) \qquad (5.88)$$

By denoting the scalar term by variable *A* and by applying vector definitions from Eq. 5.87, we obtain

$$\vec{F}_{RF} = A \cos(\theta(t) - \omega t) \, \bm{k} \qquad (5.89)$$

By substituting angle $\theta$ from Eq. 5.57, this reads

$$\vec{F}_{RF} = A \cos\left(\sum_i a_i \sin(i\omega t) - \omega t\right) \bm{k} \qquad (5.90)$$

The average force over one cycle of rotating field ($t=2\pi/\omega$) can be specified as

$$F_{av} = A \frac{\omega}{2\pi} \int_0^{2\pi/\omega} F_{RF}(t) dt \qquad (5.91)$$

The result of this integral for a configuration where $J_C = \tau_C/I = $ 1e4 m/s² and for a range $\omega$ is shown in table 5.7.

| No | $J_C$ | $\omega$ (rad/s) | $\theta_0$ (rad) | $\theta'_0$ (rad/s) | $|\theta|$ (rad) | $F_{av}/A$ |
|---|---|---|---|---|---|---|
| 1 | 10000 | 60π | π | -55.666 | 0.2750 | 0.13688 |
| 2 | 10000 | 65π | π | -51.15 | 0.2358 | 0.11775 |
| 3 | 10000 | 70π | π | -47.31 | 0.2042 | 0.10183 |
| 4 | 10000 | 75π | π | -43.99 | 0.1784 | 0.08926 |
| 5 | 10000 | 82π | π | -40.05 | 0.1497 | 0.07472 |
| 6 | 10000 | 90π | π | -36.33 | 0.1245 | 0.06228 |
| 7 | 10000 | 105π | π | -30.945 | 0.0917 | 0.04623 |
| 8 | 10000 | 120π | π | -26.96 | 0.0703 | 0.03553 |
| 9 | 10000 | 135π | π | -23.89 | 0.0555 | 0.02809 |
| 10 | 10000 | 150π | π | -21.45 | 0.0450 | 0.02275 |
| 11 | 10000 | 170π | π | -18.88 | 0.0350 | 0.01793 |

Table 5.7. Solutions of Eq. 5.56 in form Eq. 5.57 for various rotating field speeds and forces obtained using Eq. 5.91.

The term $A$ actually corresponds to the force generated when magnetic moments become parallel (having same azimuthal angle). The average force $F_{av}$ is found fit to

$$F_{av} \cong \frac{1}{2}\tan^{-1}(J_C/\omega^2) \qquad (5.92)$$

And by referring to Eq. 5.58, this reads

$$F_{av} \cong \frac{A}{2}|\theta| \qquad (5.93)$$

The Fig. 5.15 shows plots of $\theta(t)$ and $F_Z(t)$ on the left figure and the dependence of $F_{av}$ on $\omega$ in the right figure. Note that these force figures fit when the floator has no translational oscillations. However in general, it is not the case because the force $F_Z$ is periodic and drives the floator to an oscillation on axis-z. This is true ($F_{av} \neq 0$) even if the floator don't have the angular oscillation ($\theta(t) = 0$) for a reason. The translational motion has parametric excitation because the force $F_{RF}$ also varies with distance $r$ between moments by the inverse of the fourth power, present in scalar term $A$. The time evaluation equation for this motion can be approximated as

$$m\frac{d^2r}{dt^2} + kr + F_S - \frac{3\mu_0 m_R m_F}{4\pi r^4}\cos(\omega t - \theta(t)) = 0 \qquad (5.94)$$

where $m$ is the mass of the floator, $k$ is optional stiffness parameter which can be introduced by presence of a static field. $F_S$ is a static force like gravity can act on floator on axis-z. The term $\theta(t)$ represents the angular oscillation of floator on axis-z which can be introduced by Eq. 5.57 if the amplitude of oscillation variable $r$ is small enough neglecting its effect on $\theta(t)$. Since phase lag condition holds also here, it is expected that this motion would induce a net force on the floator in direction of the weak field similar to ponderomotive force.

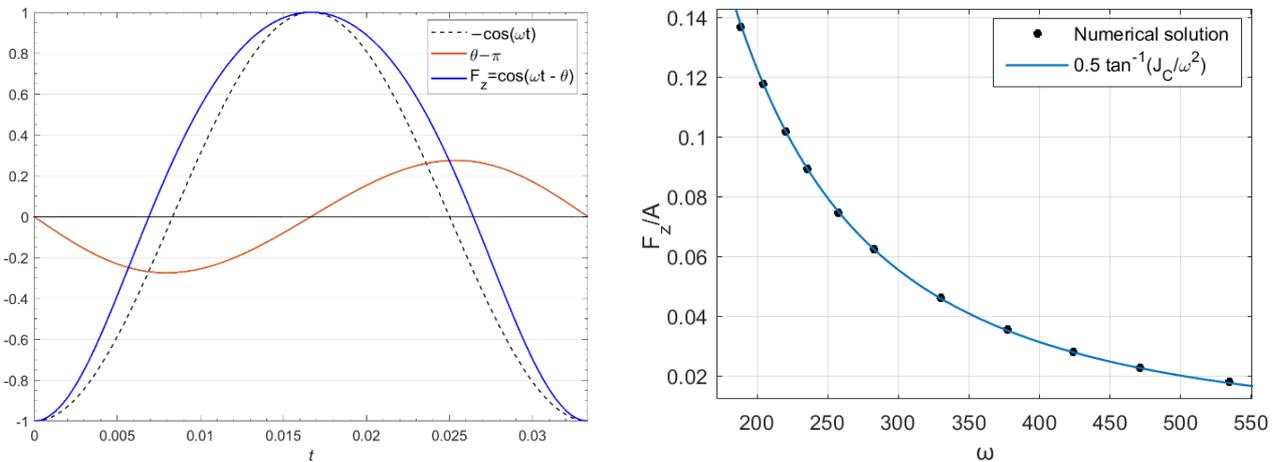

Fig. 5.15. (L) Plot of floator azimuthal motion $\theta(t)$ and the $z$ component of the force $F_z(t)$ it receives for the configuration 1 at table 5.1. Red curve corresponds to angle $\theta$ which varies between $\pi \pm 0.275$. The force $F_z$ marked by the blue curve

is a deformed cosine curve which leaves its negative region early and the positive region late. As a result, the integral of this force becomes positive, corresponding to a force in direction of the rotator weak field. A reference cosine curve is plotted with a dashed line. The value +1 of the $F_z$ in this plot corresponds to the force when dipoles are parallel ($\theta = \omega t = \pi$). (**R**) A curve showing variation of $F_Z/A$ by $\omega$ which fits well to the formula $F_Z/A = (1/2)\,tan^{-1}\,(J_C/\omega^2)$.

As mentioned above, in experiments and in simulations, it is seen that the body's conical motion axis deviates from axis-$z$ in absence of the static field $B_S$. The torque responsible for this deviation is explained by non-zero time integral of the component-$x$ of torque receives the body when the axis of the conical motion deviates from axis-$z$ by a rotation on axis-$x$ (Eq. 5.48) and called torque-phi and it is rather a weak torque. This deviation continues until the body's magnetic moment meets the $xy$ plane. Obviously, this common deviation axis of this rotation and of the torque axis can be any axis on the $xy$ plane. This plane is also the plane of rotating field vector $B_C$. This allows to describe the effect as alignment of the body to the plane confining the rotating field vector. When the rotating magnetic field is belong a dipole, there is no flat surface that floator can find equilibrium, but still there are path similar to force lines that the floator can be aligned where this average torque is zero. According experimental observation, these lines are similar to force lines of a dipole in the standard cross section except it holds for any cross section plane covering the axis-$z$. These alignment lines called *guide-lines* here, have no polarity direction and the floator can align to them in both directions. The subject is sketched in Fig. 5.17 where a floator can find an angular equilibrium when it is aligned to these elliptical lines similar to alignment of a dipole to force lines, except there is no polarity. From this figure, it can be seen the floator orientation changes by 180° in this cross section, going from an axial position to a radial position while it keeping its alignment with guide-lines. The polarity symmetry of the rotating field is broken in presence of a static field. In this case, floator can find equilibrium in an angle between guide-lines and force lines of the static field by the balance of the *torque-phi* and the torque of the static field which is polarity dependent. Since the floator is forced be aligned to guide-lines and orientation of guide-lines varies spatially, this relation can also force a floator get a position along guide-lines when its orientation is forced by an external static field. For example, if an external field enforces the floator orientation orthogonal to axis-$z$, the *torque-phi* causes to floator take a position on axis-$z$ where it get aligned with the guide-lines. This mechanism is experimentally widely observed and plays role in trapping bodies in air stably over a rotating dipole lying on the rotating plane in presence of an overhead dipole oriented parallel to rotating plane which force the floator orientation as the same.

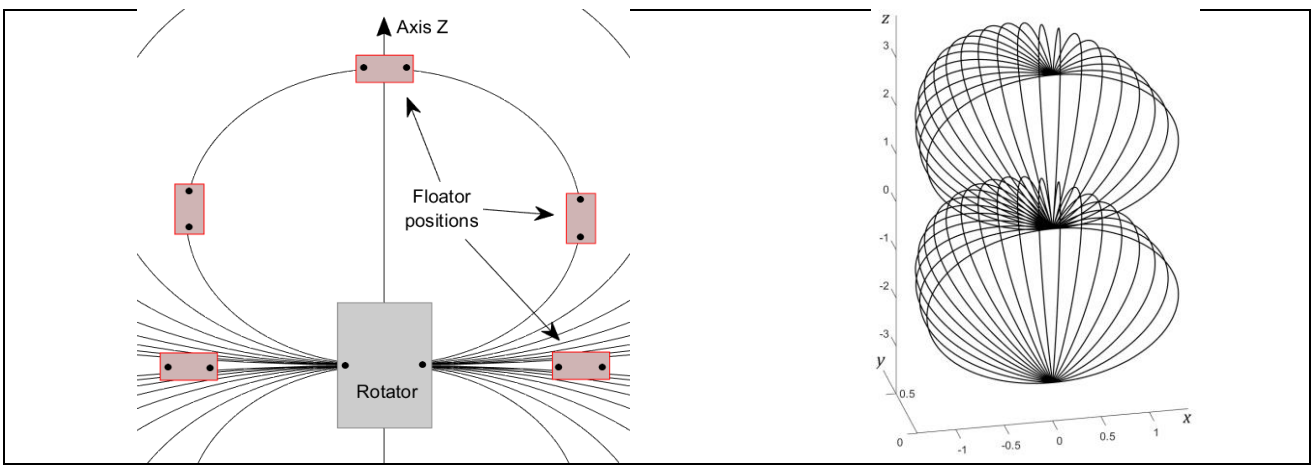

Fig. 5.17. (**L**) Diagram about alignment of floator with 'guide-lines'. Pole positions are marked with dots. As the rotator spins around the axis-$z$, marks showing its polar orientation just means poles are on the side but not at the top/bottom. This is true also for the floator position aligned with axis-$z$, because in this position it has freedom to rotate on axis-$z$. (**R**) A 3D visualization of a single layer of guide-lines under $xz$ cross section. The rotator (not shown) resides at the origin, poles in horizontal alignment. When the floator moment is aligned with a guide-line, the torque-phi becomes zero, therefore an angular equilibrium is found in absence of external fields.

Here, forces of PFR and of this dipole are upward and they are balanced by the weight of the floator as shown in realization at Fig. 5.16. This is a way to trap bodies in absence of static component of the rotating field and in presence of gravity. Here, the angular enforcement of the overhead magnet is translated to positional enforcement by the *torque-phi*. The mechanism is that when the floator is not aligned to guide-lines by means of an external torque opposing the torque-phi, body would experience a force in direction of the guide-lines where its orientation will be parallel to body orientation. If this force get also opposed by an external force, floator may find an equilibrium position and orientation where torques and forces are balanced. In absence of the overhead dipole but in presence of gravity, the lateral stability could not be obtained and floator slides to a side since the spatial profile of PFR (isosurfaces) is curved around the rotator. This is similar to a ball sliding to a side when placed on top of a convex surface.

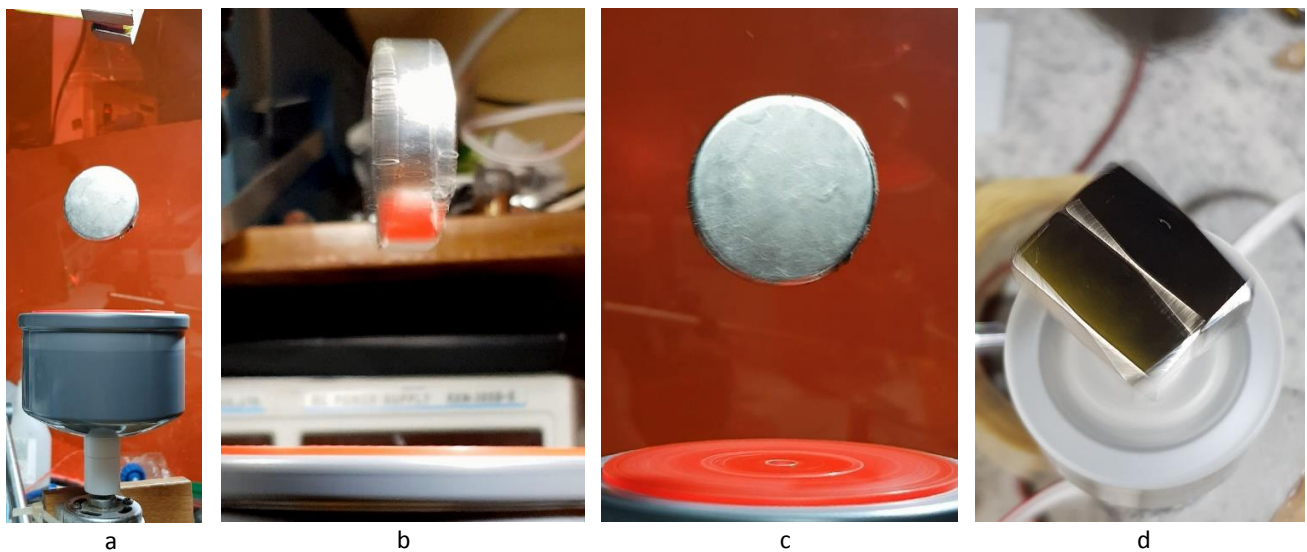

Fig. 5.16. (**a, b, c**) Configuration and motion characteristics of a cylindrical (⌀35×10) magnet (floator) trapped in air over a rotating magnet assembly (a stack of four 10×25×50) and by the help of a static field of an overhang magnet having horizontal dipole orientation (partially visible in (a), at top). The trapped magnet which having also horizontal dipole alignment is subject to angular oscillation on the axis-*z*. This oscillation can be seen on (b) from the fuzzy profile and from the elliptical light traces. On these traces, the major (horizontal) axis corresponds to an angular oscillation (peak-to-peak ~0.14 rad.) and the minor (vertical) likely to a translational oscillation (peak-to-peak ~1.5 mm). These oscillations further increase when floator get lowered by decreasing the pulling force of the static field. (**c**) The close-up of (a). On this camera view, angular oscillations become almost invisible, allowing to check presence of oscillations on the other DoF. (**d**) A similar configuration shown from top (about axis-*z*) where the floator consist of two stacked 25×10×10 magnets. Angular oscillations on axis-*z* can be clearly identified. Dimensions are in mm.

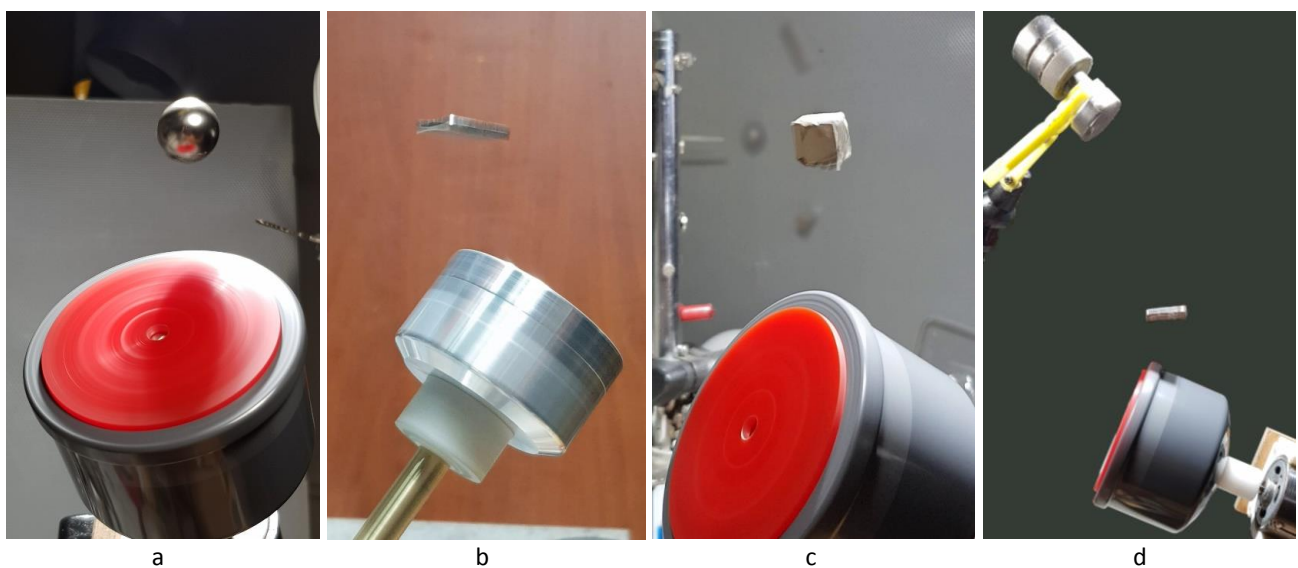

Fig. 5.18. Trapping configurations where various shaped floators are kept above rotators by the help of overhead magnets (shown on (d)) and the gravity in configurations where the rotator axis is not vertical. Rotators magnetic configurations do not generate static magnetic components and these can be seen in Fig. 5.35(g, h). In the configuration (d), overhead magnet position and orientation is hard to predict. Note: Background of this picture is erased for clarity.

Trapping solutions in Fig. 5.16 can be extended as shown in Fig. 5.18. Here, floators find equilibriums at various positions relative to the rotator's orientation while respecting guide-lines shown Fig. 5.17. The complex combination of PFR, torque-phi, static magnetic field and gravity also allows equilibriums where the overhead dipole can have orientation and positions hard to predict.

As a note, it can be shown that any cyclic field, moment, dipole or another magnetic entity $X(t)$ having zero time integral over one cycle ($\int X(t)dt = 0$) which can result in a force by interaction with another magnetic entity Y having zero time derivatives, the time integral of the force $F(X,Y)$ will be zero because function $F$ should be linear allowing the integral to be carried inside the function.

$$F(X_1 + X_2, Y) = F(X_1, Y) + F(X_2, Y), \qquad F(\lambda X, Y) = \lambda F(X, Y) \tag{5.95}$$

$$\int F(X(t), Y) dt = F\left(\int X(t) dt, Y\right) \tag{5.96}$$

This conclusion is also validated by experiments supporting the given model for the PFR. That is, no force is registered on an anchored magnet when it is exposed to the field of a rotating magnet in any direction.

### 5.2.3 Experimental observation of conical motion of a body having full DoF

Experimentally, the circular angular motion of the floator can be observed by reflecting a laser beam through a flat polar surface of a cylindrical or prismatic floating dipole magnet trapped in a stable equilibrium using PFR and an attractive static magnetic force/torque aligned with the axis of the rotating field, said main axis. The axisymmetric shape is required for the angular motion to be exactly circular, otherwise the motion becomes elliptical. An exact circular trace can be obtained by aligning the originating beam with the main axis and placing the screen orthogonal to it. Such a realization is shown in Fig. 5.19 which is also an example of magnetic bound state based on dipoles. The required static field for the trapping the floator in air is generated by tilt of the rotator dipole axis from its rotation axis plane about 10° as seen in Fig. 5.19(c). Due to this tilt, the magnetic moment of a rotator have a constant *z*-component which produces the static field. This field is also a dipole field and serves basically to three purposes. One is to balance the PFR, second is to align the floator orientation on the rotation axis and the third is keep floator on the rotation axis. In this configuration, the static field is strong enough to keep the floator against gravity also in radial directions since the rotator axis is horizontal. This static moment needs be only a small fraction of the rotating moment. Depending on configurations, it can be small as 1/100 and large as 1/6. Above this ratio, stability becomes difficult because the equilibrium distance could be too short. At this point, the motion can no longer be approximated for above analyses.

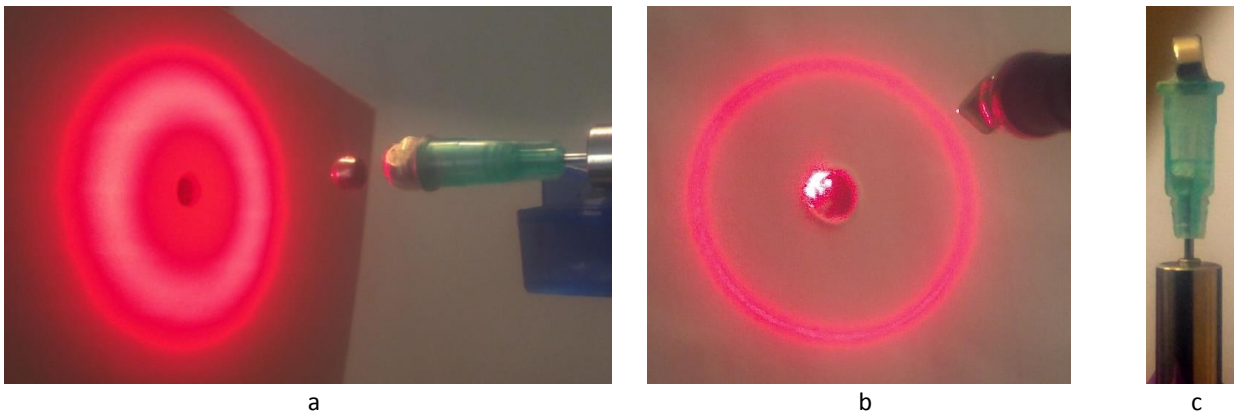

Fig. 5.19. (**a**) Circular angular motion trace obtained by reflecting a laser beam from mirror like polar face of a cylindrical magnet of dimension ⌀4×2.5 mm trapped horizontally in air, close to axis of a rotating field. Laser beam emanates from the hole at the center of the screen. The fuzzy halo is caused by the excessive intensity of light received by the camera and by the roughness of the magnet nickel coating. The rotating field is provided by a ⌀6×3 mm magnet attached to a micro motor by the plastic part of a syringe needle. This magnet provides also the required static field. Motor speed is in 22000 - 28000 RPM range. Magnets are type NdFeB/N35. As the floator is trapped horizontally, gravity pulls it in a radial direction which slightly shifts the floator off the axis; however, a deviation from the circle pattern is not noticeable here. It should be noted the circular oscillation is only one of the available of oscillation patterns which can be obtained in same or in similar configurations. (**b**) Similar configuration using floator as a 1/8" cube magnet (grade N52) which is partially visible behind the rotator. (**c**) The rotator assembly. Note that the magnet axis is tilted by angle $\gamma$ from the rotation plane, generating a static dipole aligned with the rotation axis.

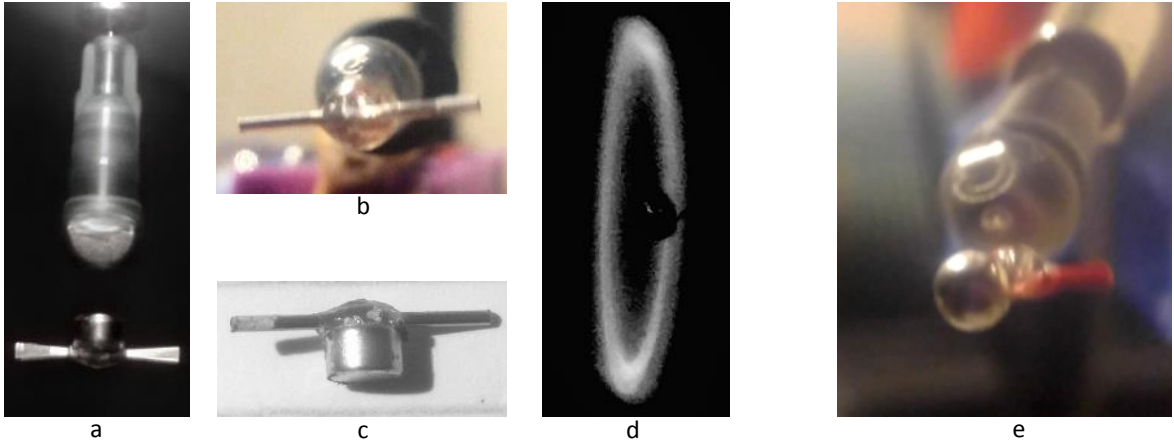

Fig. 5.20. A dipole body consisting of a small cylindrical magnet (⌀4×2.5 mm) and a non magnetic needle section glued together (**c**) is trapped horizontally by a rotating dipole field produced by a rotating magnet (⌀6×3 mm). In (**a**), the trapping is seen from the top and from front on (**b**). The body's MoI in directions orthogonal to needle orientation is increased due to the needle. From (**b**), it can be seen the body sits little below of the rotation axis because of its weight. This asymmetric position forces the needle orientation to horizontal, matching to the direction of the lower MoI. By defining xyz coordinates, *z* as main rotation axis, *x* as horizontal and *y* as vertical, angular oscillation consists of *x* and *y* components where amplitude of *x* component is several times larger than *y*. The fuzziness of the needle can be seen in (a) corresponds to the *y* component and the absence of *z* component from (b). (**d**) The elliptical pattern of angular oscillation obtained by a laser beam reflection from *xy* face of a body from a similar setup. It should be noted that oscillations on *x* and *y* axes correspond to deviation of the beam in *y* and *x* directions, respectively. (**e**) A similar configuration with a single arm based on a floator spherical magnet having diameter 1/8", NdFeB/N42.

Under unstable regime, it is observed that floator's translational motion on axis-*z* increases exponentially and ends up by a final dive into the rotator or fly away with a speed few m/s. These conditions are also include angular instability visible at final moments. Presence of the static field have also an effect to increase the amplitude of angular motion in accordance to Eq. 5.24. Typically, angular oscillations amplitudes (angle $\varphi$) up to $\pi/8$ are stable, but some configurations allow even larger angles, it is observed. On the other hand, there are other cases where floator involved in complex angular motions where the amplitude of oscillations can reach $\pi/2$. Fig. 5.20 shows another configuration where a small magnet is trapped under bound state. This trapped body has different MoI in *x* and *y* direction, resulting the angular motion becomes highly elliptic as can be seen from the laser trace at (d). In this case too, the assembly is horizontal (the axis-*z*) and the floator find equilibrium against gravity slightly off the axis-*z*, downward as seen in (b) where the camera is on axis-*z*. This asymmetry appears to have an effect to align the floator azimuthal angle in order its larger MoI becomes in the direction (vertical) of this offsetting. A remarkable effect is present in (e) where a single armed floator magnet is shown. Here, the arm is kept horizontally in opposition of the torque of gravity and the small torque on the axis-*z* which is present in general caused by the rotating field in its direction which try to rotate the floator CW, in this view angle. This effect is briefly evaluated at the end of Section 5.4.3.

The combination of angular and lateral translational oscillation of the body gives an image of angular motion where the rotation center has an offset $\delta$ from CM in direction *z*. The relation of the amplitude $\beta$ of this lateral oscillation and the amplitude of angular oscillation $\varphi$ a trigonometric relation as

$$\beta = \delta \tan \varphi \qquad (5.97)$$

Here, a quick formula is given to estimate the offset $\delta$ with small angle of $\varphi$ and with small displacement $\beta$ allowing these motions be approximated as simple DHM where force and torque are independent of motion variables and spring constants are omitted. A DHM defined this way and its solution reads

$$u'' = c \sin \omega t, \qquad u(t) = -\frac{c}{\omega^2} \sin \omega t \qquad (5.98)$$

By applying this to translational and to angular motions, we obtain amplitudes of these motions as

$$\beta = \frac{F}{m\omega^2}, \qquad \varphi = \frac{T}{I\omega^2} \qquad (5.99)$$

where $F$ is amplitude of lateral component of rotating force, $T$ the amplitude of rotating torque, $m$ and $I$ are mass and relevant MoI of the body. Then, with small angle approximation we can write

$$\delta \cong \frac{\beta}{\varphi} = \frac{F}{T}\frac{I}{m} \tag{5.100}$$

In an approximate model for an interaction where forces are received by poles, defining component-$x$ of these forces as $F_1$ and $F_2$ (the motion variables are defined in plane $xz$), the term F/T can be expressed as

$$\frac{F}{T} = \frac{2}{h}\frac{k+1}{k-1}, \qquad k = F_1/F_2 \tag{5.101}$$

where $h$ is the distance between pole points. Typically, forces are antiparallel, giving $k$ with negative sign. Applying this to Eq. 5.100, we can write

$$\delta \cong \frac{k+1}{k-1}W, \qquad W = \frac{2I}{hm} \tag{5.102}$$

This formula gives $|\delta| < W$ for $k < 0$ and $|\delta| \geq W$ for $k \geq 0$, $k \neq 1$ (case of forces are parallel but not equal). Positive $\delta$ values correspond to offsets in direction from CM toward $F_2$ point. Term $W$ is equal to 1/6 for a body in a rod shape, 1/3 for a cube, 1/2 for a disk and 7/24 for a cylinder having equal diameter and height.

### 5.2.4 Summary

As a summary, analytical results are supported by integrated magnetic field and rigid body dynamics simulations and by experimental observations where magnetic bodies are held autonomously in air in stable fashion, satisfying stability requirements, making qualitative evaluation of these experimental results worthwhile. While the mathematical model used here is restricted to simple configurations, experiments shows that the repulsive force and the stability can be obtained regardless of orientation and position of magnets, using different shaped magnets, spanning large range of form factors and sizes, including irregular shaped magnet fragments and assemblies of magnets and combined with other materials. By applying external static fields in various directions and strength, it was possible to obtain stable trapping solutions where floator position can be set at any position around the rotator (except positions occupied by rotator driving assembly).

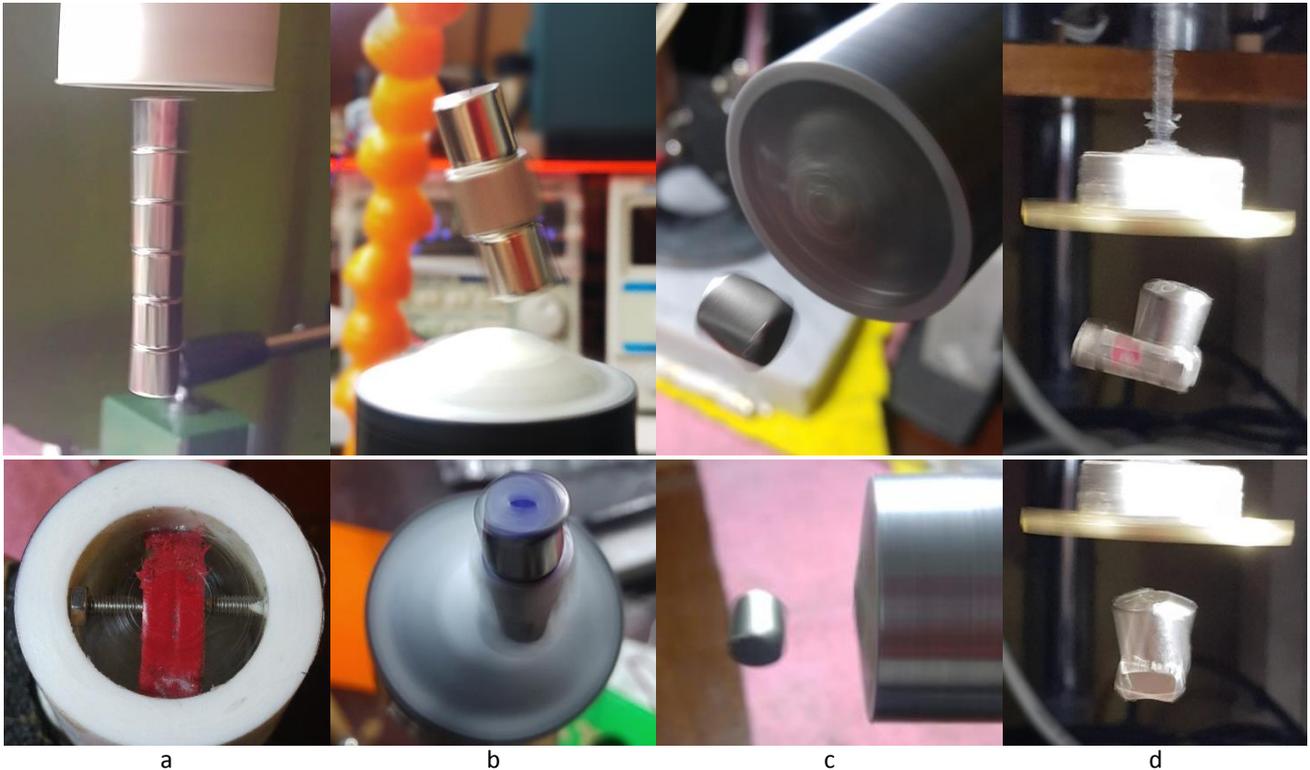

Fig. 5.21. Miscellaneous configurations where magnets held in air by interaction of a rotating magnets. (**a**) A stack of six ⌀10×10 mm magnets is trapped under a rotator assembly with a gap less than 4 mm. Here, the repulsive action is mostly

generated by the lateral translational oscillation of the floator because the angular oscillation is too small to generate an effective repulsive action due to the large MoI of the floator in radial direction. Bottom figure shows the rotator assembly where two stacked ⌀25×4 mm magnets covered by a tape are housed in a plastic frame. (**b**) A floator assembly consisting a 10 mm cube magnet sandwiched between two ⌀10×10 mm magnets hovers over a rotator assembly embedding an disk magnet ⌀30×5 mm rotating at 5530 RPM (579 rad/s) CWW. An overhead magnet help to keep rotator upright and partially compensate its weight. The Pizza tower appearance of the floator is behavior of long floators. From these pictures, the outcome of combined angular and translational motions which set rotation center of the body off the CM can be seen. This point is about 1/6 of the height of the body higher than CM. The unique outcome of this configuration is the sustained spin of the rotator at same speed of the rotator but in the opposite direction (CW). In order to obtain this result, floator is speeded-up this way by an air-jet until it reaches this speed. After cutting the air-jet, floator keeps indefinitely this speed and even anticipates some frictions (a touch of a soft brush). This counterintuitive behavior reminds the Rattleback effect [22]. The misalignment of the floator rotation axis with rotator axis might allow to this effect be happens. The detail of the rotator assembly can be seen at Fig. 6.6(c). (**c**) A ⌀10×10 mm, diametrically magnetized cylindrical floator magnet is held in air in a horizontal setup. The separation between magnets is about 10 – 12 mm. The floator equilibrium position is off the axis-*z* since the gravity acts in radial direction. This produces the same angular alignment effect explained for the experiments shown in Fig. 5.20 since the floator is not axisymmetric with respect to dipole axis. Relatively large amplitude of the elliptical angular oscillation is visible for the minor axis (vertical). (**d**) An asymmetric assembly of floator consisting of a rectangular magnet (15×10×5 mm, magnetized in thickness) and an axially magnetized cylindrical magnet (⌀10×10 mm) held in air by a rotator assembly hung by a thread, allowing five DoF. The sixth DoF (vertical) is also effective here because elasticity of thread allows small vertical motions of the rotator. The amplitude of angular oscillation on the axis of the camera angle at the bottom picture is larger than the top one as can be perceived. The rotator assembly consists of two stacked 20×10×10 mm magnets glued on the top face of a thin brass disk and a small ⌀9.5×1 mm disk magnet glued to the bottom face which provides the attractive static field. The brass disk (⌀44 mm) weights 13.8 g and have MoI figures as $I_R$ = 2.15e-6 Nm$^2$, $I_A$ = 4.29e-6 Nm$^2$. The purpose of this disk is to provide resistance or anticipate the magnetic torque caused by the floator. This subject is evaluated in Section 6.

Some additional trapping and bound state realizations are shown in Fig. 5.21 where peculiar characteristics of these solutions are evaluated in this section. In general, it is possible to reproduce these solutions within the provided data; however, the critical parameter is the tilt angle $\gamma$ of the rotator magnet and may need to adjust it by trial and error for specific configurations. A guidance for selecting a comfortable rotor speed is to start with high speed and lower it until instabilities emerge and go back until they vanish. In these trials, damping the motion with nearby copper or aluminum blocks might ease this process.

## 5.3 A 2D Simulation giving PFR like results

A JavaScript based 2D simulation application MagPhyx [23, 24], which can be run on a browser is configured to obtain PFR like results. While the application is not designed for this purpose, nevertheless allows a rotating spherical dipole magnet modeled as a free body having a dipole moment, inertia and MoI be scattered from a point dipole field belonging to another anchored spherical magnet of the same. Two factors contribute to generation of repulsion in this 2D model. One is translation oscillation through the field gradient similar to ponderomotive force and the other is modulation of angular speed of the body. Here, the angular speed of the body varies by the cyclic torque generated by its rotation where this speed is high when body experiences attraction and low when in the repulsion phase. This effect decreases the attraction period and increases the repulsion period allowing more time for the body to accelerate in the direction of the weak field. Such a variation of angular speed is also present in the PFR schemes where angle $\varphi$ is not constant or equal to π/2. Fig. 5.22 shows two MagPhyx simulation screens where the path of the free body is traced in red.

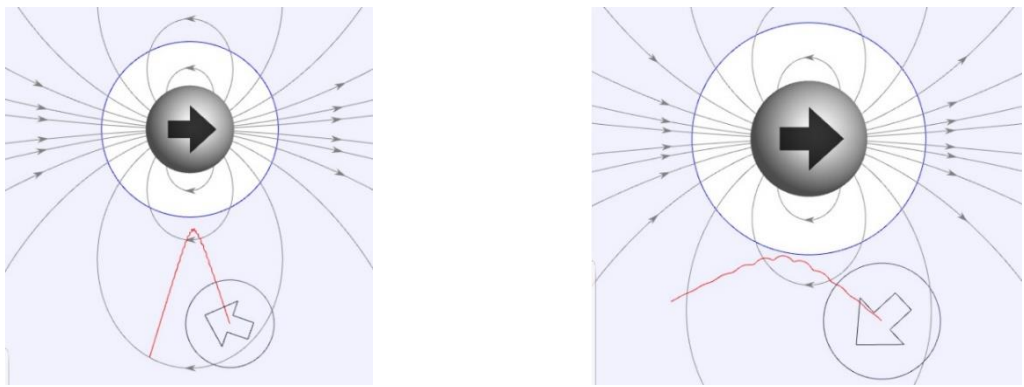

Fig. 5.22. 2D simulations of a spinning dipole body scattering from a static field using MagPhyx application. While the static field is produced by the central point dipole, rotating dipole body is shown by the other circle. The embedded arrows show the instantaneous direction of the dipole moment. Red traces show the travelled path of the spinning free body.

## 5.4 Characteristics of PFR

PFR is repulsion of a magnetic body having DoF (the floator) from a cyclic field having gradient. When this field belongs to a rotating dipole (the rotator, NdFeB permanent magnets in experiments), it is observed that this repulsion exists regardless of position and orientation of the floator with respect to the rotator while repulsion strength may varies, but kept in same magnitude. This characteristics can be explained by three complementary motional factors contributing the PFR. These are angular, lateral and through the gradient translational motions. These ratios of contributions varies by positions and orientation but overall the repulsion does not vary significantly, it is observed. This allows the floator move one position and orientation to another one smoothly when the speed of this motion is small compared to speed of the rotating field.

### 5.4.1 Energy aspects of PFR

In a configuration where the PFR finds a stable equilibrium with an external static force field, it may be possible to alter this equilibrium for obtaining a cyclic motion of the floator around the equilibrium position. When the motion on this cycle is about two orders of magnitude slower than the driven motion of the floator, this may correspond to an adiabatic process [25]. After repeating this test by obtaining different trajectories by varying initial conditions, if these cyclic motions are found stable, that is if they keep their amplitude constant, this can point out conservative character of the PFR under adiabatic condition. This would be equivalent to obtaining a closed loop under a conservative force or a conservative vector field where potential and kinetic energy is restored at the end of the loop [26]. This behavior is generally observed with experiments involving trapped bodies using PFR in air, however, results are qualitative.

Simulations of angular motion of the floator (where forces are absent) show its kinetic energy $E_K$ can be conserved. $E_K$ is constant when the angular motion of the axisymmetric body is symmetric about the rotating field axis; that is, the angle $\varphi$ is constant otherwise oscillates but its average and its amplitude would be constant if the motion is stable. Fig. 5.23 shows such two plots where $E_K$ oscillates. The first plot corresponds to simulation at Fig. 5.1.7 where $E_K$ oscillates twice the frequency of the rotating field with constant amplitude of 4.6e-5 J around 1.74e-4 J. In the second plot, the first harmonic is not zero.

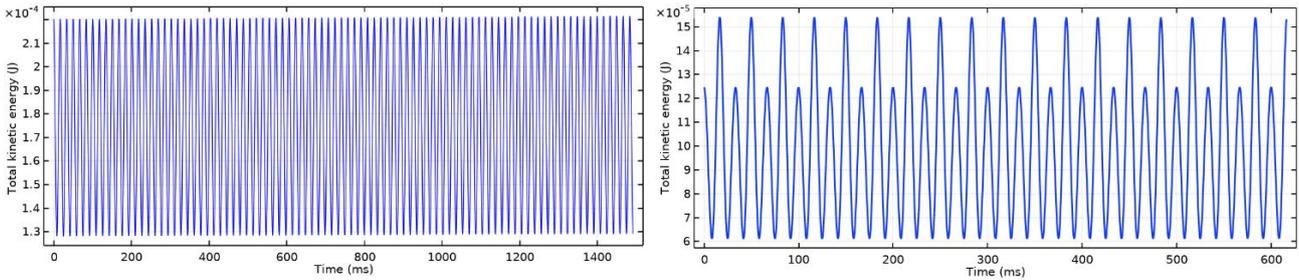

Fig. 5.23. (**L**) The plot of kinetic energy of the floator corresponding to simulation at Fig. 5.7 where body is an ellipsoid. Kinetic energy oscillates around 1.74e-4 J with a constant amplitude 4.6e-5J during the simulation time (1.5 s). This oscillation corresponds to harmonic angular motion of the body. The slight increase of energy is the simulation artefact, which is corrected later by reducing the time step. (**R**) A similar plot of a simulation at Fig. 5.1.4, except the angle λ is 45° instead of 48.8°. Kinetic energy oscillates in this case due to the axis of the conical motion not aligned with axis-$z$ but having the angle λ where the conical motion is elliptic. Frequency of the rotating field in both simulations is 60 π rad/s.

In a DHM, the total energy $E_T$ of a body is the sum of kinetic $E_K$, potential $E_P$ and the exchanged energy $E_X$ with the driving force. Therefore, one can expect that $E_X$ varies when $E_K$ is not constant. However, by setting initially $E_X$ as zero, the time average $\langle E_X \rangle$ of $E_X$ should be zero when time averages $\langle E_K \rangle$ and $\langle E_P \rangle$ are constant. Note that $E_P$ depends on orientation and position of the floator's magnetic moment with respect to the rotating field and to other static fields, its weight also enters equations when it is present.

In experiments about trapping free floaters in air under stable conditions, while could be time averaged energy exchange with the rotating field, it can be assumed that total energy of the floater needs be conserved since the interaction basically is non-dissipative. Above arguments point out a possibility of considering PFR as a pseudo conservative force. That is, having the behavior of a conservative force. A condition which violates this conservation is the progressive gain of spin of the body around an axis close to its dipole axis it acquires from the rotating field. This spin can be avoided using non-axisymmetric bodies or it can be fixed automatically at certain speeds (to an integer fraction of the rotator speed). This spin locking effect is examined in Section 6.5. An experiment might directly point out the conservative character of PFR is the trapping a rotating magnet by a static field described at Fig. 5.36 and shown at Fig. 5.37. While in this experiment, the motion of the floater is sustained for compensating air friction and vibration losses caused in the fixed platform by an external rotating field, motion is stable without this enforcement until it slows down after a period. Electrically inductive energy losses might be considered in general in these experiments since NdFeB magnets are conductive but the remnant magnetization of magnets might prevent significant variation of magnetic field inside the magnet, thus limiting these inductive losses.

Another experiment for this purpose is described in Fig. 5.24. Here, a dipole magnet is attached to a string where the other end is fixed, realizing a spherical pendulum. The experiment consists of swinging this magnet while it is subjected to PFR from another magnet attached to a rotor which has no DoF, excluding its rotation axis. Swinging causes the PFR to vary since the position and the distance of the pendulum magnet continually change in time. Since the total work made by a static conservative force field on a closed path is always zero, this free pendulum test can serve to determine the conservative character of PFR. Although, the realized setup is rudimentary and does not provide ideal conditions, nevertheless, performed tests give an idea about this conservative character. In the first run of this experiment, the pendulum magnet is swung following a path where its distance to the rotator magnet varies in a factor of two. In this run and its repetitions where motion paths are arbitrary, it is observed that the motion decays close to the timing where it slows down in absence of the rotating field by the air friction. In the other runs, the magnet directly swung into the rotating magnet allowing to bounce from it at a very short distance (about 2-3 mm) vigorously like a ball bounces from a wall whereas the farther distance in this motion is about 60 mm giving a factor about 30. In this case, it is observed that there is no decay, the pendulum continues its motion indefinitely. This implies that it gains enough energy at each bounce which compensate friction losses. These results suggest the PFR might act as conservative force under adiabatic condition; that is, when the variation ratio (time derivative) of the PFR is small compared to the variation ratio of instantaneous magnetic forces.

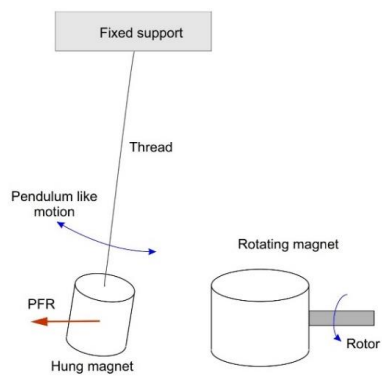

Fig. 5.24. Pendulum like configuration where a dipole magnet attached to a thread experiences PFR from another dipole magnet rotating on an axis orthogonal to dipole axis. The orientation and fixture point of the hung magnet is not important. Rotor velocity can be chosen in the 5000 - 15000 RPM range for a pendulum magnet of ⌀10×10 mm in size.

Other experiments shown in Fig. 5.29(a) and similar ones are about trapping a magnet in air using rotating field and a static field which also helps to compensate the weight of the body and allowing levitation gaps between 50-65 mm and the trapped magnet can move slowly around its equilibrium position while allowing vertical distance varies up to 12 mm. These experiments can result in stable levitation of bodies, pointing bodies do not gain kinetic energy during the interaction. This condition can be broken by lowering the rotating field velocity under some limits, causing the body to enter ever increasing vertical oscillations. Possible conservative character of PFR is further evaluated in simulations of magnetic bound state of free bodies in sections 6.0.1 and 6.0.2.

It should be reminded that in Section 5.2, PFR is associated with the component-$z$ of magnetic force between dipoles which is constant when floator perform a symmetric and synchronized motion around rotation axis of the rotator. This case might ease to evaluate the conservative character of PFR.

### 5.4.2 Overview of short-range character of PFR and a related classroom experiment

Under a stable regime where a floating body can have reversible kinetic exchange with the field, the stiffness of PFR, similar to a static force field can be evaluated. The approximated formula (Eq. 5.81) giving the profile of the repulsive force in the direction of the axis of rotating field shows that PFR has positive stiffness for this specific configuration. According to experimental observations, the interaction of a dipole body with a rotating dipole on the rotating plane always generates a repulsive force regardless of body's orientation and position. The profile of this repulsive force versus distance is about two times nonlinear than static forces between dipoles. This makes PFR very strong at short distances, vanishing quickly by increasing the distance and can be classified as a short-range force. This high power factor is because the force enters twice to the equation (or force plus torque) as a product since PFR strength varies both with the strength of the magnetic forces and the amplitude of the cyclic motion which varies also with the strength of the magnetic force or the torque. This characteristics eases finding equilibrium with attractive magnetic forces which have negative stiffness. As described above, the simple pendulum swing experiment (Fig. 5.24) can give experimenters a clear figure about the large stiffnesses of the PFR at short distances (and even at shorter distances under unstable regime). In the original experiment, magnets are NdFeB/N35, the pendulum magnet is cylindrical (⌀10×10 mm), attached from one of its polar face center to a thread having length ~100 mm. Rotating magnet is rectangular (15×10×5 mm) polarized in thickness and the longest length is in the rotation axis direction. It rotates at 9500 RPM mounted in a brass cylindrical housing in depth of 10 mm, attached to a DC motor. Centering, aligning and balancing the rotating magnet is important otherwise the motor and the assembly will experience significant vibrations preventing desired rotation speeds and may prevent the reproduction. Even a rotating magnet is precisely aligned with the rotating axis, still its magnetic moment can be off of the rotating plane due to magnetic misalignments at the factory. A misalignment will generate a static moment which may have noticeable effects when it is more than 2°. On the other hand, selection of magnets is not critical. They can be any strong magnet weighing 5 to 20 g. The setup should also allow the rotating speed to be varied between 6000–20000 RPM by noting the PFR is inversely proportional to square of the speed. In this experiment, one can observe that the swung magnet bounces from the rotating magnet without making a physical contact. Effective impulse duration can be less than 20 ms. A slow motion video would also give some interesting details about the bouncing moment. This pendulum with a refined design might also serve to investigate conservative character of the PFR as mentioned above.

### 5.4.3 Experimental results on dependency of PFR on rotation field velocity

The dependency of PFR on the velocity $\omega$ of the rotating field comes from angle $\varphi$ which makes PFR approximately proportional to $1/\omega^2$ when $\varphi$ is small. This relation can be seen from Eq. 5.81. Measurements also reflect this relation as shown in Fig. 5.25. This set of measurements gave the power factor −1.9 in presence of a dissipative mechanical interface allowing such a measurement. Stability of the motion also depends on $\omega$ by a lower limit which is determined by configurations. By approaching this limit $\omega$, the body starts to gain some low frequency or sub-harmonic angular oscillations in experiments and finally destabilizes. The lower limit of $\omega$ for a stable motion raises when the strength of the cyclic torque increases and body's MoI decreases.

Fig. 5.26 shows a stability chart where the axis-$y$ corresponds to the minimum frequency required for the stability of a levitation for a range of bodies whose masses are mapped on axis-$x$. Masses vary between 0.01 - 100 g and frequencies between 1800–100000 RPM (188.5–10472 rad/s). This gives a scalability figure of PFR in a complex relation between rotating field frequency, magnetic moment, mass and MoI of the body. Fig. 5.27 shows trapping of millimeter sized magnets and irregular magnet fragments using small rotating magnets attached to high speed micro motors which are used to realize to trap such small bodies. Two series of measurements are made in order to determine the relation of frequency to mass ratio. In the first series, the distance between floating body and the rotator is kept large and PFR to body weight ratio is about one third (since weight of the body adds up to PFR and they together balanced by the attractive force in this configuration) and tried to find lowest frequency of stability. In the second series, this distance is kept small and the PFR to weight ratio is about one but this condition does not give the lowest frequency for a stable operation. First se`ries depicts the frequency as inversely proportional to cube root of the mass and the second as close to inverse square root. The strength of the rotating field does not enter to scaling characteristics beside the difficulty to spin large rotators at high speeds. As the mass of floator is a parameter on the stability equation, this would be useful for electrically sorting neutral particles by PFR, similar to a job based on ponderomotive effect [27].

In experiments involving trapping bodies, precise initial conditions are required for stability under an undamped regime. This problem can be overcome by providing a damping at the start and by removing it progressively while establishing the stable regime.

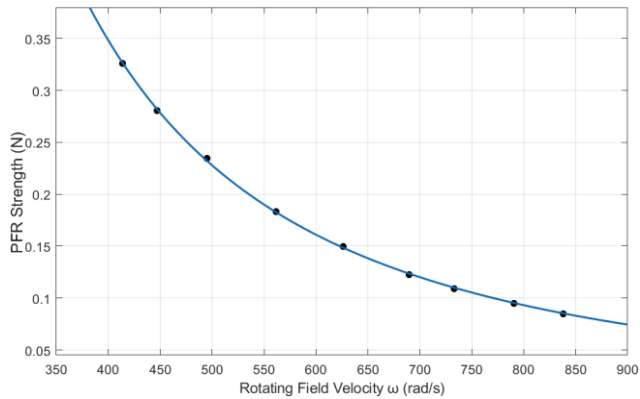

Fig. 5.25. Strength of polarity free repulsion varying with rotation speed of a dipole magnet acted on a magnet cushioned on a balance. Sample points fit well to function $y = 224 \times 10^6 \, x^{-1.9}$.

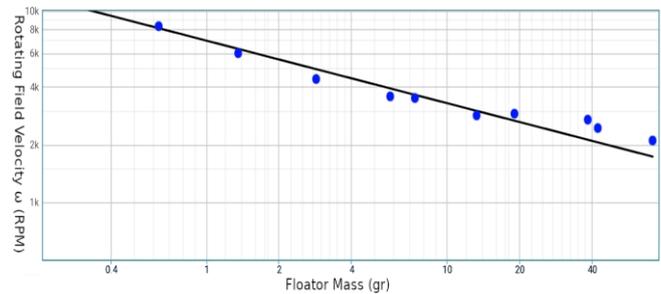

Fig. 5.26. Minimum frequency required for stability for levitating bodies obtained by tests using the same rotator configuration and using floators of different masses $m$ but having similar geometries. This relation is reflected by the function $\omega = a \, m^b$ by curve fitting method where the exponent $b = -0.326$. This shows a possible relation between $\omega$ and body mass by its inverse cube root ($b = -0.333$).

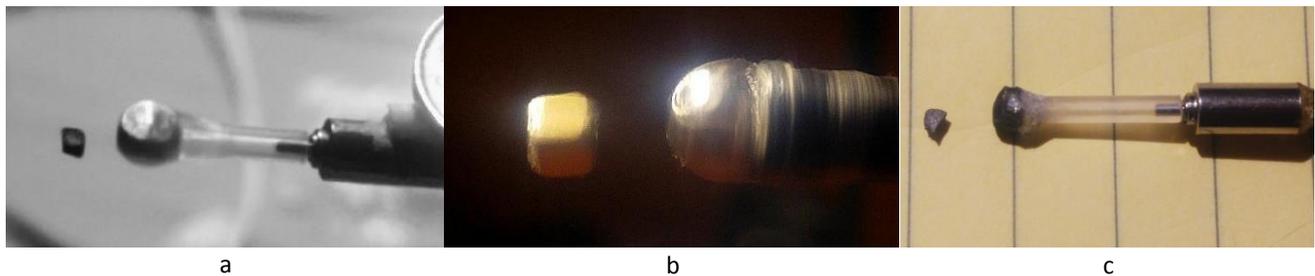

a  b  c

Fig. 5.27. Images showing a bound state of small magnets held horizontally. (**a**) A magnet fragment about 1.3 mm in size trapped by a rotating dipole magnet of ⌀4×2.5 mm in size attached to a micro motor running about 1500 rev/sec. (**b**) A 1/8" cube magnet bound to a rotating dipole magnet of size ⌀6×3 mm rotating about 420 rev/sec. (**c**) Picture of a floator and a rotator assembly mount to a micro motor having dimensions ⌀4×10 mm. Both magnets are irregular NdFeB type fragments. Vertical lines are 8 mm apart.

### 5.4.4 Dependence of PFR on body orientation

It is observed that PFR strength obtained by rotation of a dipole field varies in some extent by the angle of the body's dipole with the rotation axis while angles corresponding to maximum and to minimum remain orthogonal as they vary by the position of the body. For example, in the Fig. 5.16, a floator is shown in different positions. By varying the orientation of the floator while keeping its position fixed, it is observed that the maximum and minimum PFR strength correspond respectively to orthogonal and to parallel orientations relative to guide-lines. For example, in a configuration where the body sits on the rotation axis, PFR has a maximum when body's dipole is parallel to rotation axis. Within the variance of PFR strength with its alignment factor to guide lines, it can be said that body is forced to an orientation where PFR is minimum. In the definition of *torque-phi*, in a previous section, its relation to the rotating field velocity $\omega$ is mentioned. This relation is experimentally supported by varying $\omega$ and varying body sizes within configurations.

Another angular stiffness can be present in PFR when the MoI of the body is not axisymmetric with respect to its magnetic axis. Under this condition, body's angular motion becomes elliptic where the major axis corresponds to the orientation of the minor principal MoI. When the rotating field is asymmetric with respect to the body position; that is, the body is off the rotation axis, the axes of elliptical pattern are forced to orient in a direction according to this asymmetry. However, no detailed observations are made in order to determine whether this orientation serves to minimize the amplitude of this elliptical motion or not. Fig. 5.20 shows such a configuration and the obtained trace of an elliptic motion.

### 5.4.5 Spatial profile of PFR of a rotating dipole field

PFR induced by a rotating dipole around its center on an axis perpendicular to its moment has full space coverage, it is observed. PFR strength does not vary much by the angular position of the body with respect to rotation axis and its orientation. It is observed that different geometries of permanent magnets and arrangements used to generate rotating field can produce different PFR profiles. It is also observed that the PFR profile of a spherical dipole magnet is smooth and variations unnoticeable without instrumentation. It should be reminded that the strength of the PFR might vary by the dipole orientation of the body as mentioned in Section 5.4.

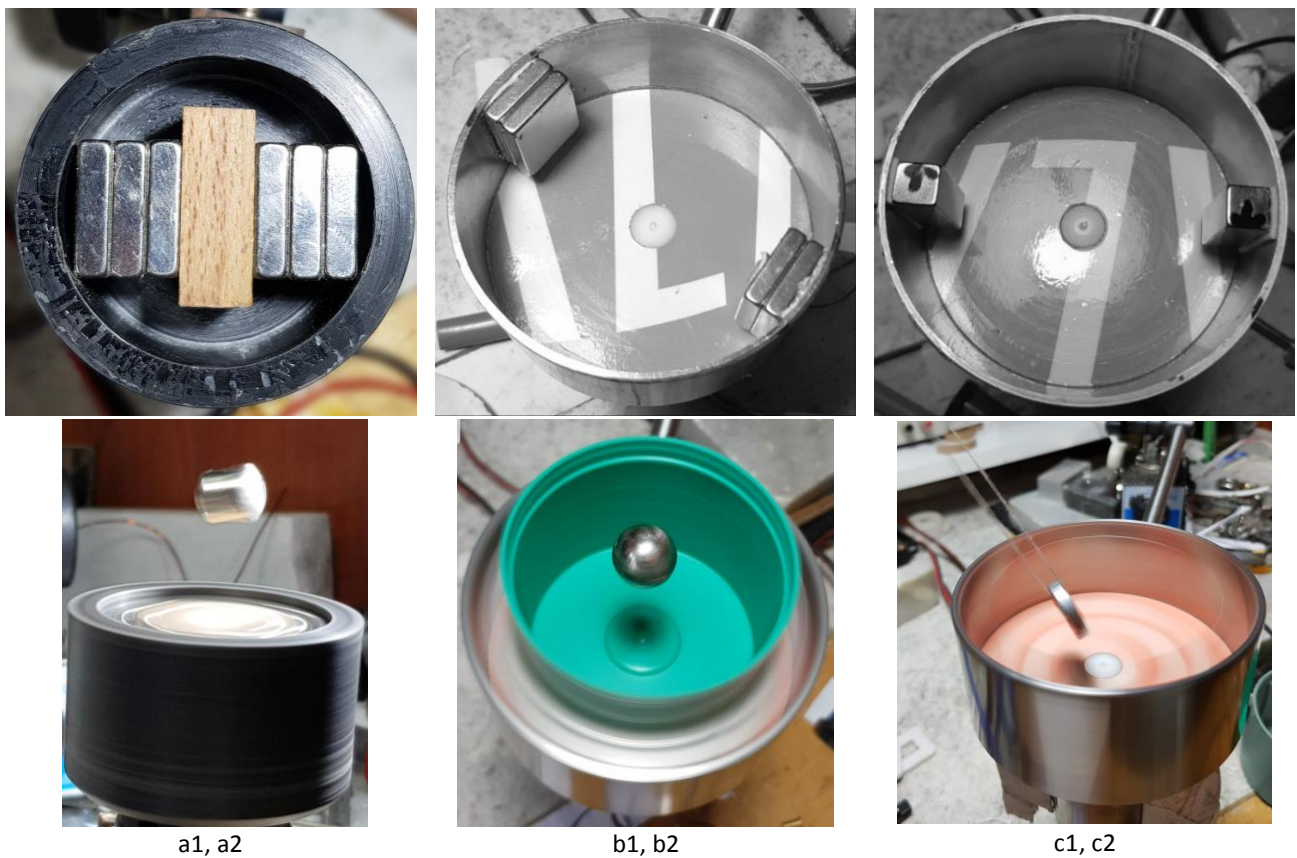

| a1, a2 | b1, b2 | c1, c2 |

Fig. 5.28. Realization (top) of split dipoles (**a**,**b**) and quadrupole (**c**) and their experimental outcomes (bottom). No levitation can be obtained with quadrupole configuration (**c**) but only a radial inward force. Magnets used in rotator setups (a, b) are 25×20×5 mm, in (c) is 25×20×5 mm, all of them magnetized in thickness. Floators dimensions are (a): ⌀10×10 mm, (b): ⌀15 mm, (c): (⌀19.7 – ⌀8)×3.9 mm.

Sometimes, it is desirable to flatten the spatial profile of the PFR in this scheme by reducing the strength in a zone close of rotation axis. This can be achieved by splitting the dipole in two parts and leaving a gap between them as seen in Fig. 5.28(a1). It is observed that the PFR might not be obtained or get inverted in a zone surrounding the gap depending on the gap length. The transition zone close to the gap where the dipole field is reversed forms a quadrupole. It is found that a dipole body can be trapped using rotating split dipole configurations and by balancing PFR with the gravity; however, equilibrium characteristics vary significantly within configurations. The split dipole scheme is also tested with a large gap where the gap length is several times larger than dipole length of split parts and also several times larger than the floating body size. Within this scheme, it is possible to define an inner and an outer zone around the rotation axis (*z*). Within these zones, the radial components of the net force acting on a body are inward and outward respectively. Inner zone has a maximum diameter as large as the gap length at proximity of rotating magnets and shrinks while extending in

±*z* directions. In the axial direction, forces are outward. Within this zone, a body can find stable equilibrium in full DoF when magnetic forces are balanced by an external force in *z* direction but not in the outer zone.

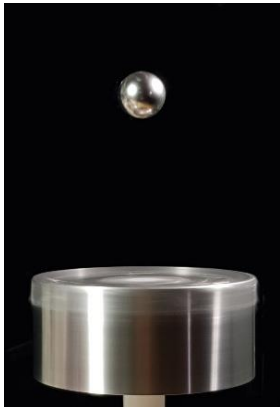 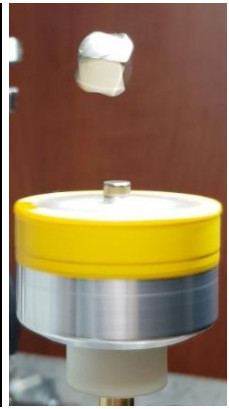 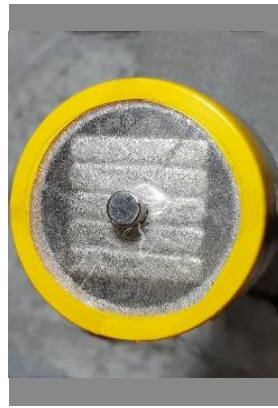 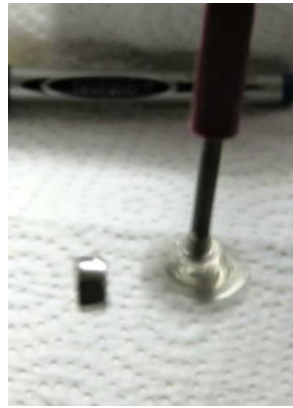 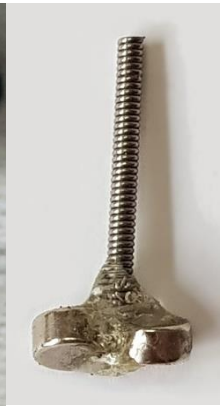

Fig. 5.29. A spherical dipole magnet ⌀15 mm is levitating over a large rotating dipole magnet assembly by the help of an overhang magnet and gravity. In this configuration, magnetic forces are upward and are balanced by the weight of the floating magnet.

Fig. 5.30. A cube dipole magnet is trapped by a rotating magnetic assembly consisting of five stacked 20×20×5 mm magnets housed in a cylindrical aluminum block and having dipole orientation orthogonal to rotation axis. A small magnet (a stack of two ⌀6×3 mm) fixed on the top face and aligned with rotation axis provides the attractive field. Stable equilibrium is possible despite the attractive magnet being closer to the trapped magnet because the slope of the PFR vs distance is still larger than of the attractive force.

Fig. 5.31. A ⌀6×3 mm dipole magnet is trapped by a rotating magnetic assembly consisting of a pair of the same magnet arranged like propeller blades and may having asymmetry with respect to rotation axis. Since magnets have opposite alignments with respect to rotation axis, this causes near zero static field on this axis. This way, trapping occurs in radial direction of the rotating field.

As the rotating field is symmetric with respect to plane *z* = 0, *z* component of the field and its gradient on this plane should be zero. However, it is observed that this does not provide a stiffness to a dipole body in order to keep it on this plane. On the other hand, this may be obtained by a modified configuration by splitting the dipole again in *z* direction (becomes four pieces) and finding equilibrium at the center. Another configuration providing inward radial force to a dipole body shown in Fig. 5.28(c). In this configuration, two dipole magnets are placed symmetrically with respect to rotation axis and dipoles oriented along the assembly rotation direction, creating a quadrupolar field on the rotating plane. When a dipole body inserted inside of this rotating assembly, it is forced radially to the center and can find stable radial equilibrium in absence or in presence of external forces on radial plane. This configuration has similarity to rotating quadrupole field based equilibrium obtained at [4] but having a different magnetic arrangement. Within the PFR model, a dipole body is pushed inward as the field weakens in this direction. Compared to configuration with split dipole scheme this quadrupolar scheme provides stronger inward force because of a deeper local minimum which reaches zero.

Rotating a dipole in any other ways than symmetric with respect to the rotation axis generates polarization; that is, the interaction of the dipole with this field depends on polarities. Fig. 6.7 shows two schemes for obtaining asymmetry with a single dipole. This logic is used to generate a static field component which is used to balance PFR. By using multiple dipoles, it is possible to obtain a large number of field geometries for both cyclic and for static components. Fig. 5.31 and Fig. 5.32 show such implementations.

### 5.4.6 Motion characteristics of bodies within PFR

PFR depends on angular and translational motions of a body. The translational motion is lateral in a symmetrical motion with respect to axis-*z*; that is, it lies on the *xy* plane. Under some conditions, the angular motion is minimized and the translational motion can do most of the job. This can happen when the floating gap is three or more times smaller than the body size. Here, the floator could be an elongated body in dipole direction such as shown in Fig. 5.21(a), Fig. 5.37 and Fig. 6.5(d). These motions are lateral with respect to the dipole center, but the whole translational motion of the body is not necessary since rotation center RC and CM may coincide. Here, it is shown that a lateral motion of a body can induce PFR similar to angular motion. In this simple model, torques and angular motions are ignored. Here, two dipole moments are present belonging to entities, at least one of them should be an inertial body having translational DoF. One entity (*A*) can generate a rotating dipole moment around axis-*z* and the other (*B*) have a moment in direction-*z*. We also consider a stable harmonic motion confined in *xy* plane, driven by the magnetic interaction which is reflected to the spatial vector *r* between moments where the amplitude and the phase of this motion is defined by the variable $\beta$. This way, these vectors can be defined as

$$\vec{m}_A = m_A \begin{bmatrix} 0 \\ 0 \\ 1 \end{bmatrix}, \qquad \vec{m}_B = m_B \begin{bmatrix} s_{\omega t} \\ c_{\omega t} \\ 0 \end{bmatrix}, \qquad \vec{r}_{AB} = \begin{bmatrix} \beta s_{\omega t} \\ \beta c_{\omega t} \\ h \end{bmatrix} \tag{5.103}$$

where *h* is a positive constant corresponding to distance *z*. By putting these vectors in Eq. 4.2 we obtain

$$\vec{F}_{AB} = \frac{3\mu_0 \, m_a \, m_b}{4\pi} \begin{bmatrix} h \, s_{\omega t}(h^2 - 4\beta^2) \\ h \, c_{\omega t}(h^2 - 4\beta^2) \\ -\beta(4h^2 - \beta^2) \end{bmatrix} (h^2 + \beta^2)^{-\frac{7}{2}} \tag{5.104}$$

Here, the component-*z* of the force does not have an oscillatory component. We may exclude the motion in direction *z* from equations since this force can be balanced statically. In experiments, the amplitude $\beta$ of the lateral motion is one or more orders smaller than the distance *h* between dipoles and $\beta$ is always less than half *h*. Therefore, we are interested in solutions where the force coefficient $h(h^2 - 4\beta^2)$ has positive sign. Since the displacement and the acting force are in opposite phases under phase lag condition, the displacement $\beta$ should be negative according to the vectors *r* and *F* defined above. This renders component-*z* of the force positive, that is, a steady repulsive force approximately proportional to the amplitude $\beta$ of the oscillation. When both of bodies have lateral DoF, accelerations due to lateral force on each body add up for obtaining relative acceleration as

$$a_R = a_A + a_B = \frac{F}{M_A} + \frac{F}{M_B} = \frac{F}{M_R}, \qquad M_R = \frac{M_A M_B}{M_A + M_B} \tag{5.105}$$

where terms $M_A$ and $M_B$ denote mass of bodies. This is also known as reduced mass formula for reducing a two body problem to single body. The repulsion force $F_Z$ can be also predicted by simple visualization where two bar magnets are arranged in T shape. When the T is symmetric, there is no vertical force, but they would repel each other in this direction when the top magnet slides horizontally in the opposite direction of the force it experiences.

A realization based primarily on lateral oscillation is shown in Fig. 5.37 and sketched in Fig. 5.36. Here, the entity *A* within Eq. 5.103 is an anchored permanent magnet and *B* is a rotating magnetic assembly having full DoF trapped in air.

### 5.4.7 A summary of the PFR model

Under the stable regime of DHM, body can follow a symmetric angular and translational (lateral) motions around the rotating field axis, synchronized with the field where the angular motion complies with the model defined in Eq. 5.7. Therefore, body is kept motionless in the frame of rotating field and its dipole obtains fixed position and orientation with the rotating field. Phase lag condition of the motion ensures the dipole experiences a constant fixed force in the direction of the weak field. Radial component of the force the body receives gets balanced by inertial forces but as the acceleration has cyclical pattern, the body can return to its previous position at the end of the cycle. On the other hand, the component in axial direction causes a steady force on the body repelling it from the rotating dipole. Body can also have a spin around its dipole axis which has only an effect on the dynamic of the motion. Angular motion for this case is evaluated in Eq. 5.62 through Eq. 5.69. The spin of the body has an effect on the stability of the motion by extending the range of parameters to obtain stability and also varies the zenith angle of the body (Eq. 5.68) which has a direct effect on PFR strength. Eq. 5.67 points out that the harmonic character of the motion is lost when it spins parallel with the rotating field (k = 0), at least when the body has uniform MoI.

The configuration, dynamics and magnetic field plot of this model is shown in Fig. 5.13 and Fig. 5.14. This static case evaluation of PFR also simplifies the evaluation of bound state equilibrium as well, since the bound state can be obtained by the equilibrium of attractive static force with PFR. The circular angular motion present there becomes elliptic when the body is not axisymmetric with respect to its dipole axis. Such a realization is shown in Fig. 5.20. Ellipticity can also occur when the body does not oscillate symmetrically with respect to the rotation axis but is offset by a zenith angle as shown in the realization in Fig. 6.5(b). This asymmetry which becomes apparent when the permanent magnet constituting the body has longer length in dipole orientation and the field component in direction of the rotating axis is relatively weak can be explained by the torque-phi effect. According to this, zenith angle $\varphi$ of the dipole is determined by the equilibrium of the static axial component of the driving field and the torque-phi. When the body is close to rotation axis, torque-phi forces the dipole (body) to be orthogonal to rotation axis and the static field forces it to parallel. However, when the body departs from the rotation axis, torque-phi changes its forcing angle and becomes parallel to rotation axis in the half way of the dipole heads to the radial position by following guide-lines. Within these two factors, the zenith angle is set depending on radial position of the body as it is experimentally observed.

As a remainder, the rotation center (RC) and the CM of the body do not overlap in general. This is the result of superposition of angular and translational motions of the body. Still it is possible to RC meets CM under specific conditions by offsetting the dipole center from the CM center.

Analyses show that PFR originates from phase lag condition. This generalization might be useful to obtain a generalized PFR model which can better cover experimental results.

## 5.5 Trapping dipole bodies with PFR

It is possible to trap dipole bodies with an interaction involving PFR. As mentioned in above sections, it is possible to find stable equilibrium with PFR, by balancing it by various static force fields. A special case called magnetic bound state where the rotating magnetic field has a static component is evaluated in Section 6. However, it is found that trapping is also possible without requirement of a static magnetic field. In the literature, there are realizations and proposals to trap bodies having magnetic moment by the static and rotating quadrupole fields; however, a generic rotating quadrupole field alone on the symmetry axis orthogonal to quadrupole plane is not sufficient to trap a dipole body in space because it does not provide a stable equilibrium in the axial direction. This problem can be solved by various ways and one solution is given [4] based on a rotating double quadrupolar scheme and may in presence of some polarity asymmetry. Within solutions involving rotating quadrupolar fields, there is no precise line to evaluate them through the literature or by guidelines of PFR. In equilibriums of dipole bodies within a rotating quadrupolar field, angular oscillations can be absent or may not contribute to the stability. On the other hand, there are experimentally found schemes in this work to trap bodies by rotating fields having quadrupolar components. These cases are not presented as a PFR solution but as trapping schemes with rotating dipole fields.

Earnshaw theorem which can be extended to cover magnetic dipoles states that the total stiffness of a static interaction based on force fields obeying inverse square law is zero. That is, if there is positive stiffness in some DoF, there should be negative stiffness in some other; therefore, stable equilibrium cannot be obtained in presence of any negative stiffness. As the PFR provides positive stiffness, it can be used to remove the negative stiffness factor of the static interaction, rendering a trapping solution stable in an equilibrium in combination of static force field and PFR. A simple way to levitate a body this way is finding equilibrium between PFR and gravity in vertical direction and obtaining lateral stability by a static dipole field. The vertical equilibrium point can be chosen in order the sum of the negative stiffness of the static field and the positive stiffness of PFR around this point is positive. Such a setup consists of symmetrically rotating a dipole magnet on a vertical axis *z* and a dipole magnet is placed in a higher position on axis-*z*. A dipole body can be levitated above the rotating field while having a horizontal dipole orientation. Dipole orientation of the top magnet should be horizontal too in order to meet this alignment according to torque-phi effect (Section 5.2.2). This magnet also works against gravity by raising the equilibrium position. Such realizations are shown in Fig. 5.16. It is also possible to obtain levitation while the rotating field axis has arbitrary orientation and by adjusting the position and orientation of the top magnet with compliance of the torque-phi effect.

Two configurations based on split rotating dipole scheme, which are mentioned in Section 5.4 are examples for levitating a dipole body with PFR only, without help of static components. Cyclic field requirement of PFR can be also obtained by a relative motion between a free body having a dipole moment and a static field spatially changing polarity in direction of the motion. This scheme is experimentally realized by building a circular track with alternately oriented magnets shown at Fig. 5.32. By rotating the circular track around its axis, PFR is obtained by a dipole magnet hovering over the rotating track. It was possible to trap the floating magnet over the track. In this trapping solution, it is likely the local minimums generated along the track by a sequence of alternating poles served a role along the PFR. Anyway, this 1D alternating array scheme might be extended to 2D in order to obtain a surface of static periodic field where a free dipole body or a structure consisting an array of dipole bodies having individual angular DoF to glide over this field in a range of speed.

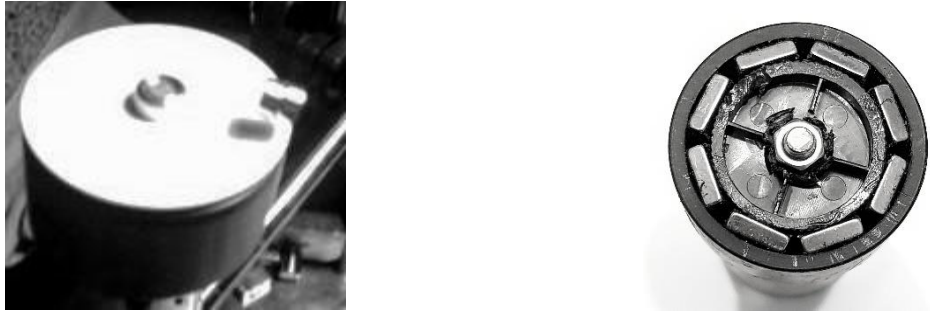

Fig. 5.32. (L) A small dipole magnet trapped in air by a rotating circular track of lined up magnets, which their dipoles are in radial direction, alternately inward and outwards shown on (R). It looks like the trapping occurs over this track due to a possible local minimum of the alternating field present at this zone.

## 5.6 Trapping dipole bodies with multiple rotating dipole fields

It is also possible to find stable equilibrium by using multiple rotating dipole fields having no static components. The basic configuration of each rotating field consists of a dipole rotating symmetrically on a plane. It is observed that a dipole body can find equilibrium at the middle of three autonomously synchronized fields in a horizontal triangular formation where axes point to the center of the triangle as shown in Fig. 5.33 and Fig. 5.34.

Rotation direction of two dipoles are the same and the third in the opposite direction. This third dipole is mainly responsible for their synchronization, acting as a magnetic gear in between. A dipole body can find equilibrium at the center while anticipating gravity. The quality of trapping becomes better when the triangle formation is equilateral or symmetric with respect to the counter rotating field although no precise alignments are required. Separation between dipoles are close to the length of dipoles. Trapping also works when rotating dipoles are tilted upward but not downward. It is expected that upward orientation helps to anticipate weight of the body.

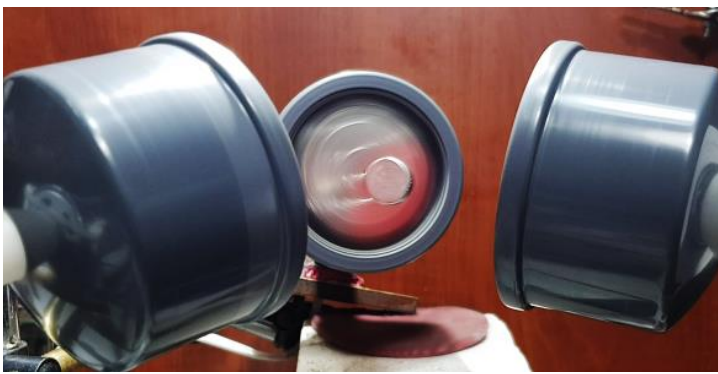 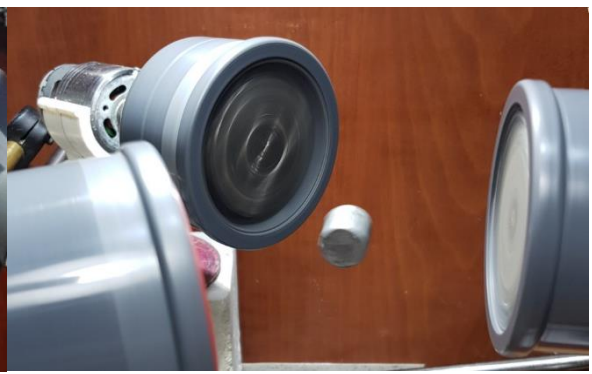

Fig. 5.33. A cylindrical magnet (the small object at the center of the frame) is floating in air at the center of the equilateral triangular formation of large rotating magnet assemblies providing only AC fields (a cyclic field without static component).

Fig. 5.34. Similar experiment of the left picture but with isosceles triangular formation. Rotating unit at right has opposite rotation direction of the units at left.

Another configuration consists of two synchronously rotating dipoles facing each other, which effectively generate a rotating quadrupolar field as shown in Fig. 5.35(a, b, e). It is observed that a dipole body can be trapped on the rotation axis between of rotating dipoles while its dipole gets oriented orthogonal to rotating axis and performs elliptic motion on the plane orthogonal to its dipole. Here, floator have both angular and translational motion on the rotators axis and the latter can be seen at Fig. (b, e). This configuration allows to anticipate gravity from any directions and allowed be extended by varying angle between axes of rotation between 0° to 150°. Some these configurations work with gross misalignments of axes as is shown in Fig. 5.35(d). It is also observed that two unsynchronized rotating fields can be used to trap a body. With this scheme, the body does not stay completely motionless but rotates in a frequency equal to difference of speed of two rotating fields. This may be accompanied by additional periodic angular and translational motions. In these multiple rotating field configurations, it is found that the floating body is trapped in a location where it is less agitated by the interaction; that is, where it's oscillation amplitude is minimum. A quadrupolar character of the field at trapping zone is expected and this may contributes to the equilibrium by providing a local minimum of the oscillating field. It is observed that oscillations of the trapped body are retained also in the magnetic equilibrium position, the position it should be if there were no gravitational force acting on it. These oscillations may have both translational and angular components varying within configurations. In the configuration with dipoles rotating synchronously on same axis, the oscillation is mainly translational and the body vibrates in small circle on the plane orthogonal to dipole axis.

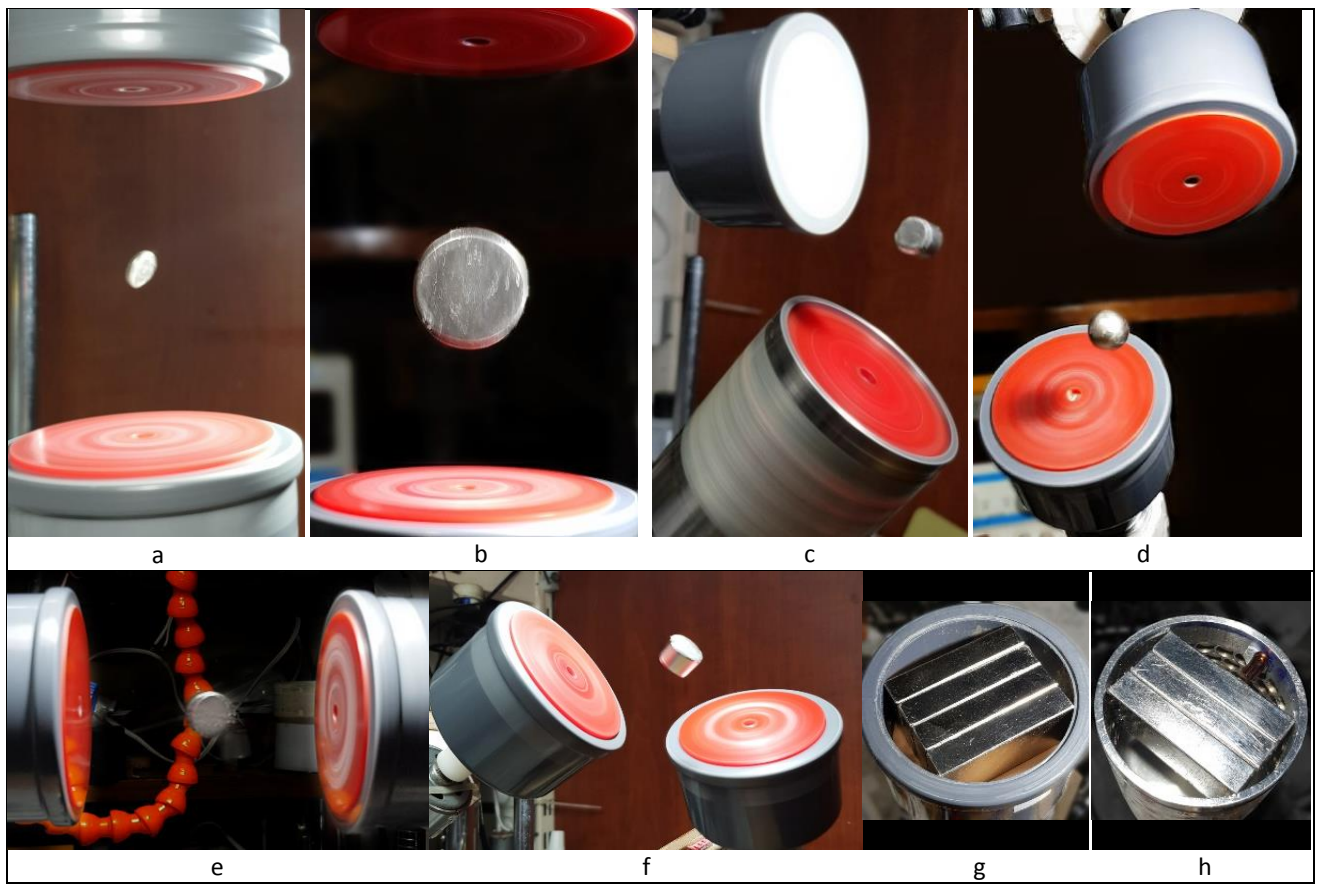

Fig. 5.35. Dipole bodies trapped by various configurations of two rotating dipole units. Floating magnets dimensions (in mm) are (**a**) = ⌀9.5×1, (**b**) = ⌀25×5, (**c, e**) = ⌀10×10, (**d**) =⌀15, (**f**) =⌀12×10. Rotating assemblies embeds stacked magnets of dimensions 10×25×50, magnetized in thickness, four for all them, except three for units at **c.** Rotating units are synchronized autonomously by the magnetic coupling mechanism. (**g, h**) Magnet arrangements of rotators.

Within the triangular configuration, it is observed that the body oscillates in angular motion on a vertical axis passing its center or on a shifted position from farther from the counter rotating dipole. It is also found that the body's dipole get aligned parallel without polarity preference to rotating planes in the configuration of two rotating dipoles on the common axis. In the triangular formation, it is aligned to the counter rotating field, by conserving bilateral symmetry of the configuration. In configurations with two rotating fields where rotating planes are not parallel, this enforcement is weak and alternative alignments may occur. This characteristic alignment can be attributed to the torque-phi effect. Within this accordance, it is also observed that forces acted on the body varies with its orientation. This variation is highest when rotating planes are parallel and trapping force in radial direction ceases when body is oriented in axial direction.

## 5.7 Trapping rotating bodies with static fields

A rotating dipole body can be trapped by a static dipole field using PFR and a static component of the field of the rotating body, which balances the PFR attractively. However, there is a problem need be solved: A dipole is forced to align parallel to the field it is exposed, however it is needed here to keep the rotating dipole almost orthogonal to static field; otherwise, the rotation of the body will no longer generate a rotating field required for PFR, but becomes a static dipole, which get attracted by the static dipole in full strength. It is found two ways to prevent this alignment and to keep proper angle between dipole and rotation axis. They are based on providing a counteracting torque against the magnetic torque which tries to change the zenith angle of the dipole. One is extending the body in the direction of the rotation axis in order to receive a gravitational torque when the axis deviates from vertical orientation. Other approach is providing a mass distribution suitable for the proper orientation of the body with respect to the rotation axis in order to keep this axis by the rotation dynamics. In general, a free body can only rotate stably on the major and minor principal axes of inertia, but not on the intermediate one, as it can be derived from Euler's equations. In basic realizations, bodies are axisymmetric and greater stability can be obtained when the major axis meets the rotation axis by extending the body in radial direction, which is called oblate. Bodies extended in axial direction called prolate. In both schemes, angular oscillation of the body is largely reduced by the increased MoI, but in the prolate scheme, the torque received by the rotating field can be used more effectively to obtain lateral motion of the dipole. That is, the body oscillates in a conical motion where the rotation center is away from the dipole as shown in Fig. 5.36 and in realizations shown in Fig. 5.37. In this scheme, the magnetic torque received by the body is balanced by the torque obtained by the pair of magnetic attractive force and gravitational forces acted on the body by an offset in vertical direction. Here, PFR is mainly obtained by lateral motion of the body instead of angular where torque acted on the body is translated to translational oscillation in presence of large MoI. This mechanism is evaluated in Section 5.4.3. The oblate scheme, which does not require gravitation assistance, is used to obtain bound state between two free bodies, detailed in sections 6.1 and 6.2.

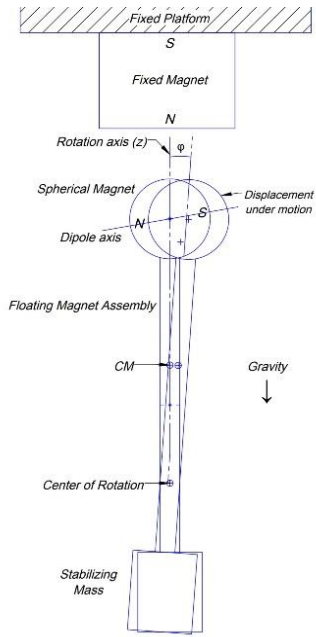

Fig. 5.36. Bound state of a floating assembly with a static magnetic field. The assembly rotates on the vertical axis by a given initial angular motion. The torque and force it receives from the magnetic field of the fixed magnet causes the assembly to oscillate around the point denoted as center of rotation. The magnet within the assembly is tilted from the rotation axis in order to obtain an attractive force from the fixed magnet. This force is balanced by PFR, resulting in the assembly kept trapped in air. The gravitational force stabilizes the orientation of the assembly. The displacement shown corresponds an instance of the motion, which can be shown in spherical coordinates where the origin is center point of rotation, $\varphi$ is the zenith angle which kept constant and the azimuthal angle $\theta$ equal to $\omega.t$ where $\omega$ is the angular speed of the system.

Another trapping solution with this scheme is trapping a complete assembly consisting of a dipole magnet attached to a small motor by a dipole magnet fixed to a motionless platform. It is found that the obtained trapping can be strong enough to anticipate external forces received from any direction. Fig. 5.38 and Fig. 5.39 show such setups.

PFR can be obtained when a body is attached to a support by a coupling which does not restrict its harmonic motions; however, this may introduce some stiffness and friction. PFR can tolerate these factors very well, hence real magnetic forces acting on the body are several times larger than the PFR and the coupling basically transfers the PFR and the weight of the body to the support.

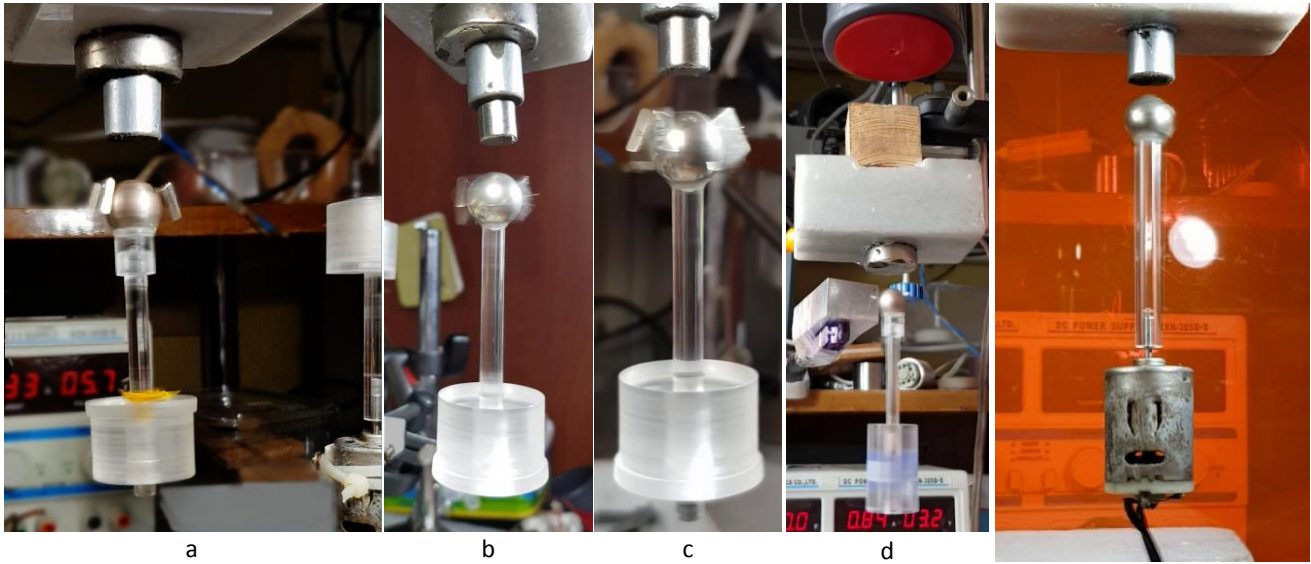

Fig. 5.37. Free spinning dipole bodies trapped by a static magnetic field of stacked permanent magnets fixed to a motionless platform. A spherical magnet (⌀15 mm) and a small magnet stuck to its side constitute the body's dipole. These side magnets are 10×10×3 mm except setup (d). Their presences cause asymmetry on the magnetic field and on mass distribution of the body along its axis. However, this factor strengthens the bound state both providing extra attractive force and extra PFR strength by increasing the lateral oscillation of the dipole (center) in the same time zeroing the lateral oscillation of the body by providing counteracting inertial force to the lateral magnetic force. As a result, the body spins on its predetermined axis without vibration in a specific rotation velocity. The steady velocity of the body is provided by its coupling to a rotating dipole field belonging to assembly at the side of the frame in (a, b, c) and at top of (d). This coupling is found to introduce instability to the system by low frequency oscillations which are damped by approaching a copper or an aluminum block.

Fig. 5.38. A spherical dipole magnet (⌀15 mm) attached to a DC motor trapped by the field of a dipole magnet (⌀15×15 mm) fixed to a motionless platform. The $z$ component of the rotating magnetic moment is large enough to keep the whole assembly in air.

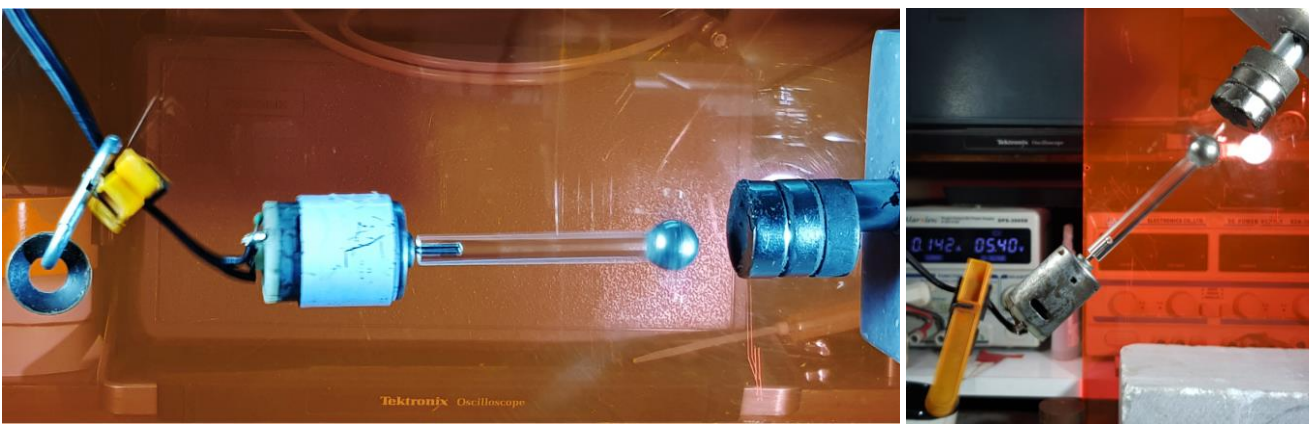

Fig. 5.39. A rotating dipole magnet attached to a DC motor hung from cables is trapped horizontally (left) and by an angle (right) by a dipole magnet stack fixed to a motionless platform. Items attached to the cable serve to suppress vibrations.

## 5.8 Notes on realizations of Polarity Free Repulsion

The simplest way to observe PFR is attaching a dipole NdFeB type permanent magnet called actuator magnet to a rotor at the center of the dipole and orthogonal to dipole axis and approaching another similar dipole magnet called floating magnet to this rotating magnet simply by holding it between fingers. As a note, spherical shaped floating magnets offer better handling. Rotating magnet provides the inhomogeneous cyclic field. Holding the floating magnet between fingers (a bit loosely) allows one to directly experience instantaneous forces and torques the magnet receives and the strength of PFR in various distances. As the strength of instantaneous torque is large, holding the magnet this way has a minor effect on its motion; additionally the damping introduced this way improves the stability by extending range of parameters.

Alternatively, the magnet can be attached to a string and hang close to the rotating magnet as shown in Fig. 5.24. This also allows observing the stability of the PFR with negligible damping. With this method, a magnet can be swung into the rotating magnet and deflections can be observed. This experiment allows also to observe the unstable state of the effect when the magnet is bounced at a very short distance (< 3mm) and receives an extra energy briefly at the bounce moment. This allows the magnet swing and bounces indefinitely by compensating friction loss by this energy. Magnet sizes suitable for these experiments can be 10 to 15 mm in their maximum extent. This should be limited by 20 mm for not causing injuries.

# 6 Magnetic Bound State

Magnetic bound state (MBS) is a special case of trapping a body in a stable equilibrium autonomously by a cyclic magnetic interaction. This interaction could be between two entities, body or field, associated either with a magnetic moment or with an inhomogeneous cyclic field having both alternating and static magnetic moments. Magnetic moment of one entity should be dipolar and the other dipolar or quadrupolar according to realizations. A cyclic field and a free body are mandatory for obtaining MBS. Such a cyclic field can be obtained by rotation of a dipole. MBS can be obtained between compact objects; that is, one object is not surrendered by the other. MBS does not require the interacting field to have a local minimum. Bound state can be obtained between two magnetic bodies, each having full DoF according to simulations and supported by experimental tests approximating this condition. In this work, MBS is evaluated as a stable equilibrium of PFR and a static magnetic attractive force. This way, the dynamic of the motion is encapsulated in PFR and MBS is evaluated in terms of PFR and a static force. This is an approximate method that makes sense when amplitude of translational motion in the direction of PFR is small and variation of forces and torques due this motion can be neglected compared to amplitude of cyclic forces and torques.

As mentioned in Section 5.1, it is possible to decompose forces acting on the floator as cyclic and static forces. Similarly, it is possible to decompose the magnetic moment of the rotator to a cyclic component $\vec{m}_c$ rotating on $xy$ plane and to a static component $\vec{m}_s$ aligned with the rotation axis $z$ (Eq. 5.27). This allows to evaluate the bound state as equilibrium of forces induced by these components where PFR is associated with $\vec{m}_c$ and the attraction force with $\vec{m}_s$.

In above sections, it is shown PFR allows a stable equilibrium against an attractive force between two magnetic moments and in Section 5.2.1, MBS of a free body and a rotating dipole moment is shown through integrated simulations (Fig. 5.14). In order to obtain a stable bound state between two bodies, several requirements need to be fulfilled. In stiffness terms, this interaction provides positive stiffnesses to floator in three translational DoF and two in angular. The remaining DoF is the rotation of the body around its dipole axis, therefore to not cause basically a stability issue. Positive stiffness can be present in this DoF too, under some conditions. It is also shown that PFR can be obtained through a DHM having zero damping parameters; that is, it is not based on a dissipative process. In this process, the kinetic energy of the body can be kept constant under a stable equilibrium. Simulations also show the stability of the motion based on the PFR model in absence of damping. On the other hand, a body can exchange kinetic energy with the field under varying parameters as PFR is not truly a conservative force. Experiments on the bound state show that the stability can be obtained under varying conditions. However in these experiments, dissipations are unavoidable, primarily by the mechanical methods to generate the rotating field. In most experiments about MBS, rotators are permanent dipole magnets or an assembly of magnets generating a dipole field. They are mounted on a rotor lacking DoF except its rotation axis. Floators are also permanent magnets having full DoF and kept in air through the interaction of rotator's magnetic field. Magnetic bound state of two free magnets are also realized experiments in an approximated scheme.

Stability of the bound state is complex due its dependence on six DoF, each enters to equations with position and velocity terms. The heuristic stability criterion given at Eq. 5.44 also points out that the stability is not invariant to torque scaling, that is stability cannot be sustained by scaling-up the torque the body receives while keeping the ratio of static and cyclic torques constant. As torques are inversely proportional to cube of distance between two point dipoles, this scaling happens when equilibrium distance of the bound state varies. Experiments also point out that stability of the bound state can fail when equilibrium distance decreases below a limit depending on configurations.

### 6.0.1 Simulation of bound state of free bodies

In rigid body dynamics simulations of MBS of two free bodies, it is first shown that a MBS configuration is stable under near equilibrium condition, then it is shown that the same configuration is also stable under non-equilibrium where distance between bodies varies with time. In both cases, kinetic energy is exchanged between bodies but energies of bodies are restored at the end of the cycle where bodies return to their initial positions. In these simulations, magnetic properties of bodies presented as point dipoles and basic torque and force equations between two magnetic moments (Eq. 4.2, 4.3) are provided to the simulation configuration. In other words, simulation works independently from models and equations developed in this work. Basic configurations of simulations are taken from experiments presented in this section.

While it is relatively easy to obtain a bound state between a free body F (floator) and a rotating dipole R (rotator) constrained with a fixed axis, an additional condition is required for a rotating body having full DoF in order it anticipates the torque received from the other body. As stated in the previous section, a rotating magnetic moment $\vec{m}_R$ can be decomposed into a cyclic $\vec{m}_C$ and a static $\vec{m}_S$ moment. In the PFR model, $\vec{m}_C$ is responsible for repulsive action while the $\vec{m}_S$ provides the attraction and keeps the axis of the conical motion of the body be aligned to it or close to. As magnitudes of these moments vary with the tilt angle $\gamma$ of the rotating dipole (Eq. 5.27), it is needed to keep this angle in a range in order to obtain an equilibrium. When the motion of the $\vec{m}_F$ is symmetric about axis-$z$, it can be also decomposed into a component $\vec{m}_{F1}$ rotating orthogonal to $z$ and the other $\vec{m}_{F2}$ aligned to $z$. As the dynamic of the interaction forces bodies motions be synchronized (Eq. 5.6), that is, they share the same azimuthal angle $\omega t$, the torques received by bodies are constant in magnitude. The torque of the $\vec{m}_{F2}$ forces $\vec{m}_R$ to align parallel to it, with the axis-$z$. In order to prevent this from happening (otherwise angle $\gamma$ will go to $\pi/2$), this torque can be balanced by an inertial torque based on Euler's equations of motion (Eq. 5.15). For this purpose, in experiments and in simulations about bound state of free bodies, the body embedding $\vec{m}_R$ (the rotator) needs to have a larger MoI in the axial direction. In detail, the rotator consists of a spherical dipole magnet housed inside of a non-conductive and non-magnetic ring (Fig. 6.1(a)). This ring ensures it rotates stably on the axis of the ring. The magnet is mounted with polar orientation having the angle $\gamma_0$ with the ring plane in order to have a static moment $\vec{m}_S$ which is needed for obtaining the bound state. As it is needed to balance the magnetic torque by the inertial torque on the rotator, this equilibrium is obtained when the ring plane gets the angle $\sigma$ with the rotating plane. The expected motion of the rotator is a simple rotation on axis-$z$ while the ring plane has a constant angle with this axis. This is similar to a motion of a disk mounted to a shaft from its center but not exactly orthogonal. This motion consists of two rotations

$$R_x = \begin{bmatrix} 1 & 0 & 0 \\ 0 & c_\sigma & -s_\sigma \\ 0 & s_\sigma & c_\sigma \end{bmatrix}, \qquad R_z = \begin{bmatrix} c_{\omega t} & -s_{\omega t} & 0 \\ s_{\omega t} & c_{\omega t} & 0 \\ 0 & 0 & 1 \end{bmatrix} \qquad (6.1)$$

where the angle $\sigma$ is the deviation of rotator symmetry axis from the rotation axis and the shaft rotation along axis-$z$ is specified by $\omega t$. This way, the rotation matrix $R_R$ reads

$$R_R = R_x R_z = \begin{bmatrix} c_{\omega t} & -s_{\omega t} c_\sigma & s_{\omega t} s_\sigma \\ s_{\omega t} & c_{\omega t} c_\sigma & -c_{\omega t} s_\sigma \\ 0 & s_\sigma & c_\sigma \end{bmatrix} \qquad (6.2)$$

It can be seen that the third column of this matrix corresponds to the vector belonging to a circular motion of the unit vector aligned with the symmetry axis $z$ of the rotator. Euler's equations of this scheme reads

$$\vec{\tau}_I = \mathbf{I}_T \vec{\omega}' + \vec{\omega} \times \mathbf{I}_T \vec{\omega} = \omega^2 s_\sigma c_\sigma (I_A - I_R) \begin{bmatrix} c_{\omega t} \\ s_{\omega t} \\ 0 \end{bmatrix}, \qquad \mathbf{I}_T = R_R \mathbf{I} R_R^T \qquad (6.3)$$

Here $\vec{\omega}$ is $\omega \mathbf{k}$ and $\vec{\omega}'$ denotes the angular acceleration which becomes zero. Terms $I_A$ and $I_R$ denote principal MoI of an axisymmetric body in axial and radial directions. This torque $\vec{\tau}_I$ should be provided externally in order the for rotator to perform the motion defined by rotation matrix $R_R$. In this equation, the inertial term is zero for a body having uniform MoI, therefore point out the motion has no harmonic motion character. The magnetic moment of rotator $\vec{m}_R$ can be defined as

$$\vec{m}_R = R_R \vec{m}_R^* = \begin{bmatrix} -\cos(\sigma + \gamma_0)\sin\omega t \\ \cos(\sigma + \gamma_0)\cos\omega t \\ \sin(\sigma + \gamma_0) \end{bmatrix}, \qquad \vec{m}_R^* = m_R \begin{bmatrix} 0 \\ \cos\gamma_0 \\ \sin\gamma_0 \end{bmatrix} \qquad (6.4)$$

where $\vec{m}_R^*$ is the magnetic moment of the rotator with respect to its symmetry axis. Here the sum $\sigma + \gamma_0$ becomes the tilt angle $\gamma$ of rotator's magnetic moment. This moment complies with its definition in Eq. 5.25 by phase difference $\pi$. By applying same phase to floator magnetic moment $\vec{m}_F$, the induced magnetic torque reads

$$\vec{\tau}_M = \tau_M(2\cos(\sigma + \gamma_0)\cos\varphi + \sin(\sigma + \gamma_0)\sin\varphi)\begin{bmatrix} c_{\omega t} \\ s_{\omega t} \\ 0 \end{bmatrix}, \qquad \tau_M = \frac{\mu_0 m_R m_F}{4\pi r^3} \qquad (6.5)$$

Here the spatial vector $\vec{r}$ connecting these moments is assumed aligned with axis-$z$ by neglecting a small lateral offset introduced by circular motions of bodies around axis-$z$ evaluated in Section 5.4.5. Since $\vec{\tau}_I$ and $\vec{\tau}_M$ have same unit vector, it might be possible to equalize these torques as

$$\tau_E = \omega^2 (I_A - I_R)\cos\sigma\sin\sigma - \tau_M(2\cos(\sigma + \gamma_0)\cos\varphi + \sin(\sigma + \gamma_0)\sin\varphi) \qquad (6.6)$$

where $\tau_E$ denotes an external torque might be present. Fig. 6.2(a) or Fig. 6.13(a) corresponds to simulation instances of this equilibrium where magnetic moments are on the $xz$ plane. For a stability, the stiffness of the system against variation $\sigma$ should be positive. That is, in an equilibrium where also an extra torque $\tau_E$ might be present, the sign of variation of the equilibrium angle by a small change of the extra torque should be the same of the torque variation. This requirement, based on Eq. 6.6 can be expressed as

$$\tau_E/d\sigma > 0$$
$$\omega^2(I_A - I_R)(\cos^2\sigma - \sin^2\sigma) + \tau_M(2\sin(\sigma + \gamma_0)\cos\varphi - \cos(\sigma + \gamma_0)\sin\varphi) > 0 \qquad (6.7)$$

This equation gives good stability figures when other parameters are kept constant, but it actually the term $\tau_M$ depends strongly on $\sigma$ due the equilibrium distance $r$ is based on forces where the attraction force between dipoles is proportional to $\sin(\sigma + \gamma_0)$. By omitting angles $\gamma_0$ and $\varphi$, the dependence of angle $\sigma$ can be derived as

$$\sin\sigma = \frac{\mu_0 m_R m_F}{2\pi r^3 \omega^2 (I_A - I_R)}, \qquad \gamma_0 = \varphi = 0 \qquad (6.8)$$

Due to dependence between $\sigma$ and $r$, in absence of $\gamma_0$, the attraction force can vary about by seventh power of inverse of the distance instead of fourth. Since PFR has also a similar power factor, stability of the equilibrium would be marginal, if not possible at all. Overall, variability of angle $\sigma$ is a negative factor for the stability and it is better to keep it as low as possible. As seen from Eq. 6.8, this can be achieved by providing a large MoI in axial direction. In simulations, rotator's MoI are chosen in order the angle $\sigma$ do not exceed angle $\gamma_0$, thus giving comfortable stability figures. This subject is further evaluated in Section 6.2.

A fifth simulation is made by setting angle $\gamma_0$ to zero and by adjusting moments of inertia of rotator in order to obtain an equilibrium. In this scheme the angle $\sigma$ is responsible for the attraction force and this can be obtained regardless of polarity of the floator. That is $\sigma$ follows the sign of $\cos\varphi$. It was possible to obtain stable equilibrium in simulation and experimentally although delicate and called bipolar bound state of free bodies. This scheme is evaluated in Section 6.2.

In cases where bodies perform symmetric motions around axis-$z$, the equilibrium of MBS in $z$ direction corresponds to zero value of component-$z$ of the force between dipoles in absence of an external force. The equilibrium should also cover lateral motions of bodies resulting in the vector $\vec{r}$ connecting dipoles having a lateral component with constant amplitude.

While parameters satisfying equilibrium of MBS can be obtained from these equations, it does not give stability figures. In essence, all DoF in this problem are linked to inertial figures and to cyclic forces or torques in a complex parametric excitation system. This allows kinetic energy exchange between bodies in various ways. It is even more difficult to evaluate the stability with non-equilibrium initial conditions. For this reason, stability of MBS of free bodies is evaluated through simulations. For this purpose four simulations are made. In the first simulation, initial conditions are provided as close as possible to the equilibrium, the second, a moderate non-equilibrium condition and the third, considerably far from equilibrium where magnitude of forces varies by two orders of magnitude and torques varies more than four times. As a consequence, angles $\varphi$ and $\gamma$ vary by 500%.

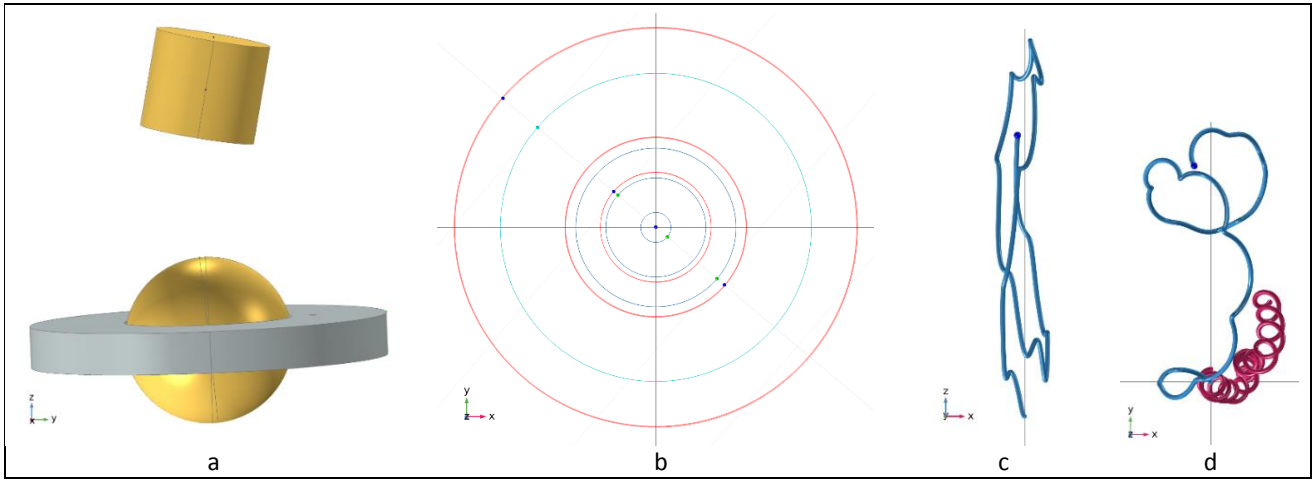

Fig. 6.1. A simulation (first run) of MBS where bodies are released very close to their equilibrium distance. During the simulation time, rotation point of the floator varied in this range: $z$: 24517 ±7 µm, $x$: -0.5 ±0.75 µm, $y$: 1.6 ±1.6µm. (**a**) Overview of the simulation. Top object is the floator, N pole looking up, bottom object is the rotator where it N pole is looking right on this instance, marked with small indentation on the ring. (**b**) Trajectories of reference points of bodies projected on $xy$ plane at the end of simulations. Red traces are belong to floator (F) and green ones to the rotator (R). From outside to inside these traces are F: bottom center, top center, CM, R: magnetic top center, body top center, CM. (**c, d**) Trajectories of floator's rotation center (blue) and rotator (red) projected on $xz$ and on $xy$ planes.

In all simulations, initial conditions are set in order to obtain symmetric motions of bodies around the rotation axis ($z$) of the rotator. These consist of spatial vectors of bodies and their first time derivatives. Initial conditions of floators are selected in order they conform to the motion based on the rotation matrix defined in Eq. 5.7 where spin velocity $v_S$ defined in Eq. 5.65 is zero. Bodies are released with zero velocity in direction $z$ and with tangential velocities of their predicted circular motions around axis-$z$. Second and third simulations are about releasing bodies at a distance in $z$ direction longer than the equilibrium distance with zero velocity where they accelerate into each other by the attractive force, bounce back at a short distance by the repulsive interaction and reach zero velocity somewhere close to their release points. The fourth simulation is similar to the first except an offset about 1 mm is given between bodies in direction-$y$. Both simulations gave good stability figures.

In the first simulation (Fig. 6.1) which aims bodies keep their initial positions, the distance between bodies varies by 14 µm in direction $x$ and body recovers its initial position by an offset of 2 µm. For $x$ direction, the variation is about 1.5 µm and the initial position is restored. For $y$ direction, the variation is 3.2 µm, while the bodies are moved back to recover their initial position simulation end before this happens. Smallness of these motions should be considered as the distance between body's centers is 24.51 mm. As expected, angles $\varphi$, $\gamma$ and kinetic energies of bodies did not change. These results give the sign of stable equilibrium. It should be noted the initial conditions figures are provided by five digits of precision in general, but the initial angular velocity of the floator on axis-$y$ was off by 0.2% causing some small irregularities on $x$ and $y$ displacements. As there is no friction/damping in these simulations, effects of initial conditions do not vanish.

In the second simulation (Fig. 6.2), bodies are released with zero $z$-velocity at distance 27.5 mm where the equilibrium position is 24.5 mm. This causes bodies run to each other, passing the equilibrium distance with a maximum velocity in $z$ direction as 0.146 m/s and bounce back at distance 22.7 mm (41.3 ms) and return their initial positions (84.37 ms) with a precision of 5 digits when their $z$-velocity becomes again zero. In this course, tilt angles of bodies smoothly varies to accommodate varying forces and toques but also contains minor fluctuations. Despite forces are very sensitive to tilt angles and affect each other's, these fluctuations did not caused instabilities and irreversible kinetic energy exchanges. On the other hand, there is a small kinetic exchange between bodies. While rotator kinetic energy is 700 mJ at the beginning, it drops to 698 mJ at the closest distance and recovered at the end of cycle. Floator starts with kinetic energy 0.57 mJ, this becomes 2.02 mJ at the closest distance and restored back at the end of cycle. By ignoring the small error of 0.5 mJ in this equation, these figures show there is about 2 mJ transferred from rotator to floator and taken back at the end of cycle. Despite fast rate of changes of forces and torques during this simulation covering only 13 rotations of bodies, stability is conserved and there was no apparent irreversible energy transfer.

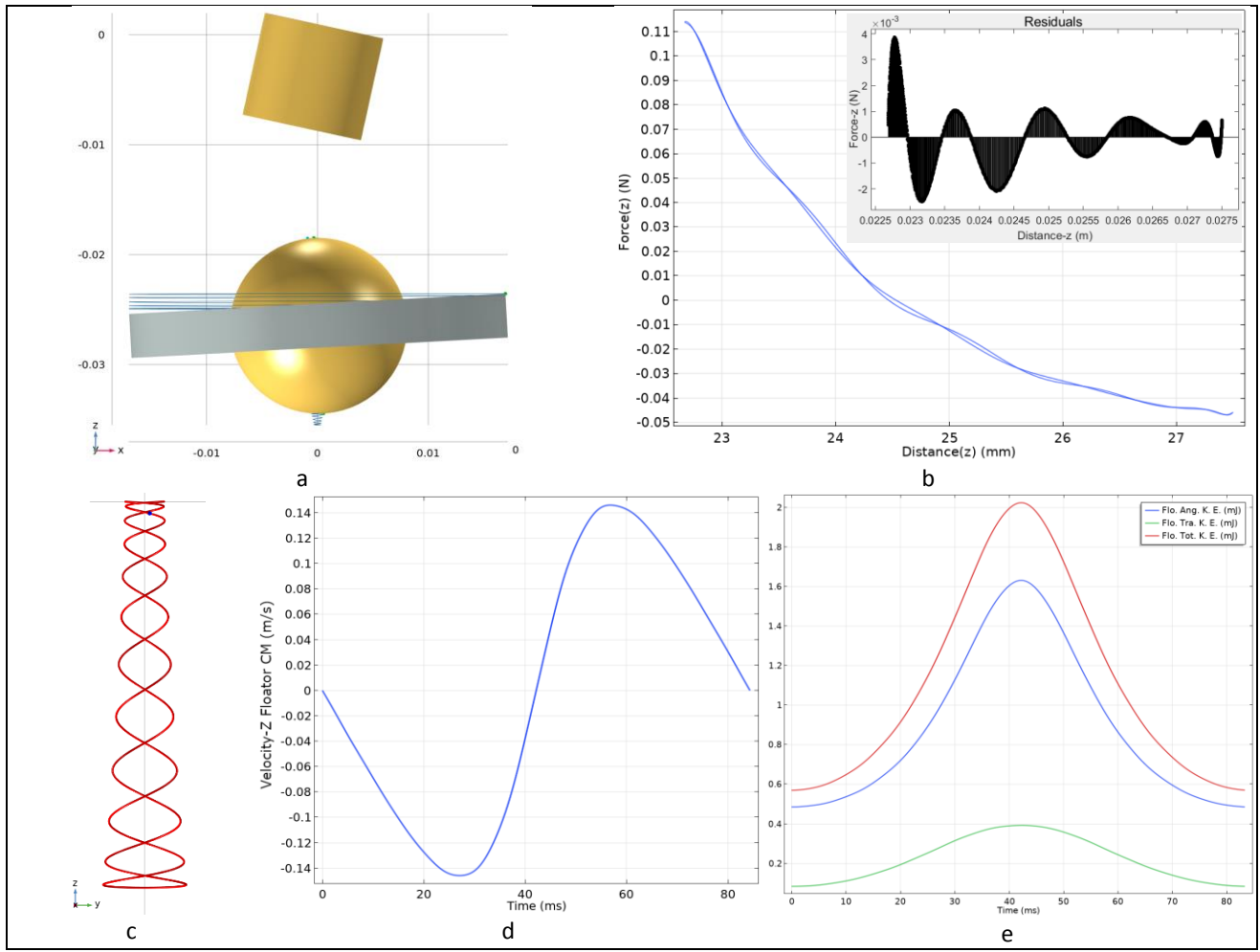

Fig. 6.2. A simulation (second run) of MBS where bodies are released at 3 mm off from their equilibrium distance in direction of rotator's rotation axis (*z*). This causes bodies to oscillate around the equilibrium point stably. (**a**) A screenshot of the simulation while bodies are in motion. The horizontal trace around the rotator is belong to trajectory of a point on the edge of ring (notice the small blue dot, marking this point) coinciding to the N pole orientation. While the trace spirals by the motion of the rotator on *z*, it keep its parallel alignment to the *xy* plane despite the ring plane is not. Floator top face is belong to its N pole. Its conical motion is synchronized with the rotator and its S pole is always looking toward S pole of the rotator despite it is attracted by the N pole. (**b**) Profile of the force-*z* between bodies vs distance while the bodies traverse this distance two times within a cycle. The undulated trace is belong the second half of the cycle where bodies are returning back to their initial position. This trace is undulated due to quick changes on torques they experience at the bounce instance. Due to this quick change, the tilt angles of bodies adapt to these changes in a oscillatory manner which is reflected on forces. Trace cross zero at equilibrium distance at 24.5 mm. The variation of the force by distance have a curve matching to $4.4e15r^{-8.5} - 1.2e8r^{-4.5}$ where residuals are shown in the embedded figure for the approaching sequence of the motion. (**c**) The projection of the trajectory of floator (CM) on YZ plane after completing its first cycle of the motion where blue dot marks its actual position. (**d**) Time profile of the velocity-*z* of the floator. The curve have significantly high slope where it cross zero at the bounce instance. (**e**) Time profile of kinetic energy of the floator detailed as angular and translational energies within one cycle of the motion along *z*. The trace shows that the initial energies levels are restored at the end of the cycle. Most of the translational kinetic energy belongs to lateral circular motion of the body and maximized at the shortest distance.

The third simulation (Fig. 6.3) is similar to the second, except the releasing distance is 35 mm, about 10 mm farther than the equilibrium distance. In comparison, this offset is 3 mm in the second. This causes bodies to obtain a peak velocity about 0.32 m/s before and after the bounce while in the second it is 0.14 m/s. This increased velocity causes a shorter bounce distance, therefore to larger forces. This also increases time variation of the forces. In this case, the magnitude of time derivative of the force on the *xy* plane $dF_{xy}/dt$ makes a peak of 121 N/s where it is 46 N/s in the second simulation. As these variations become larger, these have an effect on the motion of the body that can no longer be symmetric about axis *z*. Once this happens, the translational part of the PFR becomes effective, that is, bodies get repelled in direction of their lateral offsets from origin. This is similar to the ponderomotive action. The net force bodies experience from this effect has also short range about two times shorter than magnetic forces between dipoles. The distance profile of the force-*z* is similar to that of the second run. As the static components of magnetic moments are in parallel, this can balance the repulsion and keep the bound state at the equilibrium distance. Such an effect is visible in this third simulation where bodies gained an offset from axis-*z* after the bounce. The effect is directly visible in force plots by introduction of oscillations which slowly vanish as bodies get separated. Components *x* and *y* of the floator velocity become negative in the last 15 ms of the simulation, indicating that bodies are going to restore their alignments with the axis-*z*. On the other hand, quick changes in forces and torques may result in irreversible energy exchanges between bodies, therefore under repetition of these conditions (i.e. sharp bounce effects), the bound state should not be considered stable. In this third simulation, the bounce cycle resulted in an increase of distance by 32 μm corresponding to 0.1% change. Rotator angular velocity which is initially 857 rad/s, drops to 854 rad/s and recovers back to 856.87 rad/s at the end of the cycle corresponding to loss about 0.015%. However, this change and the variation in kinetic energy figures are too small to make a judgment.

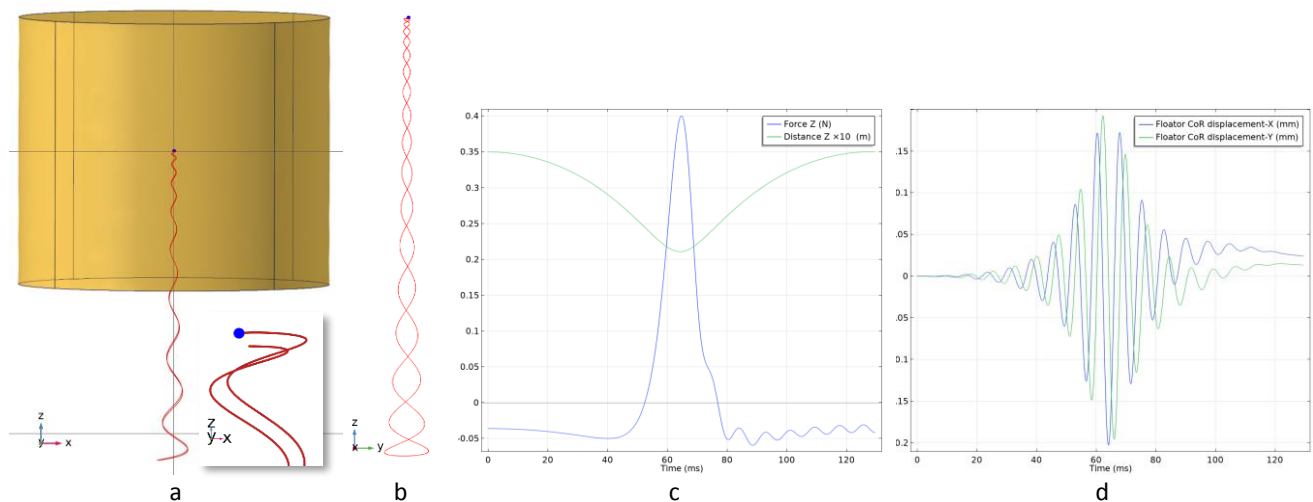

Fig. 6.3. A simulation (third run) of MBS where bodies are released at off 10 mm from their equilibrium distance in the direction of rotator's rotation axis (*z*). (**a**) The trajectory of the floator projected on *xz* plane when the floator bounced back to its initial position. The zoomed part corresponds to top of this trajectory where the floator returns to its initial position at the end of cycle with a +0.03 mm offset. Same trajectory (in a different scale) is shown on Fig. (**b**) as a projection on the *yz* plane. The undulation in the trace corresponds to a circular motion on the *xy* plane where amplitude of lateral motion gets several times larger at the bounce distance. (**c**) Time profiles of the force component-*z* and the *z* distance between bodies. (**d**) Displacement of rotation center of the floator in *x* and *y* directions. The rotation center varies as torque and forces varies by distance and this is reflected by varied amplitude of traces.

In the fourth simulation, bodies accelerates into each other in order to close their offset-*y* equal to 1 mm. Because floator is no longer on the axis-*z*, its conical motion becomes asymmetric, despite this, the stability is maintained and the equilibrium distance *z* remains almost same. Despite this simulation terminated before completing the cycle, it gave good stability figures.

Basic configuration used in these simulations is follow:

Floator: A cylindrical axially magnetized NdFeB magnet with Br 1.2 T having dimensions ⌀10.94×9.48 mm, density 7.5 g/cm$^3$, mass 6.6592 g, moments of inertia figures $I_A$ = 9.93e-8 kgm$^2$, $I_R$ = 9.95e-8 kgm$^2$, magnetic moment 0.84908 m²A.

Rotator: An assembly of a spherical dipole NdFeB magnet with Br 1.2 T with diameter ⌀16 mm, density 7500 kg/m$^3$, magnetic moment 2.0437 m²A, housed centrally in a non-conductive and non-magnetic ring with inner radius 8 mm, outer radius 17 mm, height 4 mm, density 3000 kg/m$^3$. The angle $\gamma_0$ of the dipole with the ring plane 0.07 rad. Assembly has mass 0.024529 kg, moments of inertia figures $I_A$ = 1.905e-6 kgm$^2$, $I_R$ = 1.169e-6 kgm$^2$. Angular velocity = 857 rad/s.

The time profile of component-$z$ of the force $F_Z$ also gives a clue about the mechanism of the kinetic energy exchange between bodies. While the distance between moments is constant, the conical motion of the floator is circular, resulting in a constant force. However, then this distance varies, this conical motion spirals as the angle $\varphi$ varies. This has an effect on the floator angular velocity according to Eq. 5.12. The change occurs on the component $z$ of the velocity in accordance to this equation. As the initial angular velocities are recovered at the end of the cycle, this variation is reflected on azimuthal angle of the floator by a change of +0.4 rad in the second simulation and +0.7 rad in the third. Based on Eq. 5.12, the norm (magnitude) of the vector $\vec{v}$ reads

$$|\vec{v}| = \omega\sqrt{2(1-\cos\varphi)} \qquad (6.9)$$

Since motion of the floator is forced be synchronized with rotator (as a consequence of DHM) and by neglecting the variation of rotator angular velocity, the kinetic energy of floator reads

$$K_E = \frac{1}{2}I_F|\vec{v}|^2 = I_F\omega^2(1-\cos\varphi) \qquad (6.10)$$

where $I_F$ denotes uniform MoI of floator. This equation shows that the floator gains rotational kinetic energy by increase of the angle $\varphi$ while it approaches the rotator. On the other hand, the variation of might not be smooth. Simulations show small undulation of the force in this process which is translated to (modulation) of translational motion. This is mainly caused by small oscillations of tilt angles of the bodies, since these angles vary by the distance. While simulations did not exhibit a non-reversible energy exchange due these oscillations, non-reversibility cannot be basically ruled out. Since forces equally act to bodies, this also involves the motion of the rotator, in turn, this motion also contributes to the PFR through the lateral harmonic motion.

### 6.0.2 Stability and energy aspects of MBS

In general, stabilities of bound states are evaluated by the binding energy criterion. This energy is typically negative, hence a work is needed to break the bound state. Therefore, the binding energy can be equal to the energy needed to break the bound state. There are cases where this energy can be supplied internally. Bound states allowing this are called quasi-stable. MBS can be classified as quasi-stable as the rotating field or the kinetic energy of the floator can supply energy enough to break the bound state. Above simulations show the presence of energy exchange, and under some conditions this exchange can become irreversible and the bound state could be broken. In experimental realizations, low frequency oscillations in various DoF are often encountered. There are cases where oscillations grow indefinitely in absence of damping and break the bound.

MBS exhibits a potential well where the minimum corresponds to its potential energy at its equilibrium distance. This is shown in Fig. 6.4 by integration of the effective force acted on a floating body and calculated as $V_{MBS} = -0.0035$ J on this configuration.

When the cyclic field is generated by rotating a dipole permanent magnet, there are two schemes for providing the static component of the field. One is tilting the dipole axis from the rotation plane (Fig. 6.7(a)) and the other, offsetting the dipole center from the rotation axis (Fig. 6.7(b)). These two schemes can be used together, which extend number of solutions. While the tilting scheme merely generates a static magnetic moment component aligned with the rotation, offsetting scheme can create complex field profiles. The offsetting scheme also can cause different effects which can significantly varies with the geometries of both magnets. In general this scheme generates stronger field gradients and can provide extra grip on floator in radial direction. Amplitude of translation oscillations can be increased with this scheme which also contribute to the PFR. In combinations of tilt and shift schemes, strength of the MBS can be increased, allowing longer equilibrium distances, so far it is observed. As mentioned in Section 5, the strength of PFR is three or more times weaker than static magnetic forces between dipoles in realized configurations. This provides an upper limit to the strength of static field component of the rotating dipole. It is found that tilt angle $\gamma$ should be less than 15° and the offsetting should be less than 1/4 of the dipole length.

PFR based model of MBS allows the stability to be evaluated based on PFR stability. This approximation does not work when the amplitude of motions of bodies getting large when they come too close to each other. Experiments show that a bound state can be destabilized in this condition, for example by increasing the attractive force with the tilt angle forcing bodies to find equilibrium at shorter distances. This characteristics is also reflected in numerical evaluations of motion on a 1D model. Stability of parametric excitations are covered in literature; however in this work, evaluation of the stability is limited to simulations and experiments.

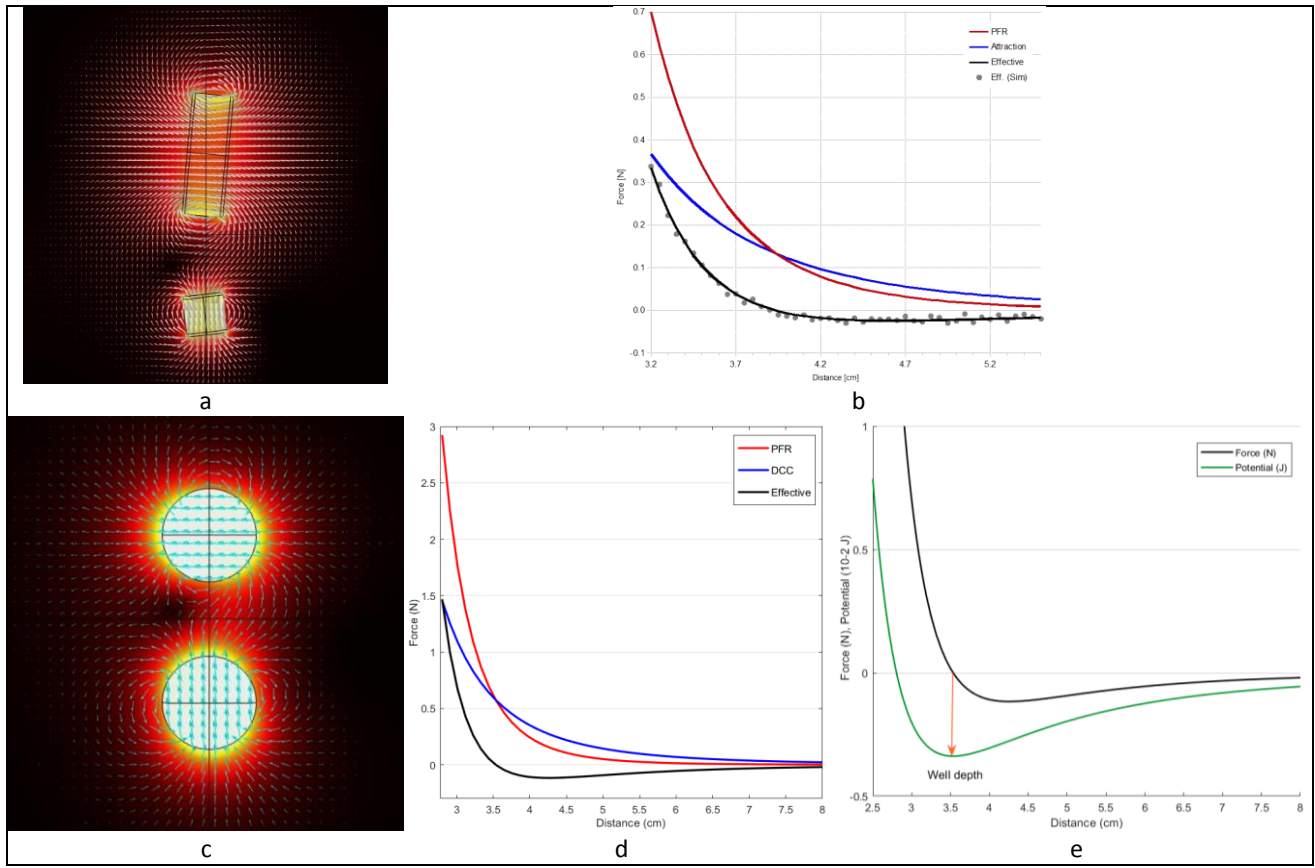

Fig. 6.4. (**a**) Configuration and 3D field plot of two interacting cylindrical dipole magnets assumed under bound state shown in cross section *xz*. In this configuration, the top magnet rotates around the *z*-axis and is tilted by an angle $\gamma = 5°$ between its dipole and *x*-axis. Bottom magnet is assumed to have full DoF and its tilt varies by the distance proportional to the *y* component of torque it receives but in the opposite direction due to phase lag. (**b**) Plots of *z* component of forces as function of distance between centers of magnets. Black curve corresponds to an external force which balances magnetic forces. The external force has positive sign when it pushes magnets into each other. It is zero at 3.95 cm, which corresponds to the equilibrium distance of the bound state in absence of external forces. This equilibrium is stable because magnets repel each other when getting closer than this and attract otherwise. Red curve corresponds to the strength of the PFR force, which is obtained by the simulation by setting the angle $\gamma$ to zero which cancels the attraction between magnets when the top magnet rotates around the *z*-axis. Blue curve is obtained by subtracting black curve from red. This curve corresponds to the attraction factor determined by angle $\gamma$. Red (PFR) and blue (AF) curves are in the form of $y = ax^b$ with *a*, *b* values (7754, −8.0) and (107, −4.88,) respectively. (**c**, **d**, **e**) Similar analysis for interacting two spherical dipole magnets having same size (⌀10 mm) and magnetizations M=1e6 A/m, corresponding to $B_r$ = 1.256 T (NdFeB/N38). Values *a*, *b* of repulsive (PFR) and attractive (AF) force functions are obtained as PFR (3950, −7.0) and AF (89.5, −4.0) when the tilt angle $\gamma$ of the rotator is set to 5°. These parameters are obtained by curve fitting method and power factor terms have an uncertainty about ±0.3 in general. Potential energy of this configuration is shown in plot **e**.

### 6.0.3 Initiation of MBS in experiments

In experiments, MBS of a free body with a rotator where its rotation is externally sustained is established by releasing the free body at proximity of its equilibrium position. This is required in general because body gains a kinetic energy while moving from release point to the equilibrium point. This energy can be only dissipated in presence of friction. Additionally, it is difficult to provide initial angular velocity to the body. This problem is solved in experiments by providing a temporary damping until body settles at its equilibrium position and obtains proper motion. One practical way to do it is releasing bodies having size larger than 8 mm while holding between two or three fingers and loosen them in order for the body to obtain its stable position and its motion dynamics while it is still in touch. The second method works with small sized bodies where bodies are laid on a soft and thick tissue like velvet and engages it to take off. A similar method which can be used for larger bodies too is trapping bodies while they are immersed in a liquid like water. This way an efficient damping can be obtained. Once the bound state is established, body can be pulled out from the liquid.

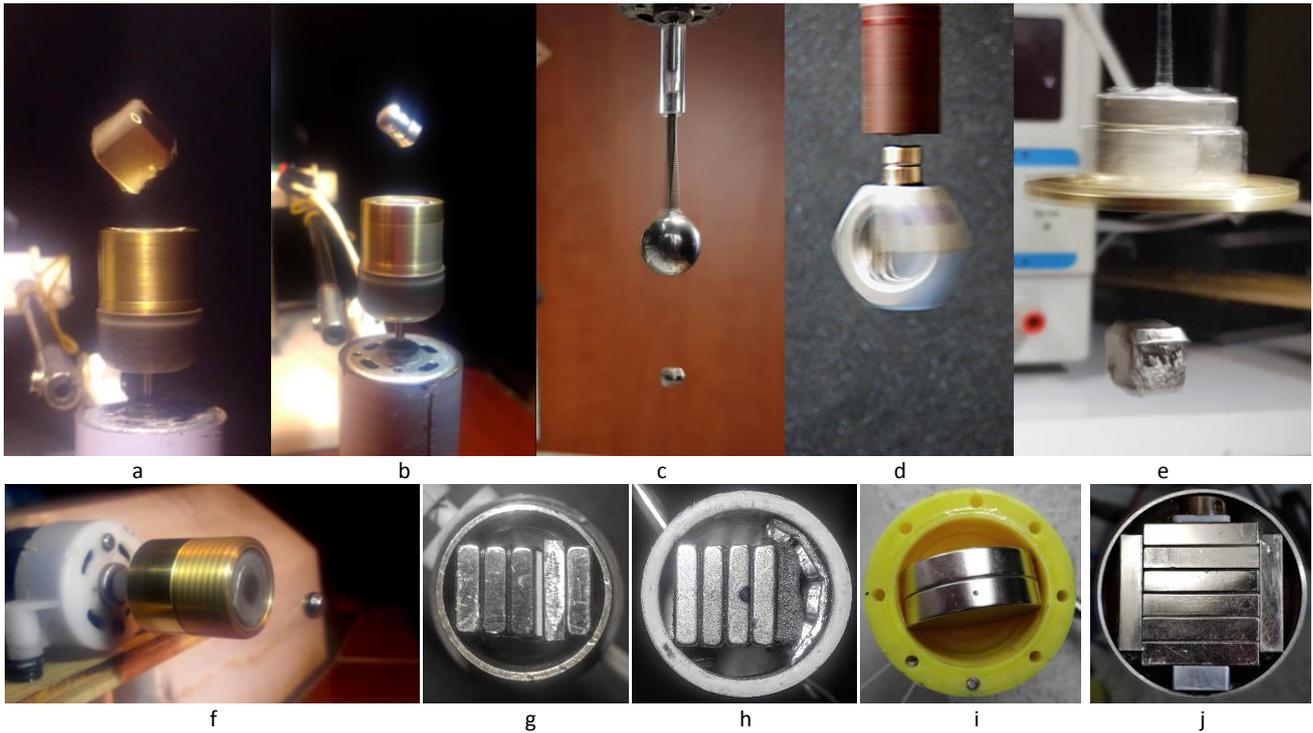

Fig. 6.5. Various bound state realizations and assemblies generating rotating fields. Assembly (**g**) which has asymmetric split dipole configuration obtained by four square magnets is used in realizations shown at (**a**, **b**). At Fig. (**c**), a small magnet (⌀6×3) is trapped below the sphere magnet (⌀15) in 'tilted' configuration. At (**d**), the floating magnet assembly consists of two stacked small magnets (⌀6×3) and a non magnetic nut is held by a rotating magnet of same size in side the cylindrical container attached to a rotor. Assembly (**g**) consists of four 10×10×3 magnets where an air gap is present between the third and fourth. Fig. (**h**) shows an off center dipole configuration consists of four stacked magnets (12×12×3) and a counterbalance weights. The round mark on the third magnet from the left shows the rotation center. Assembly shown at (**i**) is a tilted dipole configuration as the bottom face of the magnet becomes visible because of the tilt. Magnets dimensions are ⌀50×10. The tilt amount can be also seen from the small dot visible on the bottom magnet marking the rotation center. Fig. (**j**) shows a large assembly of eight 50×25×10 magnets where vertically stacked six magnets form a dipole and two side magnets in anti-Helmholtz configuration provides the static field along the rotation axis. In (**e**), a rotating body (top) is hung from a rotor by a thin string allowing five DoF. Dimensions are given in mm.

As the rotation of the rotator is sustained externally in experiments, this allows to obtain bound state in presence of damping of the floator. In this case, a continual energy transferred to the floator and its motion is sustained. In these cases, it is found that friction does not destabilize the motion and typically the stability of bound state can be extended in presence of damping. For this reason, a bound state obtained in an air environment persists also in a liquid environment where significant friction is present; however, the opposite does not always happen.

### 6.0.4 MBS force profiles and design notes

Under MBS, bodies find equilibrium in the geometry determined by the profiles of the attractive force and repulsive forces according to PFR model. While the repulsive interaction is present in any direction by the nature of PFR, attractive force profile varies with configurations. Therefore, zones where bodies find equilibrium can be primarily determined by the profile of the attractive force. Within a bound state where rotator is a dipole magnet, the tilt and shift schemes shown on Fig. 6.7 can generate static dipole fields (or virtual dipoles) for obtaining the attractive force. These fields have polar asymmetries in $z$ direction allowing a floator can find equilibrium on axis-$z$. This scheme actually might allow two floators to be bound to a rotator at the same time at opposite positions on this axis-$z$. It is also possible craft a static field geometry and cyclic field profile using two or more dipoles where floators can find equilibrium on radial positions like the realization shown on Fig. 5.31. This an interesting scheme where more solutions might be derived. Fig. 6.6 shows examples of realizations based on tilt scheme, (b) and (c) show its implementation. In Fig. 6.5, examples of tilt and other schemes are present.

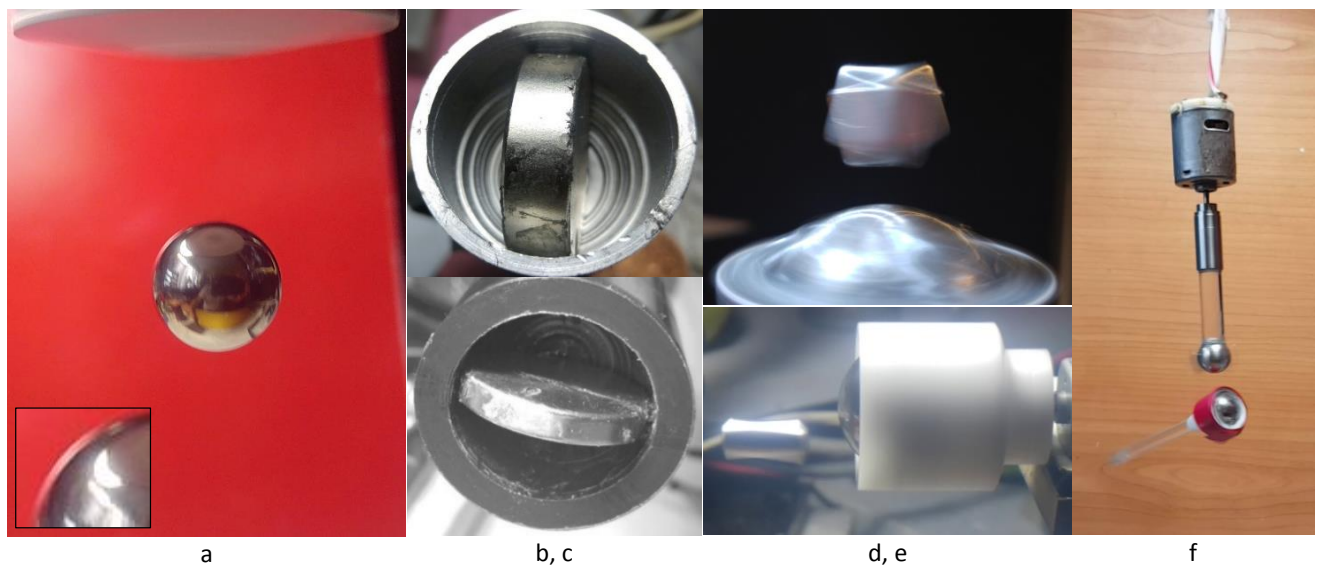

Fig. 6.6. Bound state of dipole magnets. (**a**) A spherical magnet (⌀15) is trapped by a rotating magnet assembly visible at top edge of the image consisting of a disk magnet (⌀40×10) mounted inside of an aluminum cylindrical block. (**b**) attached to a rotor. The small translational motion of the floator can be seen on its fuzzy edge, zoomed at the embedded rectangle. The small tilt angle of the rotator (~3°), suffices to balance the PFR and the weight of the trapped magnet. (**d**) The floator is a cube of edge 10 mm and the rotator is a disk of ⌀30×5 having a tilt about 4° as shown on (**c**) right picture. Floators in tests (a,d) also spin around their dipole axes and obtained a constant spin velocities equal to an integer fraction of the rotator velocity. This fraction is typically 1/2 for spherical shaped floators. (**e**) A cylindrical magnet (⌀7×15) is held horizontally by the by a rotating spherical magnet (⌀25.4) attached to a rotor using an plastic housing. The torque-phi effect is efficient on this configuration due to elongated shape of the floator and play role on it orientation combined by the force of the gravity which shift it position down from the rotor axis. This floator does not spin due to large misalignment/inhomogeneities of its magnetic field combined with its off axis position condition. (**f**) An assembly (floator) containing a spherical dipole magnet (⌀15) shown under bound state with a rotating dipole magnet of same attached to a motor swinging through its cables. Motor speed is about 100 rev/sec. The floator assembly is actually another rotator assembly but used as floator here. Dimensions are given in mm.

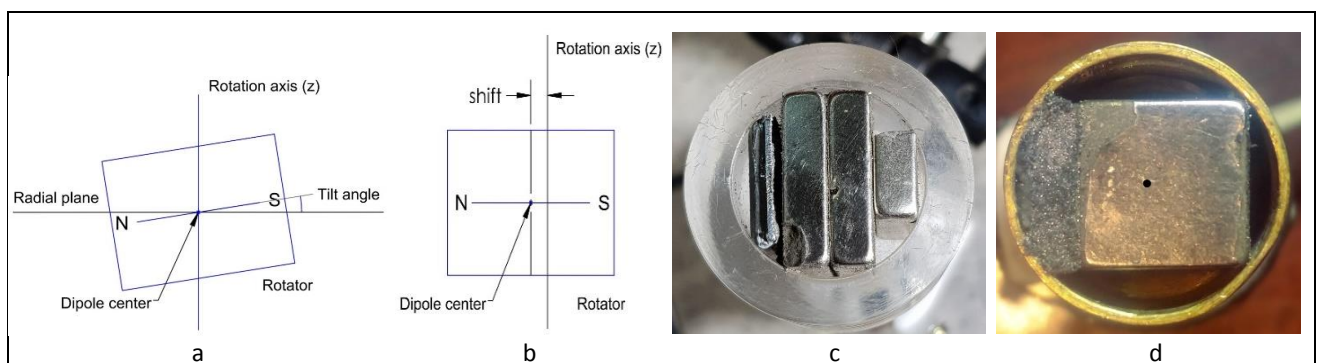

Fig. 6.7. (**a**) Tilting a dipole magnet with respect to its axis produces a dipole moment in this axis direction. This tilt produces a static $z$ component of the rotating field. (**b**) Rotating a dipole magnet on an axis off its dipole center generates a static component of the rotating field having gradient which can serve to obtain an attractive force on the floator in direction-$z$. Figures (**c**) and (**d**) show rotator assemblies based on the offset scheme. Assemblies are balanced by non-magnetic counter-weights. Magnets used in Fig. (c) are two 20×20×5 mm and one half and in fig.(d) is a ½" cube where the rotation center of the assembly is marked by a dot.

There is a tradeoff between the strength and the gap length of a bound state. For applications, larger figures of both them may be desirable. These properties can be varied by varying the strength of the attractive field, however at expense of the other. An alternative way to improve the strength of the bound state by means of stiffness is altering the slope of the attraction force without reducing the distance. This can be done by increasing the distance between floator and of the static dipole while adjusting its strength in order to obtain the same attraction force at the same gap length (between floator and rotator). This method is used in experiments (Fig. 6.8 and Fig. 6.9(a, b, d)) by providing the attractive force by a magnet placed back of the rotating magnet. This also has a tradeoff as reducing the lateral stiffness limits its applicability. Additionally, asymmetries respect to axis-z in realization at Fig. 6.8 are present which may be account for the extra stiffness on radial directions.

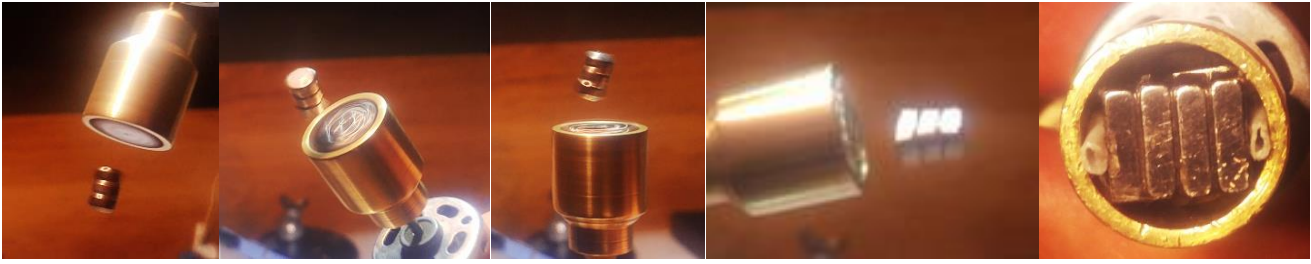

Fig. 6.8. Four sequences of a test showing a bound state anticipating gravitational force regardless of its direction. Floator is a stack of three ⌀6×3 mm magnets and rotator is an assembly of five 10×10×3 mm magnets where four of them are stacked and one at behind them. In this assembly, magnets have room to move to sides of the housing and once rotator starts to rotate in high speed, a gap about 0.6 mm develops between two middle magnets. Rotation speed is 230 rev/s.

In realizations shown in Fig. 6.9(c) and Fig. 5.30, the opposite is applied as placing the attractive magnet on the front. The effect of the offset of a dipole providing the attractive force is shown on the following. Here, the PFR force $F_R$ is associated with a rotating dipole $D_R$ which can be balanced by an attractive force either $F_1$ or $F_2$ (associated with dipoles $D_1$, $D_2$). By expressing these forces as function of the distance $r$ and following a normalization we can write

$$F_R = r^u, \qquad F_1 = -r^v, \qquad F_2 = -a(r+b)^v \tag{6.11}$$

where terms $u$ and $v$ denote the power factors of the repulsion and attraction forces which vary in range (−7, −8.5) for $u$ and (−3.5, −4.5) for $v$, $a$ denotes strength of $D_2$ relative to $D_1$ and the term $b$ the offset of the $D_2$ with $D_R$ or $D_1$. By setting $= a^{-1/v} - 1$, $F_2$ becomes equal to −1 at distance $r=1$, same as $F_1$ and $-F_R$. On the other hand, the slope of the $F_2$ at $r=1$ can be calculated as

$$dF_2/dr_{r=1} = -va^{1/v} \tag{6.12}$$

When $a$ greater than unity, $b$ becomes positive, these dipoles align on axis-z as diagram $(D_2)$ –b– $(D_R)$ –r– $(D_F)$ where floator's dipole is marked as $D_F$. For example, by setting $v = -4$, $a = 3$, the offset $b$ becomes 0.316 and the slope of $F_2$ at $r = 1$ becomes 3.04 while it is 4 for $F_1$ and −7 for $F_R$ when $u = -7$. Stiffness has opposite sign of the slope. Therefore adding these stiffness terms, the total stiffness characterizing MBS with scheme $D_1$ becomes 7 − 4 = 3 and with scheme $D_2$ becomes 7 − 3.04 = 3.96. This corresponds to an increase of stiffness by 32%. This way a positive stiffness can be obtained even $v$ and $u$ are equal, allowing a magnetostatic bound in reduced dimensions. This mechanism is used in a magnetic toy called "Inverter Magnet".

### 6.0.5 Frequency aspects of the magnetic bound state

Here, bound state of a free dipole body with a rotating dipole attached to a rotor is evaluated as its response to variation of rotor speed $\omega$. In this configuration, it is assumed the rotating dipole provides the static field required for the bound state in arrangement shown Fig. 6.7(a). This is the primary scheme used in experiments and in simulations where PFR is balanced by the attractive axial force induced by the static field. Stability of the bound state can be evaluated through PFR model by a difference where the distance $r$ between dipoles is a motion variable instead of parameter. This issue is not trivial because variation of $r$ is accompanied with kinetic energy transfer through variation of angle $\varphi$ according Eq. 6.10 and as it is observed in simulations. Simulations also indicate that this kinetic energy transfer can be reversible when variation of $r$ is slow, but might not be otherwise. Actually, the effective variable is $1/r$ since it enters equations this way. This points out that the kinetic energy transfer ratio is larger at shorter distances. Additionally, the stability criterion given at Eq. 5.44 is not invariant to scaling of torques that body receives, as mentioned in above section and it may not be possible to obtain stable equilibrium at arbitrary short distances. According to the PFR model, the repulsive force is approximately inversely proportional to square of the rotator speed according to Eq. 5.81. Under an equilibrium, repulsive force is balanced by the attractive force by setting $F_z = 0$ in Eq. 5.85. By adding the $\omega$ dependency to this equation, the relation between distance and the rotator speed can be expressed as

$$\frac{a}{\omega^2} r^{-b} - c r^{-d} = 0, \qquad r = \left(\frac{a}{c\omega^2}\right)^{\frac{1}{b-d}} \tag{6.13}$$

By rounding the term $(b-d)$ as 4.0, the equilibrium distance varies by inverse square root of rotator speed $(1/\sqrt{\omega})$. This equation and the result holds when the body is weightless. Otherwise, the term zero in the equation should be replaced by the weight of the body with a sign depending on whether the body hovers over or beneath the rotator.

While the stability range of $\omega$ is determined by configurations, it is not difficult to design a configuration where $\omega$ upper limit be at least twice of the lower limit. The dependency of this range on the body inertial figures can be based on the PFR model. In Eq. 5.24, Eq. 5.46 and many others, outcomes depend on $\omega$ and the initial figures by term $\omega^2 I$, exclusively. This means the outcome is retained by varying $\omega$ and $I$ while keeping the term $\omega^2 I$ constant. Therefore, it can be said that rotation speed $\omega$ depends on MoI $I$ by its inverse square root. MoI can be associated with a mass (m) or a volume (V) for specific shape for bodies having uniform density. This way it is possible to obtain a relation between MoI of a body and its magnetic moment such as

$$m = f(aX) = a^3 f(X), \qquad X = x_1, \ldots x_n \tag{6.14}$$

where term $a$ is a positive number corresponding to the scale factor of the object, term $X$ is the specification of the shape such as 1,W/L, H/L where symbols L,W,H, denote length, width and height. Function $f$ translates the scaled $X$ to a mass figure. It can be shown the scaling parameter $a$ can be moved out from the function. For MoI, within the same logic we can define a function $g(X)$ which relates MoI $I$ to $X$ as

$$I = mg(aX) = ma^2 g(X) = a^5 f(X) g(X) = m^{5/3} f(X)^{-2/3} g(X) \tag{6.15}$$

Functions $f(X)$ and $g(X)$ provide constant values since $X$ is the specification of the shape. For a body having uniform remnant flux density $B_r$, its magnetic moment м is proportional to its volume such as

$$\text{м} = \frac{1}{\mu_0} B_r V \tag{6.16}$$

Therefore magnetic moment is proportional to the volume or the mass. The Eq. 5.24 can be used to obtain the dependency of the rotation speed $\omega$ to body mass by keeping the angle $\varphi$ constant. Since we want to keep the same rotator but adjust its speed to match to the floater's scaling while keeping the distance between moments unchanged and by considering the proportionality of torque and magnetic moment, this can be written as

$$I\omega^2 \propto \tau \propto \text{м} \propto V \propto m \tag{6.17}$$

Since $I$ is proportional (denoted by operator $\propto$) to $m^{5/3}$ according Eq. 6.15, it can be eliminated from the equation, allowing to obtain a relation between floater mass and rotator speed as

$$\omega \propto m^{-1/3} \propto 1/a \tag{6.18}$$

This also shows that rotator speed and floater scale factor ($a$) are reciprocal. For example, if rotator speed is chosen as 4000 RPM in trapping a spherical magnet of diameter 20 mm in air at a given distance $r$, 8000 RPM is needed to obtain the same result for trapping a 10 mm diameter magnet at same distance. This holds because magnetic forces are also scaled in the same way and would also hold in presence of floater weight in the force equation since this is also scaled in parallel. This result coincides with the test result in Section 5.4, Fig. 5.26.

Besides the stability criterion of PFR, there could be low frequency resonances on MBS in all DoF, which can destabilize the system. These are likely parametric resonances fed by nonlinearity of the system. As the equilibrium position of the body within the cyclic field varies by the frequency therefore varying stiffnesses of the dynamics, resonance conditions varies by the operating frequency too. For this reason, unlike the PFR stability where there is no upper limit for operating frequency, this is not always true for MBS. Through experiments, it is observed that the MBS has a certain stability frequency range for a given configuration: First, MBS get destabilized in the lower limit in accordance with PFR stability. Secondarily, bound state becomes weaker as the frequency decreases, since the equilibrium point shifts into the weak field direction and may not tolerate external forces. Although, by increasing the attraction force within a limit, the weakness issue can be improved. About the upper limit of frequency range, it is observed that the body could be destabilized by increasing the frequency. Since the PFR strength is approximately inversely proportional to square of frequency, body finds equilibrium at shorter distances as the frequency increases in order to balance the attraction force which is almost independent from frequency. As magnetic forces and torques increase very fast by lowering the equilibrium distance, large dynamic forces and torques can stress the system, exploit nonlinearities and increase the kinetic energy of the floator, unfavorable for stability of the bound state.

Resonances are important factors to destabilize the bound state, especially on a system where damping is tried to be kept low as possible. Experiments show that a system can easily find resonances and escaping from resonances can be challenging. Addition to six DoF the floator has, the mechanical system belonging to the rotator can also have micro DoF, allowing vibrations feeding resonances. Resonance frequencies can also vary by operating parameters, some resonance observations suggested they followed the system driving frequency.

### 6.0.6 Bound state in presence of external static forces

In performed experiments, the weight of the floator enters to the equilibrium together with magnetic forces. Here, we consider net forces along vertically aligned axis-$z$. As mentioned earlier, stability of MBS requires the distance profile of the repulsive force ($F_R$) needs be steeper than of the attractive force ($F_A$) at the equilibrium distance $r_e$. This can be expressed as

$$|dF_R(r_e)/dr| > |dF_A(r_e)/dr|, \qquad F_R(r_e) + F_A(r_e) + F_E = 0 \tag{6.19}$$

where $r$ denotes the distance variable and $F_E$ an external force having no or negligible dependence on the distance like the gravitational force on the floator. When the signs of these forces are taken account as $F_R$ positive and $F_A$ negative, the stiffness $S$ of the bound state as function of distance can be expressed as

$$S(r) = -d(F_R + F_A)/dr \tag{6.20}$$

This conforms to the definition of stiffness where the reaction force generated by a system against a displacement should be in the opposite direction when the system reacts with a positive stiffness. The stability of the bound state requires the stiffness be positive at equilibrium distance. According to force profiles obtained in magnetic field simulation results shown in Fig. 6.4(c, d), force profiles $F_R(r)$ and $F_A(r)$ are found fit to

$$F_R = 3950 \, r^{-7}, \qquad F_A = -89.5 \, r^{-4} \tag{6.21}$$

where units are cm and Newton. Based on this data, the stiffness function $S(r)$ in these units is obtained as

$$S = 27650 \, r^{-8} - 358 \, r^{-5} \tag{6.22}$$

where $S$ is positive for $0 < r < 4.26$ cm. At this upper boundary, the attractive force is 0.117 N larger than the repulsive force, that it can carry a floator weighing less than 11.8 g. In turn, the floator weight is calculated as 4 g in this simulation. This stability characteristics is shown in Fig. 6.4(d). Here, red and blue curves correspond profiles of repulsive and attractive forces magnitudes and the black curve to $|F_R| - |F_A|$. Red and blue curves crosses at 3.53 cm) corresponding to the equilibrium distance when the external force $F_E$ is zero. The equilibrium distance in presence of an external force can be obtained by following the black curve where a positive value on the force axis corresponds to a force pushing the floator toward the rotator and negative values in the opposite. It can be seen that positive values are tolerated well by decreasing the equilibrium distance (to the left) but $F_E$ can only be less than 0.117 N in the opposite direction where the curve reaches minimum. At this distance (4.26 cm), red and blue curves have equal slope and beyond this distance, no stable equilibrium can be found because the slope of the attraction force is steeper than of the repulsion. When the external force corresponds to the weight of the floator, the positive zone on the black curve corresponds to a configuration where the floator floats above the rotator and the negative zone to beneath it. By following this example, within equilibrium conditions where $F_E$ is zero (1) or corresponds to weight of the floator depending if it hovers over (2) or beneath (3) the rotator, the stiffness $S$ would be 48.7, 61.9, 35.1 N/m in this order.

The tilt angle $\gamma$ of the rotator through its sine term is the basic factor of the attractive force. In this simulation, $\gamma$ is set as 5° which is a comfortable angle in general and might be optimal for configurations where the floating body stays above the rotator. The stiffness of the bound state would increase together with this angle and can better anticipate external forces, but angles larger than 8° might not be comfortable (might not work for every configurations, difficulty on establishment on bound state and might have stability issues) when the external force (i.e. the weight of the floater) adds up to the attraction force.

## 6.1 Experiments on bound state of two free bodies endowing magnetic dipoles

Realization of friction free bound state of two macroscopic objects having full DoF is difficult in terrestrial conditions because of requirement of free fall. Here, this bound state is approximated by restricting one DoF of the rotator by a counteracting force in order to avoid free fall as shown in realizations at Fig. 6.9 and Fig. 5.21(d).

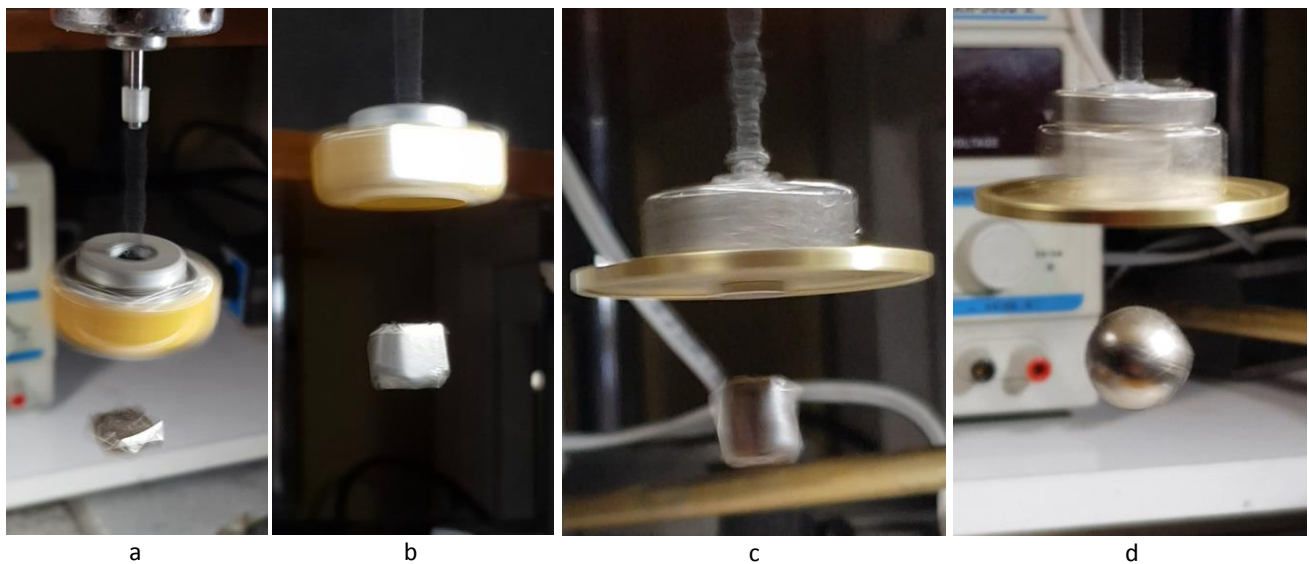

Fig. 6.9. An assembly (**a**, **b**) consists of two dipole magnets, the top, having ring shape aligned in axial direction providing DC (static) component and the other, large one in radial direction generating AC (cyclic) field. The assembly (**c**) is similar except the magnet providing attractive force is at the bottom and the attached non-magnetic metal disk is aimed to increase the MoI in the rotation axis direction. The assembly (**d**) have same magnets configurations of a, b and have an additional brass disk. As assemblies designed to have major principal MoI on rotation axis, they can spin stably on this axis with a small deviation. Assemblies are driven through threads in order to allow most of their DoF and to exhibit characteristics of bound state of two free bodies.

Presence of gravitational forces of on bodies affects following factors:

1. Restriction of free motion of the rotator along axis-$z$ when the interface providing the counteracting force has non-zero (positive) stiffness.
2. Torque on rotator when the force counteracting force has an offset with CM.
3. Damping caused by the interface which provides the counteracting force.
4. Gravitational force on floater. Gravity enforce rotator and floater system be aligned in vertical direction and when the factor 2 is present can provide extra stability to the system.
5. Torque on floater if its magnetic moment has an offset with the CM.

In realizations, the counteracting force against combined weight of rotator and stator is provided by a thin thread attached to the rotator close to its CM and the other end to a rotor. This way, one translational DoF of rotator on axis-$z$ is restricted and DoF in $x$ and $y$ directions are available when the thread is long enough. This thread is chosen as high and low elasticities (filament polyester, rubber) in term of stretching. This DoF restriction on axis-$z$ can only have effect on second and higher order derivatives of the motion since floater still have full DoF. When the floater motion is symmetric about axis-$z$ in the ideal condition, motion is free of component-$z$. That is, the dynamics of the motion does not depend on oscillations in direction of the rotation axis ($z$) of the rotator and floater, therefore the dynamics should not be affected by the DoF in axis-$z$. However, the interface can introduce some damping which can suppress secondary oscillations generated under non-equilibrium conditions. In the following, while other factors introduced by the interface are mainly resolved, this damping factor remains open and this approximation of free bodies would not be sufficient. On the other hand, simulations show that MBS of free bodies could be stable in absence of damping. In experiments, using low or high elasticity threads did not exhibited a difference, except in the establishment of the bound state where a thread with low elasticity helped to capturing floator by providing better damping, absorbing the energy released by this action.

The factor (2) stated above is not fulfilled as it requires a surface/hole at CM. Although, there are commercially available magnets having suitable geometries in ring or countersink cylinder forms, they are either axially polarized or the inner diameter is too small. The requirements for moments of inertia of rotator magnets are detailed in Section 6.0 within the simulation of bound states of free bodies. Actually, simulations are proceed by following experiments. Briefly, rotators having full DoF require a larger MoI in axial direction than the radial directions.

The factor (3), presence of damping might be the most important problem, since an unstable equilibrium can be rendered stable by damping. For this reason, simulations are valuable, since they show the possibility of MBS without damping.

The factor (4) is inevitable; however, experiments about the bound state of a free magnet with a rotating magnet on a fixed axis show that the bound state does not rely on gravity and it can be obtained regardless of its direction relative to the rotation axis.

If it is possible to eliminate the factor (2), the enforcement of gravity which align the system in its direction would have no effect if the counteracting force against gravity is applied to the CM. In this condition, rotating axis would be free and it would be governed by the net torque it receives from floator. A deviation of rotator axis from vertical would be clearly visible if this happens. Since such a deviation is not observed, it may be concluded that the upper limit of such a torque should be low despite counteracting force application point (thread attachment) have an offset with CM. This offset would generate a torque when rotation axis deviates and may limit this deviation but cannot prevent it at all. Therefore, it can be concluded that alignment of the floator to the rotator axis is the consequence of the bound state but not of an enforcement of gravity.

On the other hand, it was possible to dynamically change the direction of force acting on the rotator by the thread to some extent by swinging or jerking the system through the thread. These tests also show that the stability does not depend on the direction of gravity. In this configuration, the weight of the floator adds to the repulsion and result to extend the equilibrium distance with the attractive force. In turn, the slope of the profile of the repulsion force decreases faster by distance than the attractive force; therefore, this shift on the equilibrium distance decreases the angle between these slopes. This factor, therefore the presence of gravity can be considered as a disadvantage for the stability of the bound state in this configuration.

The factor (5) is assumed to be eliminated since bound state is achieved with floator magnets having complete symmetry (spherical, Fig. 6.9(d)) and other symmetrical shapes where dipole center coincides with CM and also with magnets/assemblies with asymmetries (Fig. 5.21(d)).

Overall, numerous experiments with different configurations show that bound states obtained by hanging rotators at the end of threads are similar to bound states with rotators rigidly attached to rotors.

In summary, tests and simulations support the availability of magnetic bound state of free bodies. Experiment configurations can be refined by reducing friction and damping factors, by using threads having large elasticity and by using rotators having geometries that allow threads to be attached to CM position. Magnets used in these experiments are NdFeB type having grade N35 or lower. Using higher grade magnets would allow to reduce weights and inertial factors dramatically since according to Eq. 5.81, PFR is proportional to fourth power of magnetic flux density of the magnets when floating distance is not too short.

## 6.2 Bipolar bound state of free bodies having dipole moment

This a special bound state of free bodies endowed with magnetic dipoles where the field generated by the rotating body does not have a static dipole moment component which is a moment aligned with the rotation axis (*z*) until it interacts with the dipole field of the other body. This moment, however, is required to keep bodies together attractively and is generated automatically when the rotating body is subjected to a magnetic field having a *z* component. When this happens, the dipole orientation, which is initially orthogonal to *z*, shifts due to the torque it receives from the field. In the case of a bound state, this field belongs to the other body's dipole. This way, a static dipole moment parallel to the dipole moment of the other body is generated. However, this shift should be limited; otherwise, no equilibrium can be found between attractive and repulsive forces. This limitation can be obtained by balancing the magnetic torque by an inertial torque. Fig. 6.10 shows such a configuration. This equilibrium is evaluated in Section 6.0.1 in detail. Here, the angle $\gamma_0$ is zero, that is, rotator's dipole lies on the rotation plane while the rotation axis is determined by the major principal MoI.

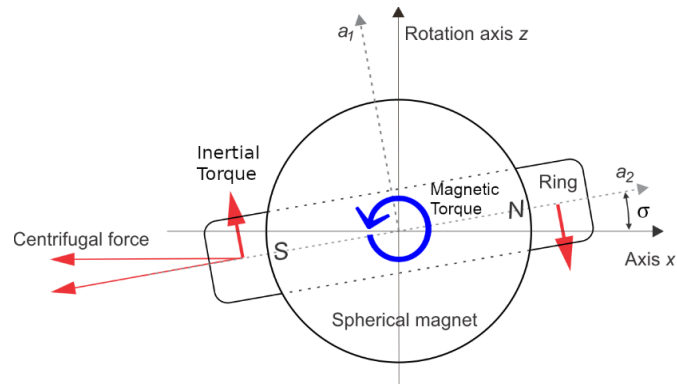

Fig. 6.10. Dynamics of free rotating body (rotator) subject to magnetic forces and torque of a floator (not shown) residing on axis *z* and N pole up. Rotator consists of a spherical magnet and an enveloping ring made by a nonmagnetic and nonconductive material. This ring determines the principal axes of inertia of the body by setting the major principal axis orthogonal to the dipole axis. In absence of external forces, bodies spin on this axis but when a magnetic torque is applied on axis-*y*, it gets tilted by angle σ and might be balanced by the inertial torque. In this tilted configuration, the *S* pole always stays below the *xy* plane and *N* pole above. It should be noted that this figure shows an instance of the motion, which can be described with Eq. 6.23. Torque vectors are on axis-*y* in this instance. Vectors denoted as inertial torques represent force vectors instead, in the logic of force × distance.

In absence of angle $\gamma_0$, the equilibrium of magnetic torque and the inertial torque using the Eq. 6.6 reads

$$\omega^2(I_A - I_R)\cos\sigma \sin\sigma - \tau_M(2\cos\sigma\cos\varphi + \sin\sigma\sin\varphi) = 0, \qquad \tau_M = \frac{\mu_0\, m_R\, m_F}{4\pi r^3} \qquad (6.23)$$

In this equation, magnetic moments are assumed aligned in direction *z* by ignoring the offset in lateral direction related to translational motions of bodies around axis-*z*. This offset is small and has a small effect on torques, but not negligible on forces. Later, by evaluation of equations, simulations and experimental results, it is concluded this offset plays a significant role in obtaining stable equilibrium in this scheme. Here, the angle σ is equal to the zenith angle $\gamma$ of the rotator's magnetic moment since angle $\gamma_0$ is zero.

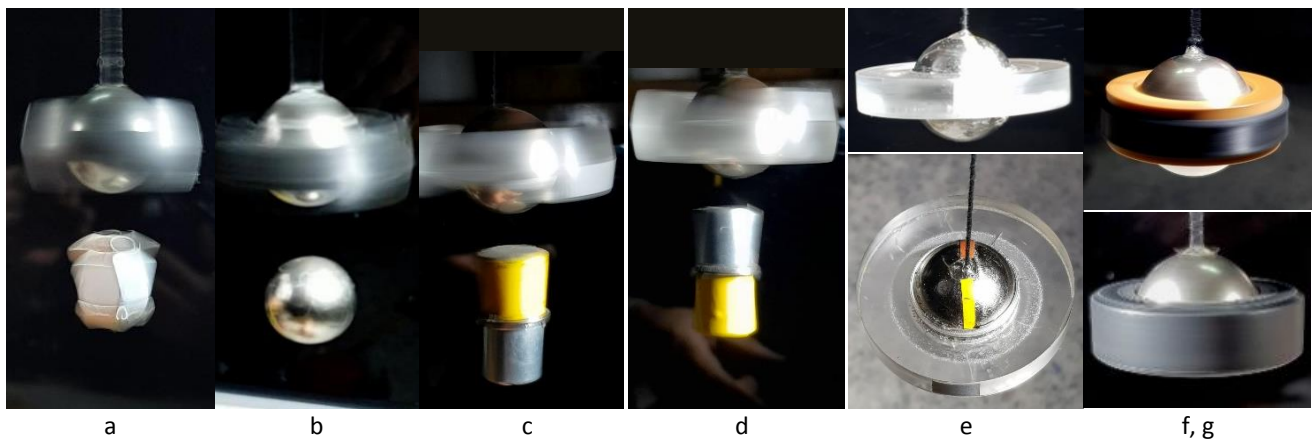

Fig. 6.11. Bipolar bound states of approximated free bodies. (**a**) A rotating assembly consists of ¾" dipole sphere magnet and a plastic ring surrounding it and attached to a rotor by a thread, approximating a free rotating body. The MoI of the magnet in direction orthogonal to its dipole is increased by this ring in order for the dipole to be aligned with the rotation plane while it spins. Under this body, a trapped magnetic assembly is found, consisting of a cube magnet of 10 mm in size

extended by two square magnets in polar direction. (**b**) A spherical magnet is trapped by a rotating assembly similar to the one used in setup (a), making BBS of two spherical magnets. The equilibrium is in the limits and bound state only persists for a short time because trapped magnet quickly gains spin causing an increase of the repulsion factor, which cannot be balanced by the attractive factor. It should be noted that rotating assemblies having spherical magnets may have small misalignments up to 0.3° between the dipole axis and the rotation plane. (**c, d**) A floator consisting of two ⌀15×15 cylindrical and a ⌀16×2 disk magnet in between is trapped in both polarities by a rotator satisfying bipolar bound state scheme. Rotator consists of a 1" spherical magnet mounted in a plexiglass ring (**e**) where its dipole is aligned with the ring plane. Stability is marginal in this test, possibly due to the weight of the floator (~40 g) plays a negative role. Additionally, rotator - floator matching might be not good enough. (**f, g**) Rotating assemblies based on ¾" and 1" diameter sphere magnets used in tests while spinning stand-alone. Rings are made from nonmagnetic and nonconductive materials. Rings of rotators are finely tuned by wrapping a ribbon (f) and tape (g) around in order to obtain proper MoI figures for matching to specific floators.

Despite this bound state is experimentally achieved (Fig. 6.11, Fig. 6.14) under approximated conditions, it is found considerably difficult to obtain, fragile and only 'matching' floator and rotator can be bound. Experimental realizations about this bound state are similar to the previous section and differ by orientation of the rotator magnetic moment with respect to the principal MoI and ratio of these MoI. The problem about this bound state can be summarized as follows: In bound state of free bodies, a destabilizing factor is the variability of the rotator's tilt angle σ. This factor can be suppressed (making σ arbitrarily small) by providing an arbitrary large MoI in the direction determined by the magnetic moment orientation and the angle $\gamma_0$ of the rotator. Since the strength of the magnetic moment component in the direction of the rotation axis is determined by the angle γ as the sum of σ and $\gamma_0$, the desired angle γ can be obtained through $\gamma_0$. However $\gamma_0$ is zero in this bipolar scheme, therefore σ cannot be arbitrarily small. Eq. 6.8 shows the relation of the angle σ and the distance $r$ in this scheme by omitting the angle φ.

In calculation forces, angle σ enters to equations. We are mainly interested in the component-$z$ of the force which provides the attraction. This term, by omitting the lateral offset between dipoles can be written as

$$F_z = (\cos\sigma \sin\varphi - 2\sin\sigma \cos\varphi)\frac{3\mu_0\, m_R\, m_F}{4\pi r^4} \tag{6.24}$$

In absence of external forces like gravity, $F_z$ should be zero under equilibrium condition and its derivative with respect to $r$ should be negative in order to obtain positive stiffness. Since the rightmost term is always positive above, we can omit it in order to evaluate these conditions as

$$\begin{aligned}\cos\sigma \sin\varphi - 2\sin\sigma \cos\varphi &= 0 \\ \tan\varphi - 2\tan\sigma &= 0\end{aligned} \tag{6.25}$$

$$d(\tan\varphi - 2\tan\sigma)/dr < 0 \tag{6.26}$$

The equilibrium of conical motion of the floator allows to relate the angle φ to γ (γ = σ) by the Eq. 5.30. By expressing this condition as

$$\frac{\sin\varphi}{\cos\varphi} = \frac{\tau_M \cos\sigma}{I_F \omega^2 - 2\tau \sin\sigma}, \qquad \tau_M = \frac{\mu_0\, m_R\, m_F}{4\pi r^3} \tag{6.27}$$

Here, the term $I_F$ denotes the uniform MoI of the floator. The substitution of the term $\cos\varphi$ in Eq. 6.25 using above Eq. reads

$$\cos\sigma \sin\varphi - 2\sin\sigma \left(\frac{\sin\varphi\,(I_F \omega^2 - 2\tau_M \sin\sigma)}{\tau_M \cos\sigma}\right) = 0 \tag{6.28}$$

Here, the term $\sin\varphi$ appears at both sides and can be eliminated. By arranging terms and using trigonometric identity, this can be expressed as

$$1 - \sin^2\sigma - \left(\frac{2I_F \omega^2 - 4\tau_M \sin\sigma}{\tau_M}\right)\sin\sigma = 0 \tag{6.29}$$

This further simplifies and gives a quadratic equation and its roots as

$$3\sin^2\sigma - 2A\sin\sigma + 1 = 0, \qquad A = \frac{\omega^2 I_F}{\tau_M} = \frac{4\pi I_F \omega^2 r^3}{\mu_0 m_R m_F},$$
$$\sin\sigma = \frac{A \pm \sqrt{A^2 - 3}}{3}$$
(6.30)

The root with the minus sign gives rotator's tilt angle required to obtain zero component-$z$ of the force between bodies when bodies' magnetic moments have zero lateral offset. This tilt angle also corresponds to the equilibrium between magnetic and inertial torques (Eq. 6.2) and can be realized by providing appropriate moments of inertia to the rotator. It is also possible to eliminate terms with $\varphi$ in Eq. 6.23 by using Eq. 6.27. First, the cosine term of $\varphi$ is substituted by a sine term and simplified as

$$\sin\varphi \, (2I_F\omega^2 - 3\tau_M \sin\sigma) = \omega^2(I_A - I_R)\cos\sigma \qquad (6.31)$$

And then the term $\sin\varphi$ is substituted using Eq. 6.27 again based on relation $sin(cot^{-1} x) = (1 + x^2)^{-1/2}$ as

$$\left(1 + \left(\frac{I_F\omega^2}{\tau_M \cos\sigma} - 2\tan\sigma\right)^2\right)^{-1/2} (2I_F\omega^2 - 3\tau_M \sin\sigma) = \omega^2(I_A - I_R)\cos\sigma\sin\sigma \qquad (6.32)$$

This equation allows resolving angle $\sigma$ from configuration parameters or choosing configuration parameters for obtaining a specific angle $\sigma$, however not accurate due to the lateral offset between moments being omitted.

The stability of a bound state in $z$ direction requires the profile of the force in this direction to have a negative slope; that is, the interaction should have positive stiffness like a spring. Equations 6.30 and 6.31 allow to evaluate the stability based on variation of angle $\sigma$. Fig. 6.12 shows plots of angle $\sigma$ as function of distance-$z$ as solution of above equations. By denoting the angle $\sigma$ of solution of Eq. 6.30 as $\sigma_F$ (related to forces) and of the solution of Eq. 6.31 as $\sigma_T$ (related to torques), the force between bodies is attractive when $\sigma_F > \sigma_T$ and repulsive when $\sigma_F < \sigma_T$ for a given distance. Therefore, for obtaining a stability around an equilibrium distance where $\sigma_F = \sigma_T$, $\sigma_T$ should have a larger slope than $\sigma_F$. However, this is difficult to obtain since these curves have similar slopes until bodies get significantly closer where the angle $\sigma$ becomes larger than 0.2 rad. This factor is reflected in experiments where distances between bodies are typically shorter than normal bound states. The Eq. 6.30 can be also applied to bound states where rotator has constant angle $\gamma$ to obtain this angle ($\sigma$) for a given equilibrium distance. Since angle $\gamma$ is fixed by the configuration, in a chart similar to Fig. 6.12, the $\sigma_T$ lines (black) would be left as is, and the red lines become simply horizontal and their levels would correspond to angles $\gamma$.

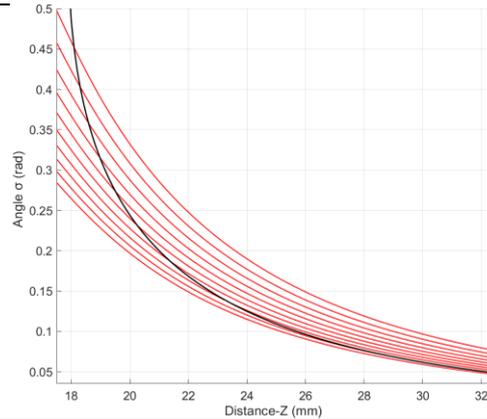

Fig. 6.12. Variation angles $\sigma_F$ (black) and $\sigma_T$ (red) vs distance for a given configuration. Red plots correspond to a range of inertia figures of the rotator where the term ($I_A - I_R$) varies between 0.181e-6 to 2.96e-6 kg m$^2$ in ten steps, top to bottom. Stable equilibrium of bodies are obtained at cross of a red and the black curve. The rotator of the last red curve cannot obtain enough tilt for the equilibrium and the interaction becomes always repulsive since red and black lines never cross. As a note, equations about these curves omits the lateral offsets of bodies caused by translational oscillations of motion. These offsets have effects on profile on forces and marginally on torques. In bipolar bound state, these effects gain importance since the equilibrium can be marginal.

In experimental realizations, weight of the floater enters also to the force equation and would affect the profile of $\sigma_F$, therefore Eq. 6.30 is not applicable. This bound state is called bipolar because there is no preferred polarity within a setup. Such setups are shown in Fig. 6.11. Each consists of a rotating permanent dipole magnet hang at the end of a thread while holding another dipole magnet in the air. This scheme works for both polar orientations of the bottom magnet and it is tested thoroughly. Alignment of the magnetic orientation with the assembly is challenging because of manufacturing tolerance on magnetic orientation about 1 - 2° with rectangular and cylindrical packages. On the other hand, the bipolarity requires an alignment tolerance less than 0.3°. For this reason, spherical magnets are mainly used on rotators as shown in Fig. 6.11(a). Tuning of alignments are made by polarity tests until no difference is found on strengths of bound state between two polarities. A cylindrical axially polarized magnet is also used in these tests (Fig. 6.14), while this magnet has also a small magnetic misalignment with respect to the geometric axis, this is neutralized by finding a proper point for the thread attachment where the misalignment angle is confined on the horizontal plane.

Presence of the gravitational force on floator could be advantageous or the opposite for the stability of the bound state. This cannot be determined by above analytical/numerical analysis since Eq. 6.30 is not applicable here. Within the general logic of this interaction, tilting the rotator (angle σ) varies the force $F_z$ in attraction direction and the tilt of the floator (angle $\varphi$) on repulsion. In this experiment configuration, weight of the floator adds up to repulsion, therefore this should be compensated by reducing the angle $\varphi$ or by increasing the angle σ. The angle $\varphi$ can be varied by varying MoI of the floator while having small effect on the magnetic torque that rotator experiences (Eq. 6.5) when $\varphi$ is small and consequently on the equilibrium angle σ according Eq. 6.6. Similarly, angle σ can be varied by the term ($I_A - I_R$). It can be said that both methods have negative impact on the stability of the equilibrium, since reducing angle $\varphi$ reduces also its variance (slope) versus distance and increasing the angle σ increases also its slope as can be seen on corresponding red lines in Fig. 6.12. Therefore, it can be said that the presence of gravity did not offered an advantage on obtaining stability in these experiments. As the bipolar bound state is succeeded in simulations (Fig. 6.13) in absence of gravity and damping, this gives sign that it can be also experimentally truly realized.

Simulations also show the contribution of lateral oscillations of bodies to the stability of equilibrium where the sum of magnitudes of these oscillations correspond to the omitted lateral offset $\beta$ in above equations. Through the PFR model, the role of lateral oscillations can be summarized as follow:

In the PFR model, the force (z-component) between floator and rotator is evaluated as sum of attractive and repulsive forces. Also within this model, when the rotating field is belong a dipole moment, this magnetic moment is separated to two components, the cyclic components lying on the rotating plane and the static component aligned with the rotation axis. At this point, the repulsion force is associated with the cyclic components and the attractive force with the static components. Similarly, the torque experienced by floator is also evaluated as a sum of the cyclic and static torques. Once, the amplitude of angular motion (angle $\varphi$) and the lateral translational motion (offset $\beta$) are determined within solutions of equations of motions, the component-z of force between bodies can be calculated with magnetostatics.

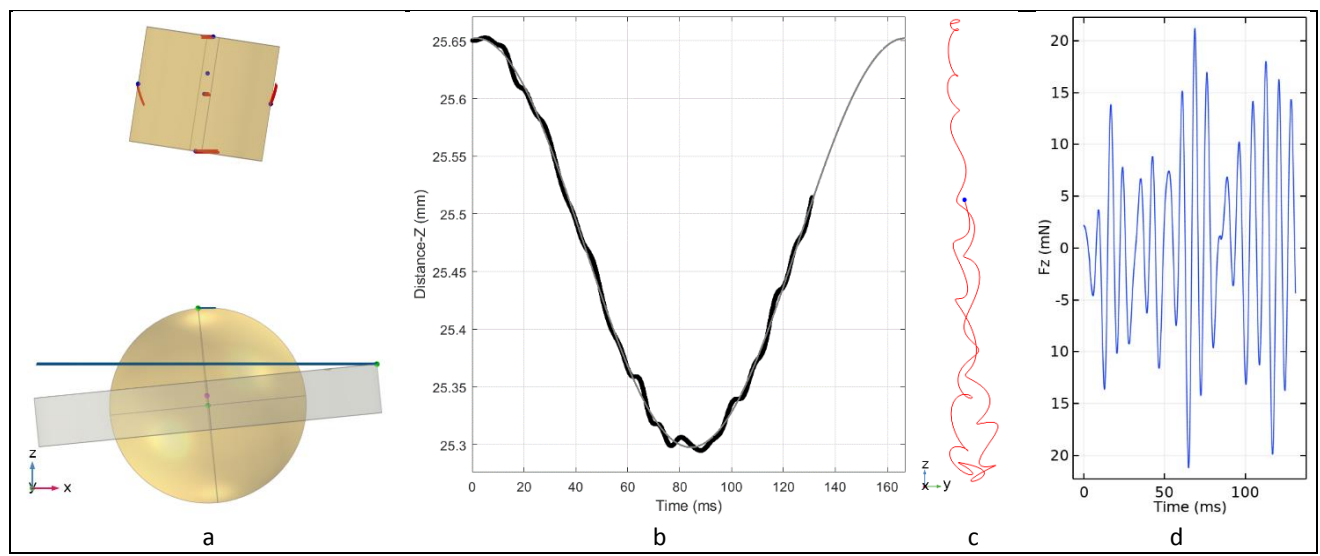

Fig. 6.13. Simulation of bipolar bound state of free bodies. Configuration parameters are $m_R$ = 20.56 g, $m_F$ = 6.666 g, $I_{R1}$ = 9.944e-7, $I_{R2}$ = 7.081e-7, $I_{F1}$ = $I_{F2}$ = 9.96e-8 kg m$^2$, ω = 857 rad/s where subscript letters R and F denote rotator and floator. Bodies are released near (25.65 mm) to the equilibrium position (25.47 mm), slowly get closer, bouncing at 25.3 mm while the simulation is ended before bodies return to their initial position. In a continuation of the simulation, the distance reached to 25.67 mm as maximum. The conical motions of bodies have some irregularities and a small drift (5e-5 m) in lateral directions due to non-precise initial and non-equilibrium conditions resulting in some fluctuations in forces. The thick horizontal line corresponds to circular trajectory (seen from edge) of a point at the edge of the ring in orientation of the rotator's N pole. The line thickens as the rotator goes forth and back on axis-$z$. (**a**) Overview of the simulation at the end (t = 131 ms). The magenta dot on the rotator marks its rotation center little above the CM (green dot). Rotation center of the floator is about 1.7 mm above the CM. (**b**) Plot of the distance in direction-$z$ between bodies by a superimposed sinusoidal curve defined as $f(x)$ = $a_0$ + $a_1$ cos ω$t$ + $b_1$ sin ω$t$, $a_0$ = 25.47, $a_1$ = 0.177, $b_1$ = 0.01048, ω = 0.03796. Based on this curve, peak acceleration of the floator is calculated as ±0.193 m/s$^2$, corresponding to forces ±0.00128 N. Due to irregularities in angular motions of bodies, actual forces varies by ±0.02 N in this course, more than 10 times than should be as shown in (**d**). (**c**) Trajectory of the floator's rotation center projected on the $yz$ plane. The amplitude of lateral force between bodies starts with 1.197 N, peaks as 1.265 N at the closest distance. Averages of angles $\varphi$ and $\sigma$ are 8.46° and 5.77°, respectively.

A remarkable point about the equilibrium of this bound state is the reciprocal roles of magnetic torques and moments of inertia on bodies. This can be summarized as strengths of attractive and repulsive factors are determined by the balance of the torque received by a body and its MoI where the balance on the rotating body determines the attractive factor and the balance on the floating body determines the repulsive factor.

As mentioned above, the distance between bodies are typically shorter in this bound state scheme compared to other schemes. This characteristics also reduces the effect of weight, since magnetic forces become larger at short distances. Another characteristics of this bound state is the low sensibility of the equilibrium distance to the rotation speed according to experimental observations. This has the effect of raising the upper limit of the speed range significantly while the lower limit is also raised compared to similar bound state configuration with fixed rotator tilt angle $\gamma$. This characteristics can be explained better with the PFR model in terms of attractive and repulsive factors. In bound states with fixed $\gamma$, attraction factor varies only by the floator tilt angle $\varphi$ though its cosine term where variation is negligible for small angles. Here, it also varies by the sine of the angle $\sigma$ which decreases by increasing the rotation speed according to Eq. 6.6 or Eq. 6.8. On the other hand, the repulsion factor mainly depends on angle $\varphi$ by its sine term which also decreases by increasing the rotation speed. In this bipolar scheme, attraction and repulsion factors vary approximately in parallel with respect to speed variation, allowing the equilibrium distance to be kept nearly constant. The presence of gravitational force on the floator in the opposing direction of the attraction is a factor for limiting the upper rotation speed according above dependencies and also observed experimentally. In absence of external forces like gravity, strength of the bound can be arbitrarily small and basically determined by the strength of the static field providing the attractive force. Since static field also provides the static torque, the minimum of the static torque required for a stable conical motion of the floator determines also the minimum of the static field in this case. However, when the floator has spin $\upsilon_S$, it can obtain stable motion also with zero or negative static torque as shown in Fig. 5.12. This option allows to extend parameters ranges of MBS and BBS. This option is not available on realized BBS experiments due to the presence of gravity.

Fig. 6.14 shows a realization of BBS where images (c) and (d) are taken using high shutter speed, corresponding two instances separated by half cycle of the rotator also shown in diagram (e) as superposition of these instances.

Since this scheme requires matching pairs of bodies, in these tests, this matching is obtained by adjusting properties (inertial, magnetic and weight figures) of the floators, by combination of stacked magnets. Tested configurations show that the rotating and floating bodies (rotator, floator) can have different proportionalities. While the floator can be small as 1/8 of the rotator in mass terms, it can be large as 4/3 as shown in Fig. 6.15(e). For small proportions of the floator, the tilt angle stiffness of the rotator should be small too in order to obtain enough tilt angle with the small magnetic torque induced by the floator. This stiffness is proportional to the term ($I_A - I_R$). For example, a cylindrical body having equal diameter and length gives an $I_A/I_R$ ratio as 7/6, which is close to unity by providing relatively low tilt stiffness. On the other hand, a rectangular block twice longer in radial direction than its axial direction gives a proportion high as 5/2, corresponding to a high tilt stiffness. With such a configuration, a BBS is obtained where the floator weights 4/3 of the rotator.

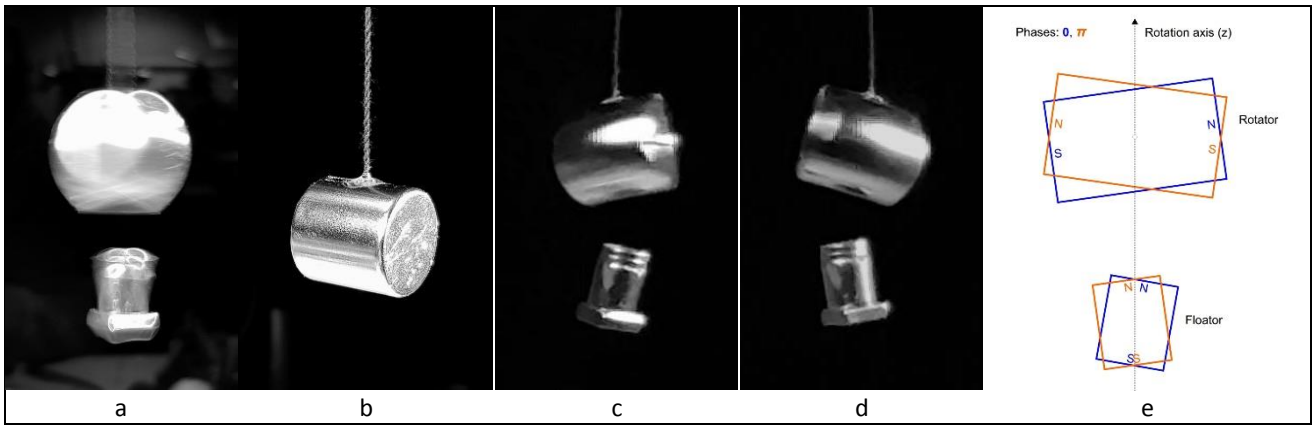

Fig. 6.14. Approximated BBS of two free bodies having dipole moment realized by permanents magnets. The upper magnet (rotator) having equal diameter and length of 20 mm is suspended by a thread attached to a rotor providing five DoF to approximate a free body (**b**). This geometry provides a larger MoI in radial directions of the cylinder corresponding to the rotation axis (*z*) of the motion by a ratio of 7/6, forcing the body to rotate on this axis, therefore orienting its dipole orthogonal to it. The lower magnet (floator) consists of four stacked pieces which allowed to adjust properties precisely, in order to obtain a stable equilibrium. The floator is held in air stably while exhibiting angular motion synchronized by the rotation of the rotator. Fig. (**a**) shows the bound state as seen by naked eye where rotation of rotator and oscillation of floator cause the motion blur. While the rotator is forced to rotate its dipole axis orthogonal to rotation axis by the inertial torque, the magnetic torque it receives from the floator cause to oppose this orientation, resulting in a tilt angle where these two opposing torques find an equilibrium. This tilt which can be seen from (**c**, **d**), produces a dipole moment in the direction of rotation axis which attracts the floator. On the other hand, magnets do not collapse because this attraction force is balanced by the PFR generated by the rotating magnetic field. However, this equilibrium can be only stable if magnetic and inertial properties of these bodies are adjusted well. From figures (**c**, **d**) which correspond two instances of motions of bodies taken by an interval of half cycle using camera 1/2000 sec. shuttle speed, it can be seen that their motions are synchronized and the floator is tilted from the vertical orientation, but its top pole pointing to the other magnet pole having the same polarity due to phase lag condition. BBS ensures that floator can be trapped regardless of the pole is facing to the rotator because rotator gets tilted automatically in order to generate a dipole moment parallel to the floator dipole. (**e**) These magnets are shown in two phases (red and blue) of rotation of the rotator by half cycle difference.

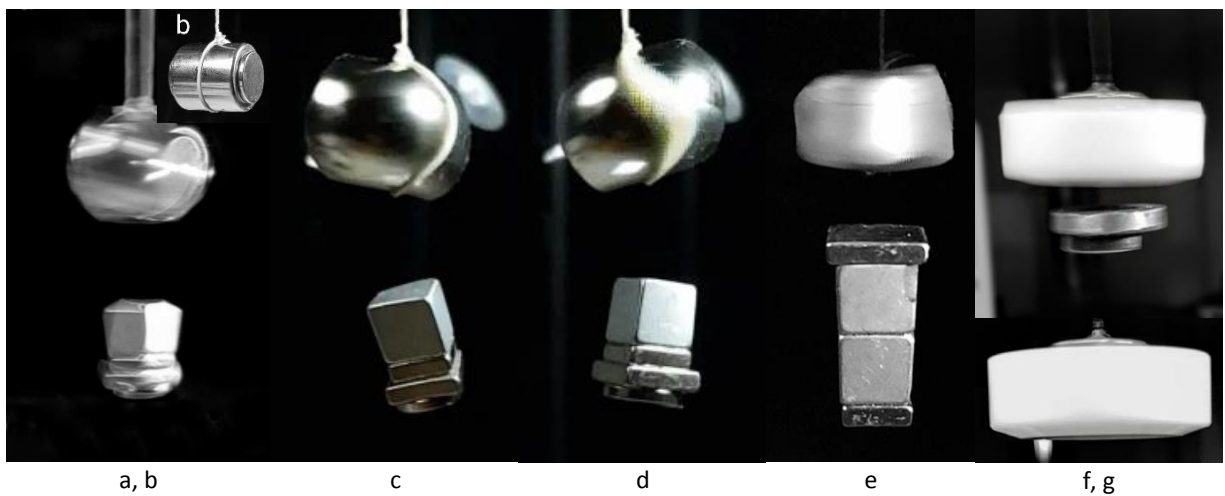

Fig. 6.15. (**a**, **b**, **c**, **d**) An approximated BBS configuration similar to Fig. 6.14 where the rotating magnet is extended axially by stacking thin disk magnets to both ends. This extension increases its MoI on the rotation axis approximately by 2.6 times and increases its tilt stiffness by this ratio. As a result, the required magnetic torque provided by the floating magnet needs to be increased by this ratio to obtain the same tilt. In this configuration, the rotating magnet is suspended by an elastic thread in order to approximate better full DoF. In this realization, a larger distance between bodies is also observed. Fig. (**c**, **d**) show two instances separated approximately by half cycle where the floating magnet performs a synchronized conical motion on the rotation axis of the upper body. (**e**) Another configuration where rotating assembly consists of two stacked cube magnets same as the ones used on the floating body which is heavier than the rotating body by 4/3 ratio. (**f**, **g**) A ring ((⌀19.7 – ⌀8) × 3.9 mm) shaped axially magnetized floator magnet having an adjustment mass is held at short distance by the rotator consisting a 1/2" cube magnet housed in a cylindrical block. This adjustment mass mainly has an effect on the moments of inertia for obtaining appropriate PFR slope vs distance for stable equilibrium. The material used in this block have density about 1.75 g/cm³ and provides rather a large MoI factor ($I_A - I_R$) resulting a small angle σ about 5° at ω = 220 rev/s and can be increased up to 10° by lowering the speed permitted by the stability in this configuration. Fig. **g** shows that the rotator does not wobble (σ = 0) in absence of the floator.

BBS could be energetically different from other types of magnetic bound states. This is because it may require to do a work to initiate the bound state. A simulation where the floator placed motionless centered on the rotator axis *z* and magnetic moment orthogonal to this axis, it keeps this orientation while gaining a kinetic energy from rotator as it involves in rotational and translational oscillations. This orientation is stable because it is enforced by the torque-phi effect. In this alignment only repulsion is present but no attraction. Therefore no bound state can be obtained unless one do some work for changing the floator orientation against torque-phi. Although a related simulation result was not conclusive, this energy profile point out presence of a potential well. It may also possible to describe the state of the rotator with shifted axis (with angle σ) as an excited state where only it exhibits the time averaged magnetic moment along its rotation axis.

## 6.3 Bound state of a body having quadrupole moment with a rotating dipole field

This scheme is similar to bound state of a body (floator) having a dipole moment but with the difference of quadrupole moment. Quadrupole field is obtained by two coaxial antiparallel dipoles called anti-Helmholtz configuration. This produces a polarity asymmetric field where there is a radial inner pole and two opposite outer poles in axial direction. Polarity is chosen in order for the inner pole to be attracted by the static component of the rotating field. First, we analyze the case without this static component and obtain the PFR. It should be reminded that similar to bodies having dipole moment, a body having quadrupole moment can experience PFR by interacting with an inhomogeneous cyclic field regardless of its position and orientation. When the body is centered on the rotator's axis in absence of external forces, it is observed that it involves in cyclic motions, basically in one angular DoF, on the plane obtained by the rotational field axis and the common axis of the dipoles (called common axis here). PFR is obtained through this motion. Amplitude of this motion can reach 30 degrees peak to peak and the center of the motion has an offset from CM in the direction of the rotation axis away from the rotating dipole, it is observed. The quadrupole body is subject to angular cyclic motion on this plane because it experiences cyclic torques from the dipole.

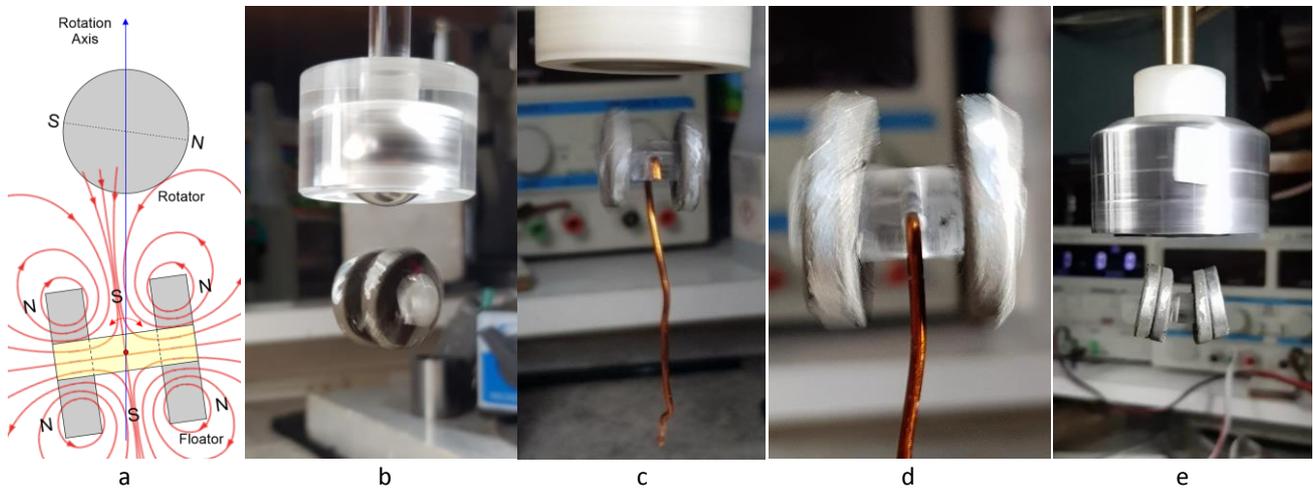

Fig. 6.16. Floating bodies having anti-Helmholtz type quadrupolar field bound to a rotating dipole field provided by three different rotator assemblies above them. This scheme generates single axis angular oscillation of floators allowing to a simple coupling that can isolate vibration. A close-up image of floator can be seen on rightmost image where its angular oscillation becomes apparent. The attached wire does not vibrate because its pass through the axis of the motion. (**a**) Positioning and magnetic diagram of the configuration on the *xz* plane at instance $t_0$. Floator magnets are arranged in anti-Helmholtz configuration and generates a quadrupole field. By calling next instances by each π/2 rotation of the rotator, at $t_1$ and $t_3$, the floator axis is aligned to axis-*x* and the diagram at $t_2$ is mirror of $t_0$. (**b**) A floator assembly consisting two axially polarized ring magnets (ring (⌀19.7 − ⌀8) × 3.9 mm) mounted on a plastic rod as anti-parallel with a distance of ~5 ⁻ 6 mm trapped by a spherical magnet (⌀3/4") mounted to a rotor. (**c, d**) The configurations of floator where the distance between ring magnets is smaller at the right configuration. (**e**) A similar configuration where floator magnets are doubled. The image is a slow motion video snapshot where the rolling shutter artefact deforms the shape of moving objects as the floator frame appears bend. At this instance, the rotator's N pole (marked by the reflective sticker) is in camera direction, and the floator frame is horizontal in accordance with the model.

Fig. 6.16 shows such realizations where two coaxial ring magnets are set in anti-Helmholtz configuration. In (a), an instance of the system is shown where another instance with a half cycle of difference corresponds to its mirror. Here, it can be seen the dipole axis of the rotator is not orthogonal to its rotation axis (*z*), but tilted by the angle γ which creates a static dipole having *N* pole looking downward. The floator is attracted to this dipole by its middle pole *S*. The cyclic torque it receives has only a component in direction through the paper (axis-*y*) due to symmetry of the quadrupole causing angular motion on axis-*y*. Due to phase lag condition of the DHM, at instances where one outer pole (*N*) of quadrupole experiences a repulsion, this pole is closer to the rotator pole *N* than its other outer pole which experiences attraction from rotator's pole *S*. This causes the quadrupole experiences a net repulsive force due the gradient of the rotating field. On the other hand, the quadrupole does not experience a torque on the rotation axis due components of two forces acting on outer poles on the rotation plane create torques in the opposite direction at any instant. These opposing torques are equal when the floator's common axis is on the rotation plane but differs otherwise. A full analysis is not provided here but according to observations, a possible angular motion due a residual cyclic torque should be one order magnitude less than the primary angular cyclic motion. The remaining angular motion around the common axis is also absent and the symmetry of the quadrupole with respect to the common axis does not permit a torque within this axis direction. Cyclic translational motions of the quadrupole are also visually not observed. Absence of translational motions on the rotation plane can be explained by the profile of total force that quadrupole experiences on this plane. Although a quadrupole field allows bodies to find stable equilibrium at the local minimum, here the trapping of the body is not related to this local minimum, rather can be explained by the PFR model. Experiments in this scheme are extended by variation of rotator and floator configurations including load carrying floators, different sizes of floator magnets, their gaps and rotator velocities. Ability of quadrupole floators to have single axis angular motion provides an advantage over dipole bodies having motion on two axes, hence this allows simple mechanical coupling in applications.

### 6.4 Bound state of rotating bodies having complex fields with static dipole fields

It is an ongoing research on obtaining bound state of rotating assemblies with static fields. One solution is rotating dipole within a static quadrupole field in anti-Helmholtz configuration (Fig. 6.17). However, not all requirements for obtaining a stable bound state have yet met.

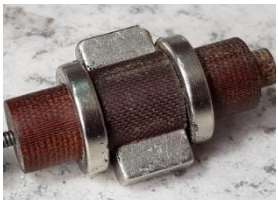

Fig. 6.17. A rotating assembly where a dipole magnet is inserted between two ring magnets in anti-Helmholtz configuration. The central magnet is polarized in thickness. When the assembly rotates it produces a rotating dipole field along a static quadrupole field. While the rotating dipole field generates PFR while interacting with a dipole field, the inner pole of the quadrupole provides the attractive force and the outer poles ensure angular stability.

### 6.5 Observational notes on magnetic bound state

It is observed that a floator under bound state receives a small torque from the rotating field on the rotation axis. In absence of a counteracting torque, this torque can spin the floator on its dipole axis in the direction of the field rotation. The cause of this torque is yet be determined but it is observed that presence of ferromagnetic material within the body significantly increases this torque while presence of high electric conductive material has minor effect. It is previously shown experimentally (Fig. 5.20 and Fig. 5.21(c) ) that a floator having unequal MoI in direction orthogonal to its magnetic moment prevents it spinning when its equilibrium position is off the rotating field axis. The spin ($v$) of the floator have some effects. One is variation of PFR strength. When the floator spins in same direction of the rotating field, the repulsion increases and decreases when body is forced to spin in the opposite direction by an external torque, for example using an air jet. This effect is observed within PFR tests and tests of bound states. In these tests, changes on PFR cause the variation of the equilibrium distance of the body according to the dependency of PFR on distance. This effect might be explained by the variance of amplitude of angular oscillation of the body (angle *φ*) with its spin, shown by Eq. 5.68. Fig. 6.18 graphically shows results of two series measurements, which may reflect PFR dependence on floator's spin. In these tests, each series has own floator configuration. Rotator configurations are common. Each measurement consists of pair of rotator angular velocity (ω) and floator spin velocity $v_S$ values satisfying the critical equilibrium condition. The test procedure is as follows: A rotator - floator configuration is chosen where floator can be trapped beneath the rotator, rotation axis aligned in vertical. The configuration allows the floator to be trapped in a range of rotation speed allowed by the electro-mechanical driving system, where the top speed can be at least twice the lower limit where the floator without a spin can be trapped critically, with near zero stiffness.

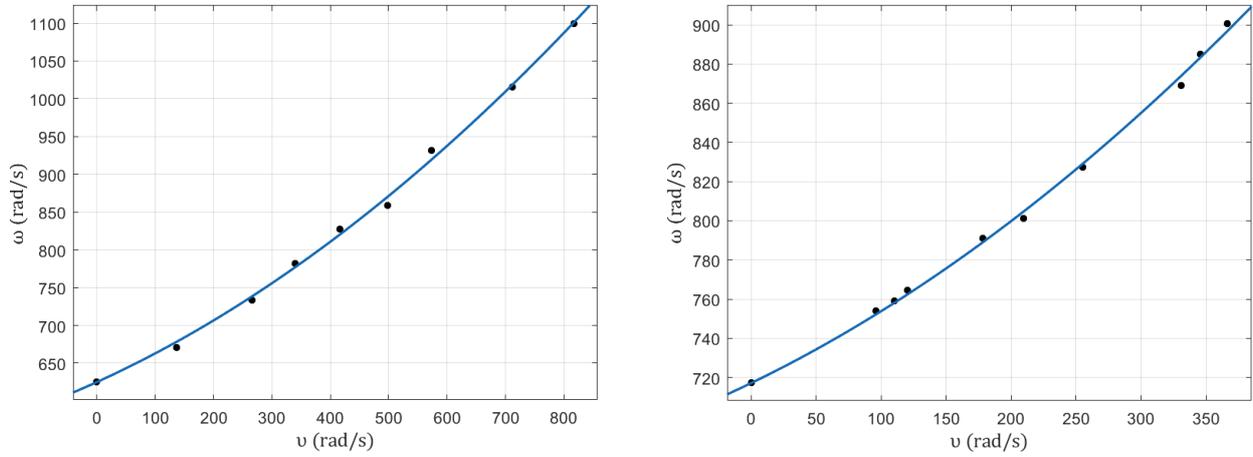

Fig. 6.18. Two series of measurements for showing relation of rotating field speed and floating body spin velocity $v_S$ (denoted as $v$ in this figure) for obtaining the same PFR strength. As the PFR is inversely proportional to square of the field rotation speed, It can be seen this relation is approximately inverse for floating body spin. First series corresponds to the bound state of a spherical magnet of 15 mm of diameter (m = 13.36 g, I = 3e-7 kgm²) and the second to a cylindrical magnet having both 10 mm diameter and length. (m = 5.64 g, $I_R$ = 8.225-8 kgm², $I_A$ = 7.05e-8 kgm²).
Series 1: $v$ = (0 136 266 340 417 498 574 712 817) rad/s, $\omega$ = (625 670 733 782 827 859 932 1022 1100) rad/s.
Series 2: $v$ = (0 96 110 120 178 209 255 331 346 367) rad/s, $\omega$ = (717 754 759 764 791 801 827 869 885 901) rad/s.

In the next run, rotator speed is set about 10% more. Increasing the speed decreases the repulsion force and causes the floator to get trapped more firmly this time. But this condition does not last too long since the floator progressively gains spin. The spin velocity is continually monitored and register the spin velocity at the moment floator is about be drop down. These tests are repeated until consistent measurements are obtained. Runs continue by increasing the rotator speed about 10% on each step. As a result, a set of pairs of values are registered consisting rotator velocity $\omega$ and floator spin velocity $v_S$. Each measurement satisfy following conditions:

$$F_A = F_R + F_G \tag{6.33}$$

On this equilibrium, the attraction force $F_A$ is balanced by the repulsion force $F_R$ and the floator weight $F_G$. The second condition is zero stiffness. That is, slopes of the $F_A$ and $F_R$ should be equal. This corresponds to a tangent condition of plots of these forces versus distance at this break-up distance $r_0$. This can be expressed as

$$dF_A/dr = dF_R/dr, \quad r = r_0 \tag{6.34}$$

If we can express these forces with power curves $a_i r^{b_i}$ where repulsion and attraction terms are denoted by indices (1, 2), it can be shown that this tangent condition can only be obtained when $F_G$ is not zero.

$$a_2 r^{b_2} = a_1 r^{b_1} + F_G, \quad a_2 b_2 r^{b_2-1} = a_1 b_1 r^{b_1-1} \tag{6.35}$$

This allows to write

$$\left(1 - \frac{b_2}{b_1}\right) a_2 r^{b_2} = F_G \tag{6.36}$$

where parameter $a_1$ is eliminated. This equation points out that repulsion power factor $b_1$ should be larger than the attraction $b_2$. We also know that $F_A$ is based on static interaction, therefore it is unlikely the power factor $b_2$ is affected by variation of $v_S$ or $\omega$. $F_A$ also depends on floator zenith angle $\varphi$ trough cosine term; however, this dependence is weak or negligible since angle $\varphi$ is smaller than 0.1 rad in these tests. In the trivial solution of this equation, all parameters are left constant; that is, they are not affected by variation of $v_S$ or $\omega$. This can be resulted when the angle $\varphi$ gets the same value for all ($\omega$, $v_S$) pairs. In this approach, the model which provides the equilibrium angle $\varphi$ in presence of spin is applied to this data using Eq. 5.69 and gives reasonable fits (>90%) but also points out a need for refine measurements or the model for a better fit. For example, in presence of an additional term as $(-\omega/3000)$ in the Eq. 5.69 as

$$-k\omega^2 = C\left(1 - \frac{\omega}{3000}\right), \qquad C = \frac{\tau_C c_\varphi + \tau_S s_\varphi}{I s_\varphi}, \qquad k = \frac{v_S}{\omega} - 1 \qquad (6.37)$$

the matches are improved for both data of both tests. In this case, the fitting value $C$ is 4.7e5 for first data and 6.63e5 for the second.

Another observation about spinning floators is a transitional reaction of the body to the crossing of its angular speed an integer fraction of rotating field speed. In these transitions, body performs a brief motion similar to finding a new equilibrium. The assembly generating the rotational field can also experience a jerk in this transition as a reaction. The ratio 1/2 is the final transition according to observations and body does not speed up further by this internal torque except in some cases including where body is a spherical magnet. On the other hand, it is possible to further accelerate the body equal to the rotating field speed and beyond by an external torque like applying an air jet. It is found that the speed of the body is kept locked to 1/1 ratio even after the external torque is removed in some configurations.

Under some circumstances, a body can lock to 1/1 speed ratio similar to the above case but rotating in the opposite direction as shown in realizations in Fig. 6.19(g) and Fig 5.21(b). This unique effect apparently requires negative torque in violation of conservation law. However, in these tests, it is observed that the body is not exactly aligned with the field rotation axis. It should be noted the effect is obtained with cylindrical bodies having length at least two times larger than its diameter.

Another remarkable related effect is subharmonic angular oscillation of the body while spinning around its dipole axis, but in some cases, it can happen without the spin. Typically, body draws a conical (angular) circular motion and the PFR is proportional to amplitude of this motion. This motion can be elliptic too in the same field conditions when body's moments of inertia orthogonal to the dipole axis are not equal (e.g., having rectangular profile) or CM does not coincide with the center of the dipole.

In some circumstances, a circular or an elliptical motion becomes epitrochoidal or hypotrochoidal on the fly (Fig. 6.19(a-f)). An epitrochoid or a hypotrochoid can be parametrically defined as

$$x(t) = \cos t + b \cos \frac{1+a}{a} t, \qquad y(t) = \sin t + b \sin \frac{1+a}{a} t \qquad (6.38)$$

where parameter $a$ can be set between $(0, 1)$ for the $x(t)$, and $(-1, 0)$ for $y(t)$. These motions mostly happen when body has a spin above a certain speed. The deviation from circular motion can develop slowly, sometimes, to up a minute to reach the final state. This oscillation might get synchronized with rotation of the field with a rate of an integer fraction of the field rotation speed mostly observed. Rates 1/7 up to 1/2 are observed. These patterns have loops count equal to the number of the inverse of the above rate. Loop count includes the main and the small loops. For example, a rate of 1/3 produces three loops, a main loop and two small ones. In some circumstances, the amplitude of these oscillations reaches close to $\pi$ peak to peak. It is remarkable that the bound state persists while the motion of the body is significantly altered although the binding becomes weaker, possibly due to the increase of the PFR. In this motion, the body spends most of its time in large angles of $\varphi$ and briefly passes from a narrow angle which corresponds to the small inner loop of the epitrochoidal pattern. In such an epitrochoidal pattern having one big and one small loop, the body returns to its initial axial orientation state after making two revolutions. It should be recalled that this angular motion belongs to the dipole axis while the body also spins typically around its dipole axis, hence this spin prepares the condition for the effect to be developed. This spin is typically not synchronized by the angular pattern but could be when the body is not axisymmetric. It should be noted that angular motion of the spin axis is different from precessional motion and has a different origin.

It is also observed through the audio noise, the harmonic content of the noise that the system produces changes even before large oscillations become visible. This noise caused by rotator assembly by the cyclic variation of force it receives from floator. While audio spectrum has peaks on harmonics of the frequency of rotating field before the change, sub harmonics are introduced than after.

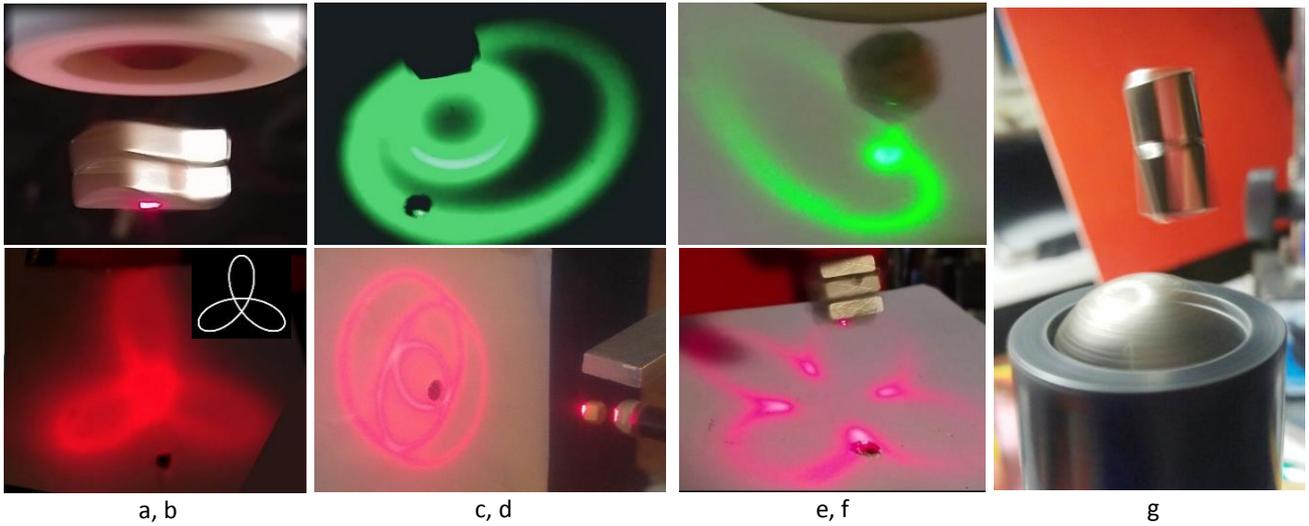

          a, b                       c, d                      e, f                      g

Fig. 6.19. (**a – f**) Various epitrochoidal/hypotrochoidal patterns of small angular oscillations generated spinning dipole bodies trapped in air. Patterns are obtained by reflection of a laser beam from a polar face of floating magnet. (**a, b**) A three armed hypotrochoidal angular pattern of a spinning rectangular floator. The small plot is obtained by settings $a = -1/3$ and $b = 0.82$ in the equation at footnote 8. (**d**) A tiny 1/8" cube magnet trapped in air in a position along axis of a rotation of a small magnet visible at right edge of the image and the reflection of laser beam emanating from the small hole on the screen reflected back from a face of the cube belong a pole. The pattern consists of one outer and two overlapping inner loops. On the center images, trapped magnets and their angular oscillations can be seen. They gave both single inner loops, large on the left image and point like on the right. At image (**f**), the pattern has four inner loops. The amplitude of the angular oscillation can be barely seen by X shaped light lines on the face of three stacked square small magnets. Here the lateral angle of the cross, which appears close to 30°, corresponds to peak-to-peak amplitude of oscillation. (**g**) A setup where the floator rotates in opposite direction of the rotator at same speed without an enforcement. This floator consist of stack of two cylinders of ⌀10×10 mm. The same effect is present in the experiment in Fig. 5.21(b).

It is found that epitrochoidal patterns can be present in all body shapes and in all body form factors. Since this effect also covers spherical bodies, the role of inequality of MoI or of torques induced by centrifugal forces can be ruled out. These patterns might be related to some resonances as they develop slowly and disappear (by returning to simple circular motion) when body is subjected to damping by presence of an electrically conductive block at proximity.

A large angle epitrochoidal motion of a trapped body is sequenced on Fig. 6.20 where angle $\varphi$ varies between 45 and 90 degrees. This pattern consists of two loops, similar to the pattern in Fig. 6.19(c) where body spends most of its time in large angles. Body also rotates around its dipole axis in a synchronized manner giving quite interesting visuals in slow motion. It should be noted that motion starts with a conical motion with small $\varphi$ without spinning and this motion develops while body obtains the required spin by itself and locks to spin ratio.

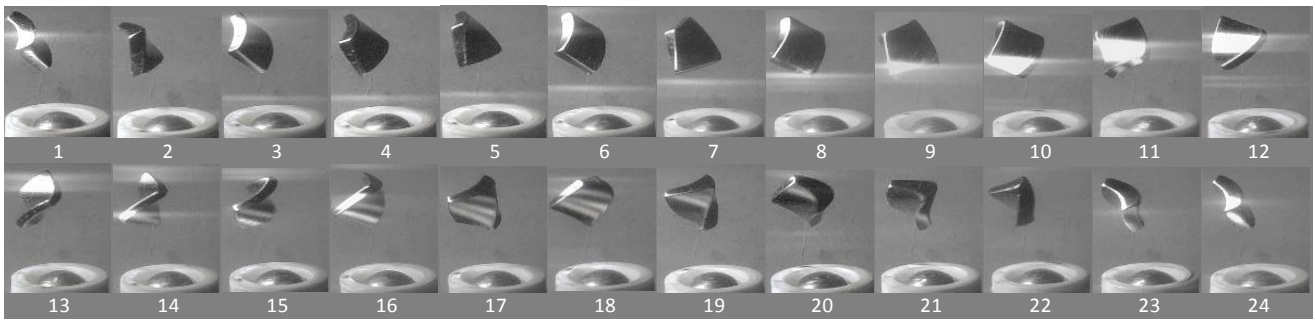

Fig. 6.20. A sequence of video frames of spinning a square floating magnet (3×10×10 mm) under an epitrochoidal motion having two loops. Each row corresponds to one cycle of rotating spherical magnet used for trapping the above magnet in the air. While frames are taken by 66 ms apart, each corresponds to motion advancement of about 4 milliseconds and by 31 degrees of rotation of driving magnet spinning 240 rev/sec. According to the epitrochoid geometry, large angles correspond to the outer loop and small angles to the inner loop. Frames also show the floating magnet is also spinning around its dipole axis, however, not synchronized with its epitrochoidal pattern. In many cases, as occurs in this experiment, the floating magnet position is off the main axis. This can be relied to the torque-phi effect which enforces the position for the average dipole moment direction of the body while the static dipole moment of the rotating field which tries to keep the body on axis becomes less effective within large angular oscillation of the body. The deformed appearance of the floating magnet is caused by the rolling shutter effect of the camera which scans the frame from top to bottom.

It is also observed that orientation of the spin axis of bodies can vary in a cyclic manner with amplitudes up to 70°. While this can be the result of inertial forces induced by spin, it is remarkable that it has no impact on stability but only a shift on the equilibrium distance in the direction of weak field. This effect may get combined with above cycloid motions, generating further highly complex motions as seen in Fig. 6.21.

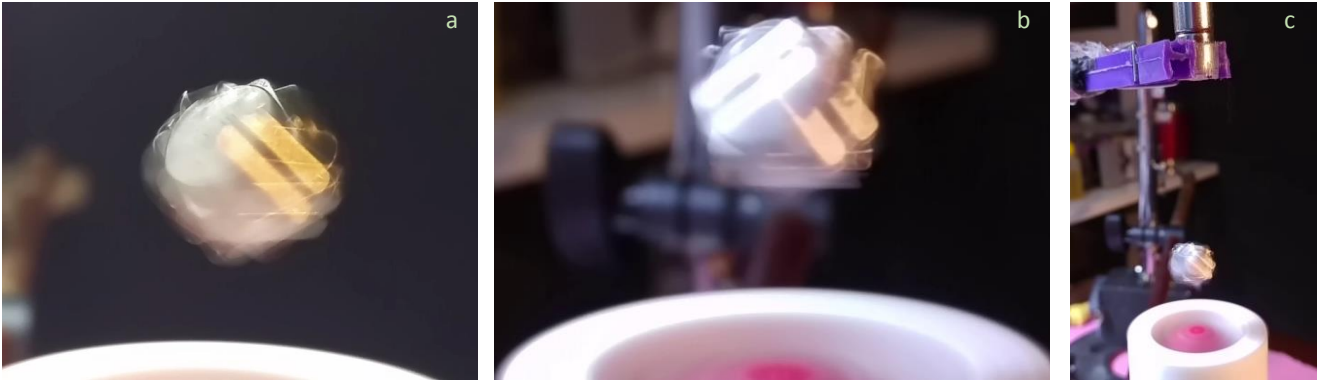

Fig. 6.21. Complex angular motions of floators involving large variations of angle $\varphi$ while they spin around their dipole axes. Floator is a stack of two square 10×10×3 and one ⌀9.5×1 magnets. This setup includes an overhead magnet assembly consisting of a stack of two ⌀10×10, visible on the top of (**c**) which allows complex motions of trapped magnets. Pictures covers a frame about 1/30 second where the floators perform several cycles of motion in this time window. Floators can ever reach to larger variations of $\varphi$ by increasing the rotation speed of the motor. Dimensions are in mm.

Another note is the availability of extra repulsion strengths in certain circumstances. It is observed that this happens when body is spinning on an axis having a large angle with its dipole axis. Actually, repulsion strength is normally a small fraction of the instantaneous force acted on the body; therefore, there is a room for increasing it. Repulsion varies with amplitude of angular oscillation of the body also by its phase. When the body spins on out of its dipole axis, the phase relation becomes difficult to predict and further tests are needed to understand the cause of this extra repulsion.

## 6.6 Realizations of PFR and Magnetic Bound State

While PFR and MBS can be obtained by a rotating dipole field using a single permanent magnet, it is also possible to obtain it by an arrangement of multiple permanent magnets. Such an arrangement should have a dipole or quadrupole moment and the axis should take place between poles. These include Halbach arrays, a stack of two or more magnets split by a gap, off-centered magnets and may by inclusion of magnets oriented in axial direction to adjust the profile of the attractive forces. By these methods, it is possible to adjust separately the torque, repulsive and attractive forces in radial and in axial profiles.

Under terrestrial conditions, MBS should have enough strength to anticipate the weight the floator. This restricts the gap length. By using NdFeB magnets, it is observed that the maximum gap is close to twice the size of the magnet providing the rotating field for magnet sizes less than 15 mm. We can use the diameter as the size for spherical magnets in the description above. For other shapes, it is possible to approximate the diameter by the equivalent volume. As the strength of the repulsion diminishes by distance faster than static interaction between dipoles, extending the floating gap is expensive. One possible way is reducing the power factor of the repulsion by reducing the slope of torque strength to distance. As the torque does not require field gradient, it is possible to provide a distance independent torque by a gradient free cyclic field. There are also other mechanisms based on static fields but in general, extending the equilibrium distance by this way costs in stability.

The difference between slopes of attractive and repulsive forces at the equilibrium point (where external forces might be present too), determines the stiffness on a DoF (Fig. 6.4(b, d)). It may be possible to alter the slope of attractive force by modifying the geometry of the rotating field. One way is introducing a dipole aligned with the rotation axis and fixing it behind the main dipole with respect to the floating body. This way, the slope of attractive force can be reduced at the equilibrium point in rotation axis direction. This, however, deforms the equilibrium surface surrounding the field and may affect stabilities or stiffnesses on other DoFs. As a note, it is also possible to place such a magnet providing attraction at the front of the magnet providing the cyclic field. This way, the slope of the attraction force will be increased but still can be less than the slope of the repulsion force profile, the required condition for the stability of the equilibrium between these forces.

Generally, the floating magnet (dipole body) is kept close to the rotation axis and anticipates external or inertial forces in radial and axial directions. Certain shapes belonging to the rotating field provide better strength against radial forces, allowing bound state in every orientation.

It should be also noted that a floator can be trapped in liquids in various viscosity and densities like it performs in the air. The Fig. 6.22 shows such a realization where the floating magnet is held at a distance about 15 mm using relatively small rotating magnets and anticipating gravitational forces received from arbitrary directions.

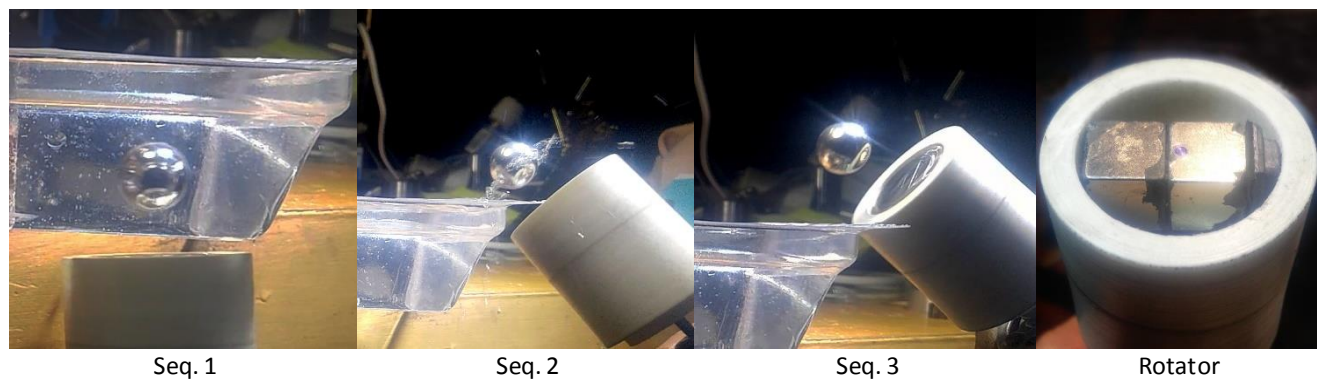

Seq. 1    Seq. 2    Seq. 3    Rotator

Fig. 6.22. Sequences from a test about capturing a spherical dipole magnet (⌀15 mm) by a rotating dipole magnet assembly from a container filled with water, initially staying at the bottom, raising it and taking out without making contact with the container. The rotating assembly consists of two stacked 1/2" cube magnets grade N50 having dipoles aligned to the rotation plane, placed asymmetrically with respect to the rotation axis inside a plastic cylindrical housing having ⌀29.6 mm inner diameter as shown on the right picture. The small circle in violet color on the right magnet marks the rotation axis. The attraction factor is obtained solely by this offset in this configuration since the dipole of these stacked magnets lies on the rotation plane.

It is found that there is no scaling limit on the MBS. It is realized with 2 mm to 80 mm diameter actuator magnets and with 1 mm to 35 mm diameter magnets used for floators. While the gap between magnets is typically proportional to diameter of the actuator magnets, minimum rotation frequency for stable operation is found inversely proportional to a figure between square root and cube root of floating magnet mass. Additionally, scaling down a system improves the anticipation of the bound state against static and dynamic external inertial forces, like floator weight or jerks on the system since inertial figures decrease faster than magnetic force figures by taking account of the reduction of bound distance parallel to the scaling.

## 6.7 Notes on realization of experiments

MBS can be obtained in an easy way by trapping a dipole NdFeB magnet by the rotating field of another magnet of the same type. This magnet is called *actuator magnet* here. The dipole axis of the actuator magnet should be off the rotation plane by a small angle called tilt angle γ. This angle can be set between 4° and 6° which corresponds to relatively weak attraction forces in order to obtain equilibrium easily in most of the cases. Increasing the tilt increases the attraction force and beyond 10°, it may become difficult to obtain stability depending on other parameters. After 15° (0.26 rad) it was not possible to obtain stability in any tested configuration. Similarly, when attraction force is obtained by shifting the rotating dipole center from the rotation, this shift also has limits. It is found that it becomes difficult to obtain stability when the shift is larger than 1/6 of the dipole length and the maximum is 1/4 found so far. Combination of tilt and shift schemes allows to shape the profile of the rotating field, its static and cyclic components and their gradients with more options and the above limits can be extended. Another method often used to obtain the attractive force required for the bound state is placing a second smaller magnet front or behind of a symmetrically placed primary magnet with dipole parallel to the rotation axis. It is found that the stiffnesses can be enhanced in certain configurations by the shift method described above and evaluated in Section 6.0.4.

The stiffness of MBS and the equilibrium position can vary in large extents within configurations. Some configurations are suitable for specific orientations of gravity and some having better stiffness figures are suitable to receive the gravitational force in any direction.

One should be aware that commercial grade magnets can have significant magnetic misalignments and inhomogeneities; therefore, the apparent tilt of a magnet does not always correspond to the tilt of its dipole.

About choosing the shape and form factor of magnets, sphere is a good choice both for handling, for aligning the CM with the rotation axis, for finding equilibrium and for its mathematical model. However, pole locations should be marked precisely. Field profile of a spherical magnet might not be the optimum profile for keeping a floating magnet close to the rotation axis obtaining stability in some conditions.

Real or instantaneous forces between actuator magnet and floating magnet can be very strong when they are getting close to each other, for example, when the gap between magnet boundaries is less than 1/2 of the actuator magnet diameter. These strong forces and torques may cause vibrational stress on the driving mechanism of the actuator magnet. There are various ways to reduce this stress. One way is increasing the inertia of the actuator assembly. Another is driving the actuator assembly by a small or micro motor and giving DoF to vibrate them together without causing stress. In addition, vibration isolation methods can be applied.

It is also possible to reduce radial forces acted on an actuator assembly by shifting its CM from the rotation axis in a proper angle in order the force caused by the eccentric motion works against the magnetic force. It is important to have a good rotational balance on actuator assemblies. It is observed that vibrations caused by an unbalance can drive floating magnet to destabilizing resonances and prevent to obtain high rotation speeds. Many mechanical losses on a system are accompanied by vibrations and noises. Therefore, a system with less noise means less mechanical losses.

# 7 Summary and concluding remarks

Polarity free magnetic repulsion (PFR) is a net short ranged force generated by a dynamic magnetic interaction based on undamped driven harmonic motion, demonstrated by experiments, mathematical methods and simulations. Because it provides positive stiffness in total of DoF, it can be used to obtain stable equilibrium between a body endowed with a magnetic moment and a field having both static and cyclic components, whilst this is not possible with a static interaction alone according Earnshaw theorem. PFR is found as a short-range force which is inversely proportional to seventh to eighth power of the distance between interacting dipole moments and present regardless of positions and orientations. As it can be obtained primarily by angular motion, bodies can be held almost at fixed positions while experiencing this force. As a result, various schemes are experimentally realized including trapping magnetic bodies with static fields, trapping with fields having no static components and bound states between compact bodies comparable to other bound states found in nature by versatility, endurance, and energy conservation criterion. PFR might be also considered as a pseudo-conservative force; that is, acts like a conservative force under adiabatic condition.

Experimentally, some configurations allowed trapped bodies to anticipate static and dynamic external forces like ones induced by gravity, by acceleration or by external magnetic fields from arbitrary directions and also allowed dynamic environment changes like submerging in liquids with various density and viscosities. Some trapping schemes allowed levitation height up to seven cm. Solutions are found scalable, it was possible to trap bodies having masses varying between 0.01 to 100 g. Bodies can also have arbitrary geometries and arbitrary magnetic alignments. Under certain circumstances, it is observed that trapped bodies involve complex and large angular oscillations where the body's dipole moment may span almost a hemisphere.

Magnetic bound state (MBS) is a solution inheriting characteristics of PFR to bind an inertial body endowed with a magnetic moment and having full DoF to an inhomogeneous cycling field. This cyclic field can be obtained by rotating a magnetic dipole on a fixed axis or by a rotation of a free inertial body having magnetic moment. Such solutions are analytically analyzed and obtained through various experiments and simulations where their characteristics are explored. The stability of the PFR and MBS is evaluated by experiments with various configurations and by simulation series and by some numerical analyses and found that stable systems can be obtained with wide ranges of parameters, their stabilities can tolerate significant variation of parameters.

Systems also found to anticipate static and dynamic forces. Through these results, MBS might be considered versatile in contrast to engineered or to finely tuned solutions. In this regard, MBS is comparable to other bound states found in nature. Compared to the orbital motion, translational motion of bounded bodies under MBS are very limited and this apparent lack of motion might conceal a possible natural occurrence of the mechanism. On the other hand, this limited motion characteristics could be valuable for building structures and for developing applications. MBS can anticipate external forces, dynamic and static. The characteristics of profile of the MBS versus distance shows that it can also anticipate external forces obeying inverse square law, attractive or repulsive. While the former is not critical, the latter reduces the depth of the potential well, therefore limits the strength of this external force can be applied without breaking the bound. This case covers electrically charged bound bodies having same polarity.

Another remarkable MBS solution is the bipolar bound state of free bodies which have no polarity preferences while the polarity symmetry is broken at the establishment of the bound state.

Overall, a realization demonstrating one or more of these solutions or characteristics can be built using components costing less than $100.

It may be worth to mention here a similarity between profiles of attractive and repulsive forces under MBS and the Lennard-Jones potential (LJ) [28], an empirical formulation for interaction between neutral atoms and molecules. LJ potential energy is defined as $V_{LJ}(r) = c_R r^{-12} - c_A r^{-6}$ where $c_R$ and $c_A$ are repulsion and attraction coefficients for an interaction and $r$ is the distance between entities. For the MBS which having power factors $p_{FA}$ and $p_{FR}$ for attraction and repulsion forces, the potential energy can be written as

$$V_{MBS}(r) = \frac{c_1}{p_{FR}+1} r^{p_{FR}+1} - \frac{c_2}{p_{FA}+1} r^{p_{FA}+1} \tag{7.1}$$

where $c_1$ and $c_2$ are some coefficients. For $p_{FR} = -7$ and $p_{FA} = -4$, potential energy power factors become $p_R = -6$ and $p_A = -3$, which are half of the LJ power factors and the ratio $p_{FR}/p_{FA}$ is retained. As a note, in a recent publication [29] it is stated that the LJ repulsion exponent $p = -6.5$ suits well for liquid Argon while $p = -12.7$ fits to gaseous form.

Finally, author would like to mention works of A. O. Barut who made progress in unification of quantum and classical electrodynamics, including nuclear forces [30] and spin [31]. The zitterbewegung, the theorized intrinsic motion of electron and its angular kind [32] might be useful to carry magnetic interactions explored in this work to subatomic physics.

## Data availability

All data of simulations presented on this study and additional visual materials about realizations are available from the author on request.

## Acknowledgements

The author wishes to thank Prof. J. L. Duarte and Prof. M. D. Simon for their support and advises.

## Author information
**Affiliations**

Independent Researcher (The author has no affiliation with an institution)

**Contributions**

H. U. conceived the idea, performed experiments and the analyses and wrote the manuscript.

**Competing Interests**
The author declares no competing interests.

**Corresponding author**
Correspondence to Hamdi Ucar (jxucar@gmail.com)